\documentclass{jfm}
\usepackage{graphicx}
\usepackage{epstopdf, epsfig}
\graphicspath{ {./figure/jpeg} }

\usepackage{amssymb}
\usepackage{latexsym}
\usepackage{xcolor}
\usepackage{amsmath,mathtools}
\numberwithin{equation}{section}
\usepackage{lineno}
\modulolinenumbers[5]
\usepackage{bm}
\usepackage{multirow}

\newtheorem{definition}{Definition}

\shorttitle{Numerical linear stability analysis of wake vortices}
\shortauthor{S. Lee and P. S. Marcus}

\title{Linear stability analysis of wake vortices by a spectral method using mapped Legendre functions}

\author{Sangjoon Lee\aff{1}
 \and Philip S. Marcus\aff{1}
  \corresp{\email{pmarcus@me.berkeley.edu}}}

\affiliation{\aff{1}Department of Mechanical Engineering, University of California,
Berkeley, CA 94720, USA}

\begin{document}
\maketitle

\begin{abstract}
A spectral method using associated Legendre functions with algebraic mapping is developed for a linear stability analysis of wake vortices. These functions serve as Galerkin basis functions, capturing correct analyticity and boundary conditions for vortices in an unbounded domain. The incompressible Euler or Navier-Stokes equations linearised on a swirling flow are transformed into a standard matrix eigenvalue problem of toroidal and poloidal streamfunctions, solving perturbation velocity eigenmodes with their complex growth rate as eigenvalues. This reduces the problem size for computation and distributes collocation points adjustably clustered around the vortex core. Based on this method, strong swirling $q$-vortices with linear perturbation wavenumbers of order unity are examined. Without viscosity, neutrally stable eigenmodes associated with the continuous eigenvalue spectrum having critical-layer singularities are successfully resolved. The inviscid critical-layer eigenmodes numerically tend to appear in pairs, implying their singular degeneracy. With viscosity, the spectra pertaining to physical regularisation of critical layers stretch out toward an area, referring to potential eigenmodes with wavepackets found by \citet{Mao2011}. However, the potential eigenmodes exhibit no spatial similarity to the inviscid critical-layer eigenmodes, doubting that they truly represent the viscous remnants of the inviscid critical-layer eigenmodes. Instead, two distinct continuous curves in the numerical spectra are identified for the first time, named the viscous critical-layer spectrum, where the similarity is noticeable. Moreover, the viscous critical-layer eigenmodes are resolved in conformity with the $\Rey^{-1/3}$ scaling law. The onset of the two curves is believed to be caused by viscosity breaking the singular degeneracy.
\end{abstract}

\begin{keywords}
% Authors should not enter keywords on the manuscript, as these must be chosen by the author during the online submission process and will then be added during the typesetting process.
% (see http://journals.cambridge.org/data/\linebreak[3]relatedlink/jfm-\linebreak[3]keywords.pdf for the full list)
\end{keywords}

\section{Introduction}
\subsection{Background}
After the introduction of heavy commercial aircraft in the late 1960s, wake vortex motion has been a major subject of flow research, which has been reviewed in several studies \citep{Widnall1975, Spalart1998, Breitsamter2011}. The focus has often been on the destabilisation of wake vortices to alleviate wake hazards \citep[see][]{Hallock2018}. The demise of wake vortices typically starts with vortex distortion, which causes either long-wavelength instability, well known as the Crow instability \citep{Crow1970, Crow1976}, or short-wavelength instability, known as the elliptical instability \citep{Moore1975, Tsai1976}. Both mechanisms are affected by vortex perturbation induced by the strain from the other vortex. The internal deformation of vortex cores, often interpreted as a combination of linear perturbation modes, plays a key role in the unstable vortex evolution process \citep{Leweke2016}.

Since Lord \citet{Kelvin1880} studied the linear perturbation modes of an isolated vortex using the Rankine vortex, analyses have been extended to other models that better describe a realistic vortex and account for viscosity with continual vorticity in space. The Lamb-Oseen vortex model can be considered as an exact solution to the Navier-Stokes equations, while assuming no axial current in vortex motion.
\citet{Batchelor1964}, however, claimed the necessity of significant axial currents near and inside the vortex core for wake vortices and deduced vortex solutions with axial flows that are asymptotically accurate in the far field. The intermediate region between the vortex roll-up and the far field may be better described by the model proposed by \citet{Moore1973}, where \citet{Feys2014} performed a linear stability study. However, the Batchelor model has frequently been taken as a base flow for linear instability studies \citep{Mayer1992, fabre2004, LeDizes2005, Heaton2007, Qiu2021}, with the support of experimental observations \citep{Leibovich1978, Maxworthy1985}. As for the Lamb-Oseen vortex, an exhaustive study on its linear perturbation was performed by \citet{Fabre2006}. \citet{Bolle2020} conducted comprehensive linear analyses of all these vortex models and concluded that linear vortex dynamics is insensitive to changes in the base flow for singular modes.

In the numerical literature, a study by \citet{Lessen1974} used a shooting method to find eigensolutions of swirling flows, where the eigenmode represents the spatial profile of the linear perturbation, and the eigenvalue corresponds to its complex growth rate in time. \citet{Mayer1992}, on the other hand, utilised a spectral collocation method with Chebyshev polynomials to generate a global matrix eigenvalue problem in the generalised form ($\mathsfbi{A}\bm{x} = \lambda \mathsfbi{B} \bm{x}$). Although a shooting method may be accurate and less likely to yield spurious modes due to numerical discretisation \citep[pp. 139-142]{Boyd2000}, a spectral collocation method should be preferred, especially when there is no initial guess, and the overall comprehension of the whole eigenmodes and eigenvalues is required. \citet[pp. 335-339]{Heaton2007} compared these two numerical methods while investigating the asymptotic behaviour of inviscid unstable modes due to the presence of a core axial flow. Several recent studies \citep{Fabre2006, Mao2011} reported the use of spectral collocation methods to obtain a bulk of the eigensolutions at once to investigate and classify their common properties.

Given no viscosity $(\nu \equiv 0)$, there are regular eigenmodes (Kelvin modes) in association with discretely distributed eigenvalues, and non-regular eigenmodes with critical layer singularities (critical-layer eigenmodes), which occur at radial locations where the perturbation phase velocity is equal to the advection of the base flow \citep{Ash1995, Drazin2004}, or equivalently, where the coefficients of the highest derivatives of the governing equations go to zero \citep{Marcus2015}. The critical-layer eigenmodes arise from the non-normality of the governing equations (i.e., non-commutativity with the Hermitian adjoint) and are associated with continuously distributed eigenvalues in the complex plane, which are known to be significant in transient growth \citep{Mao2011, Mao2012, Bolle2020}. Throughout this paper, we refer to the region where critical-layer eigenvalues exist as a non-normal region. Note that this is in line with the quantitative definition of non-normality using the resolvent formalism by \citet{Bolle2020}, who referred to non-normality as the region where the resolvent norm of the operator representing the governing equations does not meet its lower bound. The inviscid critical-layer eigenvalues are distributed on the imaginary axis, exhibiting their neutral stability pertaining to the time symmetry in the problem \citep[see][]{Gallay2020}.

However, adding even a small amount of viscosity $(\nu \rightarrow 0^+)$ breaks this symmetry and leads to the viscous damping of eigenmodes in time \citep{Khorrami1991}. Spatially, non-zero viscosity regularises the critical-layer singularities but simultaneously triggers new singularities at radial infinity as a result of the total spatial order increase of the governing equations \citep[see][p. 268]{Fabre2006}. The impact of viscosity on the formation of viscous eigenmodes is an active research area, especially in the non-normal region. As viscosity is close to zero, the discrete spectrum becomes less prevalent in the complex plane while the non-normal region expands \citep[see][p. 11]{Bolle2020}.

\subsection{Research goals}
Our primary goal is to develop an efficient numerical method for linear stability analysis of a wake vortex using eigenmode-eigenvalue theory (or spectral theory). We carefully investigate the mathematical foundation of the method to ensure accurate and correct resolving of eigenmodes and eigenvalues. We then apply our numerical method to linear stability analysis of the Lamb-Oseen or Batchelor vortex model to classify eigenmodes in terms of physical relevance, which is an additional goal for the rest of this paper. Our work demonstrates the numerical efficiency of our method, which is essential as we plan to extend the application of the method to handle hundreds of eigenmodes for the examination of triad-resonant interactions among the eigenmodes in a follow-up paper, and encompasses the linear stability analyses under either inviscid or viscous conditions.

Our numerical work is based on a spectral collocation method. It follows the typical global eigenvalue problem-solving procedure like that of \citet[pp. 127-133]{Boyd2000} and \citet[p. 241]{Fabre2006}. However, our method is distinguished because of its use of algebraically mapped associated Legendre functions as Galerkin basis functions, which were introduced by \citet{Matsushima1997} and utilised in several vortex stability studies \citep{Bristol2004, Feys2016}. These functions are defined as
\begin{equation}
    P^{m}_{L_n} (r) \equiv P^{m}_{n} (\zeta(r,L)) = P^{m}_{n} \left( \frac{r^2 - L^2}{r^2 + L^2} \right) ~~(L>0),
    \label{maplegfun}
\end{equation}
where $P^m_n$ is the associated Legendre function with order $m$ and degree $n$ \citep[see][pp. 292-295]{press_numerical_2007}, $\zeta$ is a variable in the interval $-1\le \zeta <1$ to which the radial coordinate $r$ is mapped by the map parameter $L$. Note that $\left\{P^{m}_{L_n} (r) \; \big| \; n = |m|, \; |m|+1, \cdots \right\}$ is a complete Galerkin basis set, and a regular function approximated by their truncated sum converges exponentially with respect to the truncated degree $n_{\max}$.

Radial domain truncation is not required in our numerical method as it is designed for an unbounded radial domain. Other methods that require a radially bounded domain typically use a large truncation point to mimic unboundedness and impose arbitrary boundary conditions \citep{Khorrami1991, Mayer1992, Mao2011, Bolle2020}. Moreover, our use of Galerkin basis functions eliminates the need for such explicit boundary conditions, reducing numerical error and eliminating further treatments for boundary conditions \citep[see][]{Zebib1987, McFadden1990, Hagan2013}. The distribution of numerical eigenvalues (numerical spectra) is necessarily discrete due to numerical discretisation, even if the analytic spectra are partially continuous. To deal with this seeming paradox, we investigate how the numerical spectra change with respect to the numerical parameters, including the map parameter $L$, the number of radial basis elements $M$, and the number of radial collocation points $N$. To complement the numerical spectra's inability to portray analytically continuous regions, we also briefly consider pseudospectral analysis \citep[see][]{trefethen_spectra_2005}.

We are particularly focused on eigenmodes with a critical layer that has received little attention in previous numerical studies due to the difficulty of resolving them. Under the inviscid condition, the critical layers are mathematical point singularities. The critical-layer eigenmodes are associated with a continuous spectrum on the imaginary axis in the non-normal region. However, numerical discretisation often produces incorrect eigenvalues that appear as symmetric pairs around the imaginary axis \citep[see][]{Mayer1992}. We show that our spectral collocation method can correct these results by properly adjusting the numerical parameters to reveal the inviscid critical-layer spectrum. With non-zero viscosity, there are areal spectra that emerge in the non-normal region \citep[see][]{Mao2011}. These spectra have yet to be fully explained and may contain an unforeseen continuous spectrum. We demonstrate that our method is capable of resolving the eigenmodes associated with this unexplored spectrum, which results from the regularisation of critical layers, from the other eigenmodes in the non-normal region.

\subsection{Preliminary remarks}\label{sec:prelim} 
To classify our numerically computed eigenmodes and eigenvalues, we frequently use the following terms in the rest of the paper. Note that some of our definitions may deviate from those used by other authors.
\begin{enumerate}
    \item By ``\textit{physical},'' we mean that a ``\textit{spatially resolved}'' eigenmode (as defined below) is a non-singular solution to the linearised governing equations in an unbounded domain when computed with non-zero viscosity. An eigenmode computed with zero viscosity (i.e., with $\nu \equiv 0$) is not the target of consideration but may have physical significance if the eigenmode and eigenvalue correspond clearly to a ``\textit{physical}'' eigenmode and eigenvalue computed with small but finite viscosity (i.e., in the limit of $\nu \rightarrow 0^+$). This condition is important because we are ultimately concerned with eigenmodes that can exist in the real world, such as those that would destabilise an aircraft wake vortex. Viscous eigenmodes are generally non-singular because viscosity can regularise them; numerically computed inviscid eigenmodes that would otherwise be singular are regularised by numerical discretisation. Our numerical method aims to resolve small-scale radial structures (e.g., the viscous remnants of inviscid critical layer singularities) purely resulting from physical (not numerical) regularisation, and we are interested in identifying such ``\textit{physical}'' eigenmodes.
    \item By ``\textit{non-physical},'' we mean that a ``\textit{spatially resolved}'' eigenmode does not meet the conditions described above for being considered ``\textit{physical}.'' Any numerically computed eigenmode must first be ``\textit{spatially resolved}'' to be considered ``\textit{physical}.'' In addition, the eigenmode must satisfy the following requirements. First, if the eigenmode is numerically computed in a truncated computational domain with a finite but large radius $r_\infty$, there must be no bound at $r = r_\infty$ to ensure unboundedness. Second, the eigenmode's velocity and vorticity must approach zero at radial infinity, as we are interested in eigenmodes that develop in finite time from a wake vortex with vorticity localised in radius and not extending to infinity. Accordingly, eigenmodes that other authors have classified as part of the \textit{freestream} family \citep[][]{Mao2011} are not in the scope of this paper.
    \item By ``\textit{spatially resolved},'' we mean that the numerically computed eigenmode contains at least two collocation points in its smallest spatial structure (i.e., the radial width of the smallest wiggle). Additionally, the computed eigenvalue must either agree with its known value or converge to a fixed point. For an eigenvalue that belongs to the \textit{discrete} spectrum, its eigenvalue must approach a fixed point as the number of radial basis elements $M$ increases, and its corresponding eigenmode must converge to a fixed functional form. However, when dealing with an eigenvalue that belongs to the analytically continuous spectrum (where the spectrum lies along a curve, e.g., \textit{critical-layer} spectrum; or where the spectrum fills an area in the complex plane, e.g., \textit{potential} spectrum), there is ambiguity in numerically identifying a fixed point. This is because a finite matrix has only discrete eigenmodes. As $M$ increases, the number of computed eigenmodes typically increases, and it is unclear whether a specific eigenvalue/eigenmode computed with $M$ basis elements will correspond to \textit{any} eigenvalue/eigenmode computed with $M+1$ basis elements. This is because the eigenvalues and eigenvectors of a matrix can be sensitive to the distances between the locations of the collocation points (which depend on $M$) and the radial structures of the eigenmodes. For discrete spectra, this sensitivity generally does not prevent us from tracking the evolution of as specific eigenvalue/eigenmode as $M$ increases, but for continuous spectra, it does because eigenvalues are infinitesimally close to their neighbors. Therefore, in such cases, we determine whether the eigenvalue can be found within the expected range based on analytic results or reliable literature. In particular, for an inviscid eigenmode with a critical-layer singularity, its numerical solution will often suffer from the slow decay of spectral coefficients or the Gibbs phenomenon, especially around the singularity. Nonetheless, since our interest lies in identifying ``\textit{physical}'' eigenmodes, we do not present a numerical method that exactly handles singularities. Our objective in analysing inviscid critical-layer eigenmodes is only to resolve their spatial characteristics outside the singularity neighbourhood by using a sufficiently large value of $M$.
    \item By ``\textit{spurious},'' we mean that a numerically computed eigenmode is not ``\textit{spatially resolved},'' regardless of the value of $M$ used in the computation. This definition of ``\textit{spurious}'' originates from its historical usage \citep[cf.][]{Mayer1992}. We can confirm that an eigenmode is ``\textit{non-spurious}'' by increasing $M$ until it becomes evident that the eigenmode is ``\textit{spatially resolved}.'' However, in practice, we cannot prove that an eigenmode is ``\textit{spurious}.'' It is always possible that, after increasing $M$ to a large value and observing no evidence that the solution is approaching a fixed point, we abandon the increase in $M$ due to computational budget constraints, and the solution would have converged with a further increase in $M$. Therefore, if we discuss whether some viscous eigenmode families are ``\textit{spurious},'' the discussion will be suggestive rather than definitive.
\end{enumerate}

The remainder of the paper is structured as follows. In \S \ref{problemformulation}, the governing equations for wake vortex motion are formulated and linearised . In \S \ref{numericalmethod}, the spectral collocation method using mapped Legendre functions is presented. In \S \ref{spectrum}, the Lamb-Oseen and Batchelor vortex eigenmode spectra and pseudospectra are described. In \S \ref{inviscidlinearanalysis}, the eigenmodes and eigenvalues of the inviscid problems are compared to the analytic results. In \S \ref{viscouslinearanalysis}, the eigenmodes and eigenvalues in consideration of viscosity are presented, including a new family of eigenmodes in the continuous spectra that evolved from the family of critical-layer eigenmodes associated with the inviscid continuous spectrum. In \S \ref{conclusion}, our findings are summarised.

\section{Problem formulation}\label{problemformulation}
\subsection{Governing equations}
\begin{figure}
  % \vspace{0.1in}
  \centerline{\includegraphics[width=3.5in,keepaspectratio]{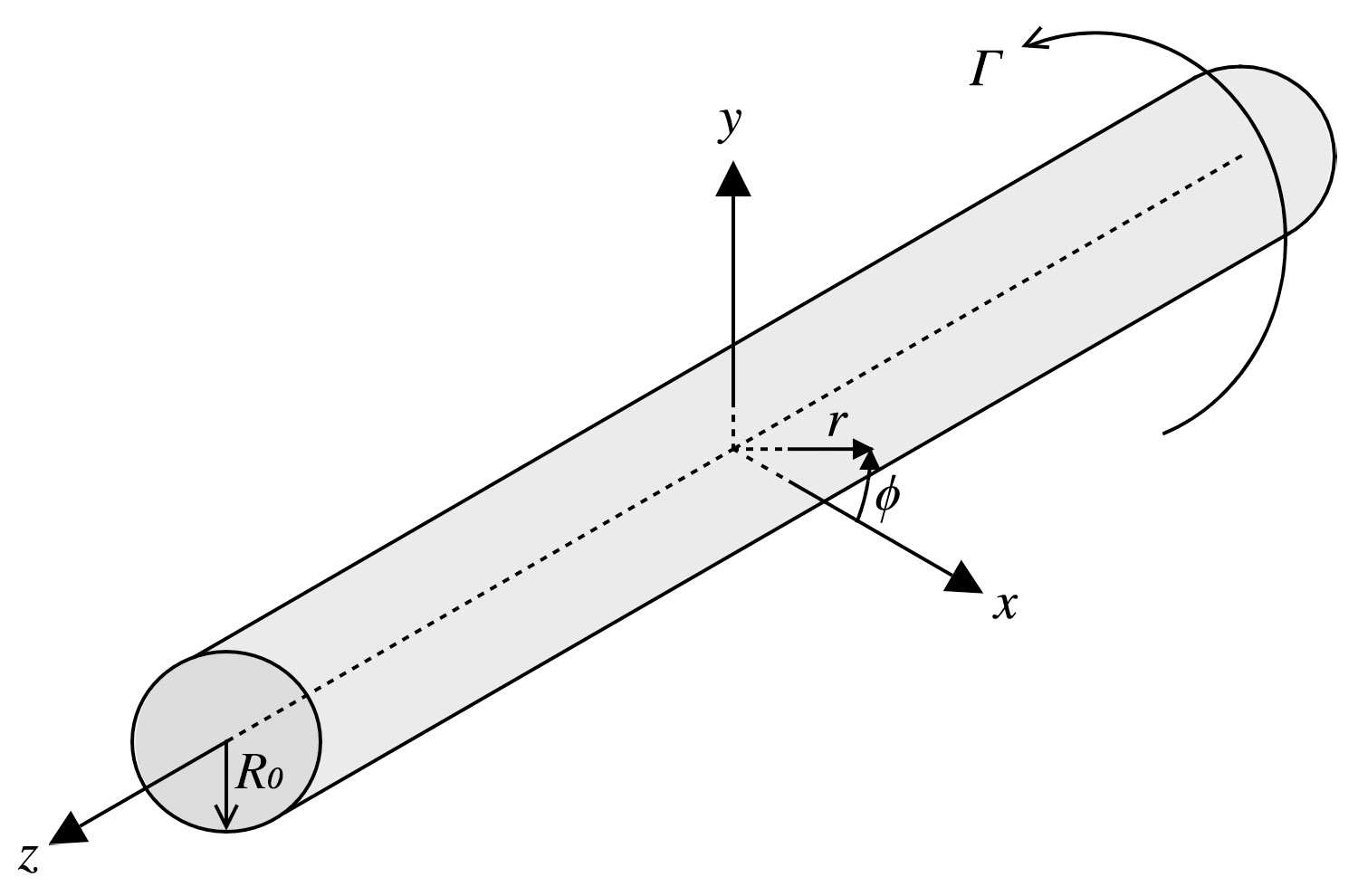}}
  \caption{Vortex with circulation $\Gamma$ of length scale $R_0$ and coordinate systems.}
\label{fig:vortex_coord}
\end{figure}

In this paper, we investigate the linear perturbation eigenmodes and eigenvalues of a swirling flow in an unbounded domain $\mathbb{R}^3$. We express the velocity and pressure eigenmodes in cylindrical coordinates $(r, \phi, z)$, as
\begin{equation}
    \bm{u}'= \Tilde{\bm{u}}(r;m,\kappa) e^{\mathrm{i}(m\phi + \kappa z) + \sigma t} , ~~ p' = \Tilde{p}(r;m,\kappa) e^{\mathrm{i}(m\phi + \kappa z) + \sigma t} ,
    \label{pertvelpre}
\end{equation}
where $m$ and $\kappa$ represent the azimuthal and axial wavenumbers of the eigenmode, respectively, and $\sigma$ denotes the complex growth (or decay) rate of the eigenmode. Here, $m \in \mathbb{Z}$ since the fields must be periodic in $\phi$ with a period of $2\pi$, while $\kappa \in \mathbb{R}\setminus {0}$ since there are no restrictions on the axial wavelength $2\pi/\kappa$. The real part of $\sigma$ represents the growth/decay rate, while the imaginary part represents its wave frequency in time. Although \citet{Khorrami1989} formulated a more general problem, we employ a more specialised form of the steady, equilibrium, swirling flow $\bm{\overline{U}}$, i.e.,
\begin{equation}
    \bm{\overline{U}} (r) = \overline{U}_\phi (r) \hat{\bm{e}}_\phi + \overline{U}_z (r) \hat{\bm{e}}_z,
    \label{baseflow}
\end{equation}
which is only $r$-dependent and has no radial velocity component $\overline{U}_r$. The unperturbed base flow profile we consider for a wake vortex model is Batchelor's similarity solution adapted by \citet{Lessen1974} with
\begin{equation}
    \frac{\overline{U}_\phi}{U_0} = \frac{1-e^{-r^2 / R_0^2}}{r / R_0}, ~~~~
    \frac{\overline{U}_z}{U_0} = \frac{1}{q}e^{-r^2/R_0^2},
\end{equation}
where $R_0$ and $U_0$ are the length and velocity scales defined in \citet[p. 755]{Lessen1974}, and $q \neq 0$ is a dimensionless swirl parameter. This flow is often called the $q$-vortex, which is steady, axisymmetric, and analytically tractable as the far-field asymptotic solution under the viscous light-loading condition \citep[see][pp. 257-260]{Saffman1993}. When the axial flow component vanishes, i.e., $1/q \rightarrow 0$, this flow is equivalent to the Lamb-Oseen vortex. A schematic of the geometry is shown in figure \ref{fig:vortex_coord}. The unperturbed vortex is oriented along the $z$-direction with a circulation over the entire plane $\Gamma \equiv 2\pi R_0 U_0$. $R_0$ is referred to as the characteristic radius of the vortex. As for the vortex profile, we consider the azimuthal velocity $\overline{U}_\phi$, which is maximised at $r = 1.122 R_0$ \citep{Lessen1974}.

To establish governing equations, we assume the fluid has constant density $\rho$ and constant kinematic  viscosity $\nu$. The total velocity $\bm{u} \equiv u_r \bm{\hat{e}}_r + u_\phi \bm{\hat{e}}_\phi + u_z \bm{\hat{e}}_z$ obeys
\begin{equation}
    \bm{\nabla} \cdot \bm{u} = 0 ,
    \label{continuity}
\end{equation}
\begin{equation}
    \frac{\partial \bm{u}}{\partial t} = -  \left( \bm{u} \cdot \bm{\nabla} \right) \bm{u} - \frac{1}{\rho} \bm{\nabla} p + \nu {\nabla}^2 \bm{u} = - \bm{\nabla} \varphi + \bm{u} \times \bm{\omega}  + \nu {\nabla}^2 \bm{u},
    \label{momentum}
\end{equation}
where the total pressure is $p$, the vorticity is $\bm{\omega} \equiv \bm{\nabla} \times \bm{u}$ and the total specific energy is $\varphi \equiv u^2 /2 + p/\rho$ where $u^2 \equiv \bm{u} \cdot \bm{u}$. We non-dimensionalise the equations using $R_0$ as the unit of length and $R_0/U_0$ as the unit of time. After non-dimensionalising and linearising \eqref{continuity}--\eqref{momentum} about the unperturbed flow (indicated with overbars $\overline{*}$), we obtain the following equation for the perturbations (indicated with primes $*'$):
\begin{equation}
    \bm{\nabla} \cdot \bm{u}' = 0,
    \label{continuity_pert}
\end{equation}
\begin{equation}
    \frac{\partial \bm{u}'}{\partial t} = - \bm{\nabla} \varphi' + \bm{\overline{U}}(r) \times \bm{\omega}' - \bm{\overline{\omega}}(r) \times \bm{u}' + \frac{1}{\Rey} {\nabla}^2 \bm{u}',
    \label{momentum_pert}
\end{equation}
where the Reynolds number, denoted $\Rey$, is defined to be $U_0 R_0 / \nu$. Note that the non-dimensionalised $q$-vortex is
\begin{equation}
    \overline{\bm{U}}(r) = \left(\frac{1-e^{-r^2}}{r}\right) \hat{\bm{e}}_\phi + \left(\frac{1}{q}e^{-r^2}\right) \hat{\bm{e}}_z .
    \label{qvortdimless}
\end{equation}
The established governing equations are essentially the incompressible, linearised Navier-Stokes equations, which, in combination with the $q$-vortex, were also used in recent vortex stability analyses, such as \citet{Qiu2021}. By putting \eqref{pertvelpre} to \eqref{continuity_pert}--\eqref{momentum_pert}, we obtain the equations that govern the perturbations:
\begin{equation}
    \bm{\nabla}_{m \kappa} \cdot \Tilde{\bm{u}} = 0,
    \label{continuity_linpert}
\end{equation}
\begin{equation}
    \sigma \Tilde{\bm{u}} = -\bm{\nabla}_{m \kappa} \Tilde{\varphi} + \bm{\overline{U}} \times \Tilde{\bm{\omega}} - \bm{\overline{\omega}} \times \Tilde{\bm{u}} + \frac{1}{\Rey} {\nabla}_{m \kappa}^2 \Tilde{\bm{u}},
    \label{momentum_linpert}
\end{equation}
where $\sigma$ is a function of $m$ and $\kappa$ (i.e., it obeys the dispersion relationship), $\Tilde{\bm{\omega}} \equiv \bm{\nabla}_{m \kappa} \times \Tilde{\bm{u}}$, and $\Tilde{\varphi} \equiv \bm{\overline{U}} \cdot \Tilde{\bm{u}} + \Tilde{p}$.
In the equations above, the subscript $(\ast)_{m\kappa}$ attached to the operators means that they act on modes of fixed azimuthal and axial wavenumbers $m$ and $\kappa$. Therefore, the differential operators $\partial / \partial \phi$ and $\partial / \partial z$ inside these operators are replaced with the simple multiplication operators $\mathrm{i} m$ and $\mathrm{i} \kappa$, respectively (see Appendix \ref{app}).

\subsection{Boundary and analyticity conditions}\label{sec:boundaryandanalyticity}
We require both the velocity and pressure fields to be analytic at $r = 0$ and to decay rapidly to 0 as $r \rightarrow \infty$. The conversion of these conditions to numerical boundary conditions can be found in previous works such as \citet{Batchelor1962}, \citet{Mayer1992}, and \citet{Ash1995}. In this section, we briefly discuss these conditions and how they will be treated in our method, where functions are treated as a truncated sum of the mapped Legendre functions.

The analyticity at the origin is equivalent to the pole condition that correctly removes the coordinate singularity \citep[see][]{Canuto1988, Matsushima1995, Lopez2002}. The pole condition for a scalar function $f(r,\phi, z)$ to be analytic at $r=0$ is that it asymptotically behaves as a polynomial in $r$, with the degree dependent on the azimuthal wavenumber $m$ \citep[see][p. 323]{Matsushima1997}, that is,
\begin{equation}
    f(r,\phi, z) = \sum_{m = -\infty}^{\infty} e^{\mathrm{i}m\phi} r^{|m|} \left(\sum_{n = 0}^{\infty} {C_n(z;m) r^{2n}} \right)~~~\text{as}~r\rightarrow 0,
    \label{scalar_pole}
\end{equation}
for some set of functions, analytic in $z$, $C_n(z;m)$. Although the pole condition for velocity fields in polar or cylindrical coordinates is rather complicated because of the velocity coupling of $r$ and $\phi$ at the origin \citep[see][pp. 328-330]{Matsushima1997}, we use toroidal and poloidal streamfunctions, given in \eqref{tpdecomp}, instead of the primitive velocity components so that the analyticity can be determined by making these streamfunctions obey the requirements of scalars (see Appendix~\ref{app:b}).

On the other hand, the rapid decay condition as $r \rightarrow \infty$ is relevant to the feasibility of linear perturbations. Since a perturbation lasting even at radial infinity would require infinite kinetic energy, decay should be necessary to consider it physical (see our definition in \S \ref{sec:prelim}). The simplest description is $\Tilde{\bm{u}},\; \Tilde{p} \rightarrow 0$ as $r \rightarrow \infty$ \citep{Batchelor1962}. Several numerical methods that require the domain truncation at large $r$ mimic this condition by imposing the homogeneous Dirichlet boundary condition for $\Tilde{\bm{u}}$ and $\Tilde{p}$ at the outer boundary of the radially truncated domain $r=r_\infty$. In other words, $\Tilde{u}_r=\Tilde{u}_\phi = \Tilde{u}_z = \Tilde{p} =0 $ at $r=r_\infty$ \citep[see][]{Khorrami1989, Khorrami1991}. However, this approach involves two problems. First, it cannot preclude non-physical eigenmode solutions that do not decay properly but incidentally end up being zero at $r=r_\infty$ (i.e., wall-bounded modes). Such non-physical solutions may also appear with non-zero viscosity, triggering non-normalisable singularities at radial infinity, where more information can be found in \citet[][p. 268]{Fabre2006} or \citet[][pp. 17-21]{Mao2011}. Second, it does not explicitly take into account how rapidly the perturbation decays. Considering the velocity field, it must decay faster than algebraic decay rates of $O(r^{-1})$ for kinetic energy to be finite as $r \rightarrow \infty$ \citep[cf.][]{Bolle2020}. Mathematically, the restriction is even more strict, requiring exponential or super-exponential decay rates \citep{Ash1995}. Our method is free from domain truncation and explicitly forces solutions to decay harmonically, i.e., $O ( r^{-|m|} e^{\mathrm{i}m\phi} )$ as $r \rightarrow \infty$, due to the decaying nature of the basis functions.

By utilising the current method, it can be ensured that any scalar functions, represented by the sum of mapped Legendre functions that serve as Galerkin basis functions, comply with the aforementioned conditions. This is precisely how each basis function behaves as the radial distance approaches zero and approaches infinity. Therefore, an advantage of employing the mapped associated Legendre functions is that there is no need for additional treatment for numerical boundary conditions. For further information regarding the properties of the mapped Legendre functions, please refer to \S \ref{numericalmethod}.

\subsection{Poloidal-toroidal decomposition}\label{sec:tpdecomp}
The governing equations \eqref{continuity_linpert} - \eqref{momentum_linpert}, along with the correct boundary conditions and given values of $m$ and $\kappa$, formally constitute a set of four equations that make up a generalised eigenvalue problem in terms of $\Tilde{p}$ (or $\Tilde{\varphi}$) and the three components of $\Tilde{\bm{u}} \equiv \Tilde{u}_r \bm{\hat{e}}_r + \Tilde{u}_\phi \bm{\hat{e}}_\phi + \Tilde{u}_z \bm{\hat{e}}_z$, which are often referred to as primitive variables, with $\sigma$ as the eigenvalue. The formal expression of the eigenvalue problem can be found in \citet[p. 7]{Bolle2020}. Some previous studies have taken additional steps to eliminate $\Tilde{p}$ from the momentum equations or even reduce the problem in terms of, for example, only $\Tilde{u}_\phi$ and $\Tilde{u}_z$, resulting in the generalised eigenvalue problem form $\mathsfbi{A}\bm{x} = \lambda \mathsfbi{B} \bm{x}$ \citep{Mayer1992, HEATON2007b, Mao2011}. However, such variable reduction inevitably increases the spatial order of the system and, consequently, requires a higher resolution for computation, which undermines the advantage of having a smaller number of state variables \citep{Mayer1992}. To avoid this issue, we use a poloidal-toroidal decomposition of the velocity field to formulate the matrix eigenvalue problem while preserving the spatial order of the governing equations. Moreover, the use of the poloidal and toroidal streamfunctions is advantageous because the formulation results in the standard eigenvalue problem of the form $\mathsfbi{A}\bm{x} = \lambda \bm{x}$.

To begin with, we apply the poloidal-toroidal decomposition to the governing equations of wake vortices that are linearised about the $q$-vortex. The basic formulation was performed by \citet[p. 339]{Matsushima1997}, and we provide more details of its mathematical foundation in this section. Although the poloidal-toroidal decomposition of solenoidal vector fields is mainly discussed in spherical geometry \citep[pp. 622-626]{Chandrasekhar1981}, it can be employed in the cylindrical coordinate system while preserving some essential properties of the decomposition \citep{Ivers1989}. When we select the unit vector in the $z$-direction $\bm{\hat{e}_z}$ as a reference vector, a solenoidal vector field $\bm{V} (r, \phi, z ) = V_r (r, \phi, z) \bm{\hat{e}_r} + V_\phi (r, \phi, z) \bm{\hat{e}_\phi} + V_z (r, \phi, z) \bm{\hat{e}_z}$ can be expressed as
\begin{equation}
    \bm{V} = \bm{\nabla} \times \bigl\{ \psi (r, \phi, z) \bm{\hat{e}_z} \bigr\} + \bm{\nabla} \times \Bigl[\bm{\nabla} \times \bigl\{ \chi (r, \phi, z) \bm{\hat{e}_z} \bigr\} \Bigr],
    \label{tpdecomp}
\end{equation}
where $\psi$ and $\chi$ are the toroidal and poloidal streamfunctions of $\bm{V}$. Such a decomposition is feasible if $\bm{V}$ has zero spatial mean components in the radial and azimuthal directions over an infinite disk for all $z$ \citep[cf.][]{JONES200845}. This zero-mean condition is satisfied in our study because our velocity fields are spatially periodic perturbations of the base flow. \citet{Ivers1989} concluded that the toroidal and poloidal fields are orthogonal over an infinite slab $a < z < b$ if $\psi$ and $\chi$ decay sufficiently rapidly as $r \rightarrow \infty$. The decay condition of $\psi$ and $\chi$ requires $\bm{V}$ to decay sufficiently rapidly to zero for large $r$.

In what follows, we find more rigorous statement for the decay condition of $\bm{V}$ as $r \rightarrow \infty$ where $\psi$ and $\chi$ are well-defined. The $z$-component of \eqref{tpdecomp} is
\begin{equation}
    \frac{1}{r} \frac{\partial }{\partial r} \left( r \frac{\partial \chi}{ \partial r} \right) + \frac{1}{r^2} \frac{\partial^2 \chi}{\partial \phi^2} = - V_z .
    \label{poisson1}
\end{equation}
Taking the curl of \eqref{tpdecomp}, we obtain
\begin{equation}
    \bm{\nabla} \times \bm{V} = \bm{\nabla} \times \bigl\{(-\nabla^2 \chi) \bm{\hat{e}_z}\bigr\} + \bm{\nabla} \times \bigl\{\bm{\nabla} \times (\psi \bm{\hat{e}_z})\bigr\},
\end{equation}
with its $z$-component equal to
\begin{equation}
    \frac{1}{r} \frac{\partial }{\partial r} \left( r \frac{\partial \psi}{ \partial r} \right) + \frac{1}{r^2} \frac{\partial^2 \psi}{\partial \phi^2} = - \left( \bm{\nabla} \times \bm{V} \right)_z.
    \label{poisson2}
\end{equation}
Solving \eqref{poisson1} and \eqref{poisson2}, which are the two-dimensional Poisson equations, can yield the solution to $\psi$ and $\chi$. By ignoring the gauge freedom with respect to $z$, we can determine the solution using two-dimensional convolution as follows:
\begin{equation}
    \psi = - \left( \bm{\nabla} \times \bm{V} \right)_z \ast \Phi,
    \label{convol1}
\end{equation}
\begin{equation}
    \chi = - V_z \ast \Phi,
    \label{convol2}
\end{equation}
where $\Phi$ is Green's function for the entire plane $\mathbb{R}^2$ equivalent to
\begin{equation}
    \Phi (r, \phi) = \frac{1}{2\pi} \ln r.
\end{equation}
In order for the convolutions in \eqref{convol1} and \eqref{convol2} to be meaningful everywhere, there exist $p_1 > 0$, $p_2 > 0$ and $p_3 >0$ such that
\begin{equation}
    \begin{cases}
    V_r \sim O \left( r^{-1-p_1} \right)\\ 
    V_\phi \sim O \left ( r^{-1-p_2} \right) \\
    V_z \sim O \left( r^{-2-p_3} \right)
    \end{cases}
    ~~\text{as } r \rightarrow \infty ,
    \label{decayrigor}
\end{equation}
given that $\bm{V}$ decays algebraically. Otherwise, $\bm{V}$ may decay exponentially or super-exponentially. If $\bm{V}$ is referred to as a velocity field, it has finite total kinetic energy over the entire space since all components decay faster than $O(r^{-1})$ as $r \rightarrow \infty$. The finite kinetic energy condition is physically reasonable, especially when dealing with velocity fields representing small perturbations \citep[cf.][]{Bolle2020}. On the other hand, \citet{Matsushima1997} considered the case where $\psi$ and $\chi$ could be unbounded by including additional logarithmic terms in $\psi$ and $\chi$, providing a more comprehensive extension of the poloidal-toroidal decomposition to more general vector fields, including the mean axial components. However, in the present study, we choose $\bm{V}$ as a linear perturbation of no bulk movement, and therefore the logarithmic terms do not need to be considered.

Suppressing the gauge freedom by adding restrictions that are independent of $z$ to $\psi$ and $\chi$ , e.g.,
\begin{equation}
    \lim_{r\rightarrow\infty} \psi (r, \phi, z) = \lim_{r\rightarrow\infty} \chi (r, \phi, z)= 0,
    \label{killgauge}
\end{equation}
we can define the following linear and invertible operator $\mathbb{P}: \mathcal{U} ~\rightarrow~ \mathcal{P}$ as
\begin{equation}
    \mathbb{P}(\bm{V}) \equiv \begin{pmatrix} \psi (r, \phi, z) \\ \chi (r, \phi, z) \end{pmatrix},
\end{equation}
where $\mathcal{U}$ is the set of sufficiently rapidly decaying solenoidal vector fields from $\mathbb{R}^3$ to $\mathbb{R}^3$ ($\mathbb{C}^3$) that satisfy \eqref{decayrigor} and $\mathcal{P}$ is the set of functions from $\mathbb{R}^3$ to $\mathbb{R}^2$ ($\mathbb{C}^2$) that satisfy \eqref{killgauge}. Using Helmholtz's theorem, we may extensively define $\mathbb{P}$ on more generalised vector fields which are not solenoidal but their solenoidal portion can be decomposed toroidally and poloidally. If we expand the domain of $\mathbb{P}$, however, it should be kept in mind that the operator is no longer injective because for any $\bm{V} \in \mathcal{U}$, $\mathbb{P}(\bm{V}) = \mathbb{P}(\bm{V} + \bm{\nabla}v)$ where $v$ is an arbitrary scalar potential for a non-zero irrotational vector field. On the other hand, it is noted that $\mathbb{P} ( \nabla^2 \bm{V} ) = \nabla^2 \mathbb{P}(\bm{V}) \equiv \left( \nabla^2 \psi , \nabla^2 \chi \right)$ for $\bm{V} \in \mathcal{U}$ because
\begin{equation}
\begin{gathered}
    \nabla^2 \biggl[ \bm{\nabla} \times \bigl\{ \psi (r, \phi, z) \bm{\hat{e}_z} \bigr\} + \bm{\nabla} \times \Bigl[\bm{\nabla} \times \bigl\{ \chi (r, \phi, z) \bm{\hat{e}_z} \bigr\} \Bigr] \biggr] \\
    = \bm{\nabla} \times \bigl\{ \nabla^2 \psi (r, \phi, z) \bm{\hat{e}_z} \bigr\} + \bm{\nabla} \times \Bigl[ \bm{\nabla} \times \bigl\{ \nabla^2 \chi (r, \phi, z) \bm{\hat{e}_z} \bigr\} \Bigr].
\end{gathered}
\end{equation}
Applying the operator $\mathbb{P}$ to both sides of \eqref{momentum_pert}, we obtain
\begin{equation}
    \frac{\partial \mathbb{P}(\bm{u}')}{\partial t} = \mathbb{P} \left( \bm{\overline{U}}(r) \times \bm{\omega}' \right) - \mathbb{P} \left( \bm{\overline{\omega}}(r) \times \bm{u}' \right) + \frac{1}{\Rey} {\nabla}^2 \mathbb{P}(\bm{u}'),
    \label{op_momentum}
\end{equation}
because $\mathbb{P}(\bm{\nabla}\varphi) = \mathbb{P}(\bm{0} + \bm{\nabla}\varphi) = \mathbb{P}(\bm{0}) = \bm{0}$. Assuming $\bm{u}'$ to be solenoidal, $\bm{u}'$ automatically satisfies the continuity equation and can be determined from $\mathbb{P}(\bm{u}')$ by taking the inverse of it using \eqref{tpdecomp}. 

Since we are interested in the perturbation velocity field as in \eqref{pertvelpre}, we define two $r$-dependent scalar functions $\Tilde{\psi}(r;m,\kappa)$ and $\Tilde{\chi}(r;m,\kappa)$ such that
\begin{equation}
    \mathbb{P} \left( \Tilde{\bm{u}}(r;m,\kappa) e^{\mathrm{i}(m\phi + \kappa z)+\sigma t} \right)
    = \begin{pmatrix} \Tilde{\psi} (r;m,\kappa) e^{\mathrm{i}(m\phi + \kappa z)+\sigma t} \\ \Tilde{\chi} (r;m,\kappa) e^{\mathrm{i}(m\phi + \kappa z)+\sigma t} \end{pmatrix} 
    .
    \label{tp-kelvin}
\end{equation}
The fact that the poloidal and toroidal components in \eqref{tp-kelvin} preserve the exponential part can be verified by substituting the perturbation velocity field formula into $\bm{V}$ in \eqref{poisson1} and \eqref{poisson2}. For convenience, we simplify the expression in \eqref{tp-kelvin} to
\begin{equation}
    \mathbb{P}_{m \kappa}(\Tilde{\bm{u}}(r;m,\kappa)) \equiv \begin{pmatrix} \Tilde{\psi}(r;m,\kappa) \\ \Tilde{\chi}(r;m,\kappa) \end{pmatrix} .
    \label{tp-kelvin-simple}
\end{equation}

Finally, putting \eqref{pertvelpre} into \eqref{op_momentum} leads to the standard eigenvalue problem form in terms of $\mathbb{P}_{m \kappa}(\Tilde{\bm{u}}(r;m,\kappa))$:
\begin{equation}
    \sigma  \bigl[ \mathbb{P}_{m\kappa}(\Tilde{\bm{u}}(r;m,\kappa)) \bigr] = \mathcal{L}_{m\kappa}^{\nu} \bigl[ \mathbb{P}_{m\kappa}(\Tilde{\bm{u}}(r;m,\kappa)) \bigr],
    \label{final_evp}
\end{equation}
where the linear operator $\mathcal{L}_{m\kappa}^{\nu}$ is defined as
\begin{equation}
    \mathcal{L}_{m\kappa}^{\nu} \bigl[ \mathbb{P}_{m\kappa}(\Tilde{\bm{u}}) \bigr] \equiv \mathbb{P}_{m\kappa} \left( \bm{\overline{U}}(r) \times \Tilde{\bm{\omega}} \right) - \mathbb{P}_{m\kappa} \left( \bm{\overline{\omega}} (r) \times \Tilde{\bm{u}} \right) + \frac{1}{\Rey} {\nabla}^2_{m \kappa} \mathbb{P}_{m\kappa}(\Tilde{\bm{u}}).
\end{equation}
Excluding the viscous diffusion term, we additionally define the inviscid operator $\mathcal{L}_{m\kappa}^{0}$ as
\begin{equation}
    \mathcal{L}_{m\kappa}^{0} \bigl[ \mathbb{P}_{m\kappa}(\Tilde{\bm{u}}) \bigr] \equiv \mathbb{P}_{m\kappa} \left( \bm{\overline{U}}(r) \times \Tilde{\bm{\omega}} \right) - \mathbb{P}_{m\kappa} \left( \bm{\overline{\omega}}(r) \times \Tilde{\bm{u}} \right),
\end{equation}
for the inviscid linear analysis solving 
\begin{equation}
    \sigma  \bigl[ \mathbb{P}_{m\kappa}(\Tilde{\bm{u}}(r;m,\kappa)) \bigr] = \mathcal{L}_{m\kappa}^{0} \bigl[ \mathbb{P}_{m\kappa}(\Tilde{\bm{u}}(r;m,\kappa)) \bigr] .
\end{equation}

\section{Numerical method}\label{numericalmethod}
\subsection{Mapped Legendre functions}
Associated Legendre functions with algebraic mapping are used as basis functions to expand an arbitrary function over $0 \le r < \infty$, ultimately discretising the eigenvalue problems to be solved numerically. The expansion was first introduced by \citet{Matsushima1997} and applied to three-dimensional vortex instability studies by \citet{Bristol2004} and \citet{Feys2016}. The algebraically mapped associated Legendre functions $P_{L_n}^m (r)$, or simply mapped Legendre functions, are equivalent to the mapping of the associate Legendre functions $P_n^m (\zeta)$ with order $m$ and degree $n$ defined on $-1 \le \zeta < 1$ where
\begin{equation}
    \zeta \equiv \frac{r^2 - L^2}{r^2 + L^2} ~\Longleftrightarrow~ r = L \sqrt{\frac{1+\zeta}{1-\zeta}}.
    \label{mapping}
\end{equation}
An additional parameter $L>0$ is the map parameter, which can be arbitrarily set. However, when it is used for a spectral collocation method, change in $L$ affects the spatial resolution of discretisation and the value should be carefully chosen to achieve fast convergence or eliminate spurious results. \citet{Matsushima1997} showed that $    P_{L_n}^m (r) \sim O ( r^{|m|} ) ~\text{as}~ r \rightarrow 0$ and $P_{L_n}^m (r) \sim O ( r^{-|m|} ) ~\text{as}~ r \rightarrow \infty$, which leads to the fact that any polar function $P_{L_n}^m (r) e^{\mathrm{i} m \phi}$ behaves analytically at the origin \citep[see][pp. 243-244]{Eisen1991} and decays harmonically to zero at radial infinity. These asymptotic properties are suitable to apply the correct boundary conditions for the present problem.

Next, we prove that a set of some mapped Legendre functions can constitute a complete orthogonal basis of spectral space. Since the associate Legendre functions $P_n^m (\zeta)$ are the solutions to the associate Legendre equation
\begin{equation}
    \frac{d}{d\zeta} \left[ \left( 1 - \zeta^2 \right) \frac{d P_n^m}{d \zeta} \right] + \left[ n(n+1) - \frac{m^2}{1-\zeta^2} \right]P_n^m (\zeta) = 0,
\end{equation}
the mapped Legendre functions satisfy the following second-order differential equation
\begin{equation}
    \frac{d}{dr} \left[ r \frac{d{P_{L_n}^m}}{dr} \right] - \frac{m^2}{r}P_{L_n}^m (r) + \frac{4 n (n+1)L^2 r}{(L^2 +r^2)^2} P_{L_n}^{m} (r) = 0.
    \label{sl_mapleg}
\end{equation}
As \eqref{sl_mapleg} is the Sturm-Liouville equation with the weight function
\begin{equation}
    w(r) \equiv \frac{4L^2 r}{(L^2 + r^2)^2},
\end{equation}
the mapped Legendre functions $P_{L_{|m|}}^m (r)$, $P_{L_{|m|+1}}^m (r)$, $P_{L_{|m|+2}}^m (r)$, $\cdots$ form an orthogonal basis of the Hilbert space $L^2 \left( \mathbb{R}^{+}, w(r)dr \right)$. Thus, for two integers $n$ and $k$ larger than or equal to $|m|$,
\begin{equation}
\begin{gathered}
    \left\langle P_{L_n}^m, P_{L_k}^m \right\rangle = \int_{0}^{\infty} P_{L_n}^m(r) P_{L_k}^m (r) w(r) dr \\
    = \int_{-1}^{1} P_{n}^m (\zeta) P_{k}^m (\zeta) d\zeta = \frac{2(n+|m|)!}{(2n+1)(n-|m|)!} \delta_{nk},
\end{gathered}
\end{equation}
where $\delta_{nk}$ denotes the Kronecker delta with respect to $n$ and $k$.

Considering a polar function $f_m (r) e^{\mathrm{i} m \phi}$ where $f_m \in L^2 \left( \mathbb{R}^{+}, w(r)dr \right)$, it can be expanded by the mapped Legendre functions as
\begin{equation}
    f_m (r) e^{\mathrm{i} m \phi} = \sum_{n = |m|}^{\infty} f_n^m P_{L_n}^m (r) e^{\mathrm{i} m \phi},
    \label{expansion}
\end{equation}
and the coefficient $f_n^m$ can be calculated based on the orthogonality of the basis functions:
\begin{equation}
\begin{gathered}
    f_n^m = \frac{\left\langle f_m, P_{L_n}^m \right\rangle}{\left\langle P_{L_n}^m, P_{L_n}^m \right\rangle } = \frac{(2n+1) (n-|m|)!}{2(n+|m|)!} \int_{0}^{\infty} f_m (r) P_{L_n}^m (r) w(r) dr \\
    = \frac{(2n+1) (n-|m|)!}{2(n+|m|)!} \int_{-1}^{1} f_m \left( L \sqrt{\frac{1 + \zeta}{1 - \zeta}} \right) P_{n}^m (\zeta) d \zeta.
    \label{integral}
\end{gathered}
\end{equation}
When we expand an analytic function on $0 \le r < \infty$ that vanishes at infinity, the expansion in \eqref{expansion} is especially suitable because they are able to serve as Galerkin basis functions. Even if we use the truncated series of \eqref{expansion}, analyticity at the origin and vanishing behaviour at infinity remain valid.

\subsection{Mapped Legendre spectral collocation method}
In order to discretise the problem, we use a spectral collocation method using the mapped Legendre functions as basis functions. Given the azimuthal and axial wavenumbers $m$ and $\kappa$, we take a truncated basis set of first $M$ elements $\{ P_{L_{|m|}}^{m} ,\; \cdots ,\; P_{L_{|m|+M-1}}^{m} \}$ and expand $f_m(r) e^{\mathrm{i}(m\phi + \kappa z)}$ as
\begin{equation}
    f_m (r) e^{\mathrm{i}(m \phi + \kappa z)} = \sum_{n=|m|}^{|m|+M-1} f_n^m P_{L_n}^{m} (r) e^{\mathrm{i} (m\phi + \kappa z)},
    \label{finiteexpand}
\end{equation}
so that the function is represented by $M$ discretised coefficients $( f_{|m|}^{m} ,\; \cdots ,\; f_{|m|+M-1}^{m} )$. The coefficients are numerically obtained by applying the Gauss-Legendre quadrature rule to \eqref{integral}. Let $\zeta_j$ and $\varpi_j$ be the $j$th root of the Legendre polynomial $P_N$ of degree $N$ in $(-1,1)$ with its quadrature weight defined as
\begin{equation}
    \varpi_j = 2 (1-\zeta_j^2)^{-1} \left[ \dfrac{dP_N}{d\zeta} \bigg|_{\zeta = \zeta_j} \right]^{-2}, ~~ j = 1,\; \cdots, \; N,
\end{equation}
and with radial collocation points $r_j$ determined from \eqref{mapping} as
\begin{equation}
    r_j \equiv L\sqrt{\frac{1 + \zeta_j}{1-\zeta_j}}, ~~ j = 1,\; \cdots, \; N,
    \label{collocdef}
\end{equation}
which means that half of the collocation points are distributed in the inner high-resolution region $0 \le r < L$ whereas the other half are posed in the outer low-resolution region $r \ge L$ \citep{Matsushima1997}. In order to describe spatial resolution, we define the characteristic resolution parameter $\Delta$ as
\begin{equation}
    \Delta (N,L) \equiv \frac{2L}{N},
    \label{resolution}
\end{equation}
which represents the mean spacing between the collocation points in $0 \le r < L$.

A quadrature algorithm presented by \citet[pp. 179-194]{press_numerical_2007} is implemented and all abscissas and weights are computed with an absolute precision error less than $10^{-15}$. The quadrature converts the integration formula to the weighted sum of the function values evaluated at the collocation points and consequently the integral of \eqref{integral} finally becomes the discretised formula
\begin{equation}
    f_n^m \simeq \frac{(2n+1) (n-|m|)!}{2(n+|m|)!} \sum_{j=1}^{N} \varpi_j f_m(r_j) P_n^m(\zeta_j).
    \label{inverse}
\end{equation}
It is convenient in practice to conceal the factorial coefficient term by defining the normalised mapped Legendre functions and coefficients as follows:
\begin{equation}
    \hat{P}_{L_n}^{m}(r) \equiv P_{L_n}^{m}(r) \sqrt{\frac{(2n+1) (n-|m|)!}{2(n+|m|)!}}, ~~~~
    \hat{f}_n^m \equiv f_n^m \sqrt{\frac{2(n+|m|)!}{(2n+1) (n-|m|)!}}.
\end{equation}
Using these normalised terms, \eqref{inverse} can be expressed as
\begin{equation}
    \hat{f}_n^m \simeq \sum_{j=1}^{N} \varpi_i f_m(r_j) \hat{P}_n^m(\zeta_j),
    \label{inverse2}
\end{equation}
and, moreover, \eqref{finiteexpand} at $r = r_j$ maintains the identical form
\begin{equation}
    f_m (r_j) e^{\mathrm{i}(m \phi + \kappa z)} = \sum_{n=|m|}^{|m|+M-1} \hat{f}_n^m \hat{P}_{L_n}^{m} (r_j) e^{\mathrm{i} (m\phi + \kappa z)}.
    \label{finiteexpand2}
\end{equation}

As a preliminary step of the mapped Legendre spectral collocation method, we need to compute (1) the Gauss-Legendre abscissas $\zeta_i$, (2) weights $\varpi_i$, (3) radial collocation points $r_i$ and (4) normalised mapped Legendre functions evaluated at the collocation points $\hat{P}_{L_n}^m (r_i)$. The normalisation procedure may require temporary multiple-precision arithmetic to handle large function values and factorials if one uses $N$ larger than about 170. There have been several multi-precision arithmetic libraries available recently and we consider using the \textsc{FM} multiple-precision package \citep{smith_using_2003}. All essential computations ahead, however, can be performed under typical double-precision arithmetic.

It is noted that the number of the abscissas (or collocation points) $N$ must be equal to or larger than the number of the basis elements $M$ for the sake of proper transform between physical space $( f_m(r_1) ,\; \cdots ,\; f_m(r_N) )$ and spectral space $( \hat{f}_{|m|}^{m} ,\; \cdots ,\; \hat{f}_{|m|+M-1}^{m} )$. On the other hand, due to the even and odd parity of the associate Legendre functions, taking even $N$ and $M$ can reduce the work by half in the transform procedure \citep{Matsushima1997}. Consequently, we set both $N$ and $M$ to be even and $N = M + 2$ in further analyses unless otherwise specified.

Finally, we discuss how to apply the mapped Legendre spectral collocation method to the present problem. Recalling \eqref{tp-kelvin-simple} where $\mathbb{P}_{m \kappa} ( \Tilde{\bm{u}} ) = (\Tilde{\psi}, \Tilde{\chi})$, we write
\begin{equation}
    \Tilde{\psi} (r;m,\kappa) e^{\mathrm{i} (m\phi + \kappa z)} = \sum_{n=|m|}^{|m|+M-1} \Tilde{\psi}_n^{m \kappa} \hat{P}_{L_n}^{m} (r) e^{\mathrm{i} (m\phi + \kappa z)} ,
\end{equation}
\begin{equation}
    \Tilde{\chi} (r;m,\kappa) e^{\mathrm{i} (m\phi + \kappa z)} = \sum_{n=|m|}^{|m|+M-1} \Tilde{\chi}_n^{m \kappa} \hat{P}_{L_n}^{m} (r) e^{\mathrm{i} (m\phi + \kappa z)} .
\end{equation}
We point out that when $\Tilde{\psi}$ is expressed in the partial sums above, it obeys the boundary conditions of an analytic scalar at the origin, i.e., as $r \rightarrow 0$,
\begin{equation}
    \Tilde{\psi}(r; m, \kappa) \rightarrow r^{|m|} \, \sum_{i =0}^{\infty}  \, a_i^{m\kappa} \, r^{2 i}, 
    \label{expand_near_0}
\end{equation}
where $a_0^{m\kappa}$, $a_1^{m\kappa}$, $\cdots$ are constants \citep[see][]{Eisen1991, Matsushima1995}.
Similar analyticity conditions are obeyed by $\Tilde{\chi}(r; m, \kappa)$, and therefore the
perturbation velocity field $\Tilde{\bm{u}}(r) e^{\mathrm{i} (m\phi + \kappa z)}$ is also analytic at the origin (see Appendix \ref{app:b}). Due to the properties of the mapped Legendre functions, the perturbation vorticity also decays as $r \rightarrow \infty$ \citep{Matsushima1997}. As a consequence, $\mathbb{P}_{m \kappa} ( \Tilde{\bm{u}} )$ can be uniquely represented by $2M$ spectral coefficients of $\Tilde{\psi}_{|m|}^{m\kappa}$, $\cdots$, $\Tilde{\psi}_{|m|+M-1}^{m\kappa}$ and  $\Tilde{\chi}_{|m|}^{m\kappa}$, $\cdots$, $\Tilde{\chi}_{|m|+M-1}^{m\kappa}$. We may discretise the eigenvalue problem for viscous cases in \eqref{final_evp} as
\begin{equation}
    \sigma 
    \begin{psmallmatrix}  
        \Tilde{\psi}_{|m|}^{m\kappa} \\ \vdots \\
        \Tilde{\psi}_{|m|+M-1}^{m\kappa} \\[5pt]
        \Tilde{\chi}_{|m|}^{m\kappa} \\ \vdots \\
        \Tilde{\chi}_{|m|+M-1}^{m\kappa}
    \end{psmallmatrix}
    = \mathsfbi{L}_{m\kappa}^{\nu}
    \begin{psmallmatrix} 
        \Tilde{\psi}_{|m|}^{m\kappa} \\ \vdots \\
        \Tilde{\psi}_{|m|+M-1}^{m\kappa} \\[5pt]
        \Tilde{\chi}_{|m|}^{m\kappa} \\ \vdots \\
        \Tilde{\chi}_{|m|+M-1}^{m\kappa}
    \end{psmallmatrix},
    \label{matrix_evp}
\end{equation}
where $\mathsfbi{L}_{m\kappa}^{\nu}$ is a $2M \times 2M$ complex matrix representing the linear operator $\mathcal{L}_{m\kappa}^{\nu}$. In a similar sense, we can define $\mathsfbi{L}_{m\kappa}^{0}$ representing $\mathcal{L}_{m\kappa}^{0}$ for the inviscid analysis and
\begin{equation}
    \mathsfbi{L}_{m\kappa}^{\nu} = \mathsfbi{L}_{m\kappa}^{0} + \Rey^{-1} \mathsfbi{H}.
\end{equation}
$\mathsfbi{H}$ is a matrix representation of the Laplacian $\nabla_{m\kappa}^2$ acting on the spectral coefficients $\Tilde{\psi}_{|m|}^{m\kappa}$, $\cdots$, $\Tilde{\psi}_{|m|+M-1}^{m\kappa}$ and  $\Tilde{\chi}_{|m|}^{m\kappa}$, $\cdots$, $\Tilde{\chi}_{|m|+M-1}^{m\kappa}$, respectively. For a scalar function expanded by the mapped Legendre functions $a(r) = \sum_{n\ge|m|} a_n^m P_{L_n}^m (r)$, if we expand its Laplacian as $\nabla_{m\kappa}^2 a (r) = \sum_{n\ge|m|} b_n^m P_{L_n}^m (r)$, then the coefficients $a_n^m$ and $b_n^m$ constitute the following relationship for all $n \ge |m|$
\begin{equation}
\begin{aligned}
    b_n^m = & -\left[\tfrac{(n-|m|-1)(n-|m|)(n-2)(n-1)}{(2n-3)(2n-1)L^2} \right]a_{n-2}^{m} 
              +\left[\tfrac{2n(n-|m|)(n-1)}{(2n-1)L^2} \right]a_{n-1}^{m} \\
            & -\left[\tfrac{2n(n+1)(3n^2 + 3n - m^2 - 2)}{(2n-1)(2n+3)L^2} + \kappa^2 \right]a_{n}^{m} \\
            & +\left[\tfrac{2(n+1)(n+|m|+1)(n+2)}{(2n+3)L^2} \right]a_{n+1}^{m} 
              -\left[\tfrac{(n+|m|+1)(n+|m|+2)(n+2)(n+3)}{(2n+3)(2n+5)L^2} \right]a_{n+2}^{m},
\end{aligned}
\label{laplacianfuncs}
\end{equation}
under the assumption that $a_n^m \equiv 0$ if $n$ is less than $|m|$ \citep[p. 344]{Matsushima1997}. $\mathsfbi{H}$ can be formulated by \eqref{laplacianfuncs}.

The formulation of $\mathsfbi{L}_{m\kappa}^{0}$ involves the vector products in physical space and is conducted using a pseudospectral approach based on the Gauss-Legendre quadrature rule. Reconstructing $\Tilde{\bm{u}}$ from $\mathbb{P}_{m \kappa} ( \Tilde{\bm{u}} )$ via \eqref{tpdecomp}, we evaluate the vector products $\overline{\bm{U}} \times \Tilde{\bm{\omega}}$ and  $\overline{\bm{\omega}} \times \Tilde{\bm{u}}$ at $N$ radial collocation points and apply $\mathbb{P}_{m \kappa}$ again. As for the detailed algorithm including the numerical implementation of $\mathbb{P}_{m\kappa}$ as well as its inverse, refer to (69) and (70) in \citet{Matsushima1997}, providing the spectral coefficients of $\mathbb{P}_{m \kappa} (\overline{\bm{U}} \times \Tilde{\bm{\omega}})$ and  $\mathbb{P}_{m \kappa} (\overline{\bm{\omega}} \times \Tilde{\bm{u}})$. Integration in these equations can be performed numerically by the Gauss-Legendre quadrature rule, as given in \eqref{inverse2}. Following this procedure, we can compute the $i$th column vector of $\mathsfbi{L}_{m\kappa}^{0}$ by substituting the $i$th standard unit vector $\hat{\bm{e}}_i \in \mathbb{R}^{2M}$ for $( \Tilde{\psi}_{|m|}^{m\kappa}$, $\cdots$, $\Tilde{\psi}_{|m|+M-1}^{m\kappa}, \; \Tilde{\chi}_{|m|}^{m\kappa}$, $\cdots$, $\Tilde{\chi}_{|m|+M-1}^{m\kappa} )$.

A global eigenvalue problem solver with the QR algorithm for non-Hermitian matrices, based on the \textsc{Lapack} routine named \textsc{Zgeev}, is used to solve the discretised eigenvalue problem. The procedure of constructing a global matrix and finding all eigenvalues has been established in previous studies, such as \citet[p. 241]{Fabre2006}. However, as shown in \eqref{matrix_evp}, our formulation directly results in the standard eigenvalue problem rather than the generalised form. Thus, it is sufficient to construct only one matrix of dimension $2M \times 2M$, with a reduction in the number of state variables from 4 to 2.

\subsection{Numerical parameters and their effects}\label{sec:numericalparameters}
The mapped Legendre spectral collocation method comprises of three adjustable numerical parameters: $M$, $N$, and $L$. The first two parameters are commonly used in most spectral collocation methods, while the last parameter is unique to our method. This section elaborates on the impact of each parameter on the numerical method's performance and provides guidelines on their selection.

\subsubsection{Number of spectral basis elements $M$}
As shown in \eqref{finiteexpand}, $M$ determines the number of basis elements in use and is the most important parameter for the numerical method's convergence. The larger the value of $M$, the closer the mapped Legendre series is to its ground-truth, as the full basis set assuming $M \rightarrow \infty$ is complete. If the function of interest is analytic and decays properly, the convergence is exponential with increasing $M$. Even if the function contains any singularity in the interior, the convergence must occur at infinite $M$, albeit slowly, as long as the function belongs to the Hilbert space $L^2 \left( \mathbb{R}^{+}, w(r)dr \right)$.

For achieving better accuracy, it is always preferable to select a larger value of $M$. However, a too large value of $M$ may cause the resulting matrix eigenvalue problem to be excessively large, leading to an increase in the time complexity in $(2M)^3$. In practice, the availability of computing resources should limit the maximum value of $M$.

\subsubsection{Number of radial collocation points $N$}
$N$, the number of the radial collocation points defined as \eqref{collocdef}, depends on $M$ because $N \ge M$ needs to be satisfied. Increasing $N$ nominally enhances the spatial resolution in physical space, thereby reducing numerical errors in the evaluation of vector products. However, this effect is rather marginal, as most of the major computations and errors occur in spectral space. Moreover, if an increase in $N$ does not accompany an increase in $M$ by the same or nearly the same amount, it may have no benefit at all. One may consider the extreme case where $N \rightarrow \infty$ while $M$ is kept constant at unity. Regardless of how perfect the radial resolution is, none of the functions can be handled except for a scalar multiple of the first basis element $P_{L_{|m|}}^m (r)$.

Therefore, it is better to consider $N$ dependent on $M$, and any change in $N$ should only be followed by a change in $M$. This justifies why we use $N = M+2$. Similarly, an improvement in the spatial resolution by $N$ should imply the use of a larger $M$. Henceforth, $N$ is usually omitted when we state the numerical parameters, and $M$ implicitly specifies $N$ as $M+2$. In this case, we note that the resolution parameter $\Delta$ in \eqref{resolution} equals $2L/(M+2)$.

\subsubsection{Map parameter for Legendre functions $L$}
The map parameter $L$ provides an additional level of computational freedom that distinguishes the present numerical method from others. We highlight three significant roles of this parameter, two of which are related to spatial resolution in physical space and the other to basis change in spectral space.

\begin{figure}
  % \vspace{0.1in}
  \centerline{\includegraphics[width=\textwidth,keepaspectratio]{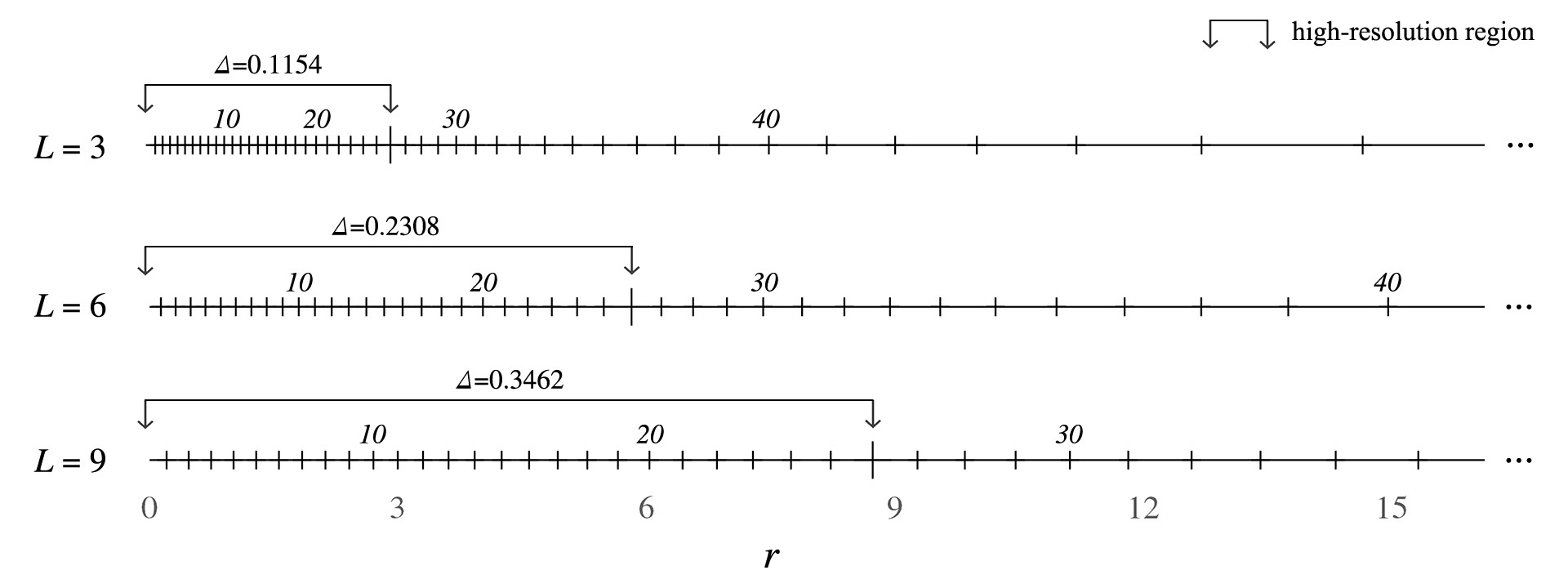}}
  \caption{Changes in distribution of the collocation points with respect to $L$ given $N = 52$. Some collocation points at large radii are omitted. The high-resolution region is $0\le r < L$, where half of the collocation points are clustered around the origin. As $L$ increases, the high-resolution region is expanded. However, the mean spacing $\Delta$ grows simultaneously. $L$ should be chosen carefully to balance these anti-complementary effects.}
\label{fig:colloc_by_l}
\end{figure}

In physical space, when $M$ (and $N$) is fixed, a change in $L$ results in two anti-complementary effects with respect to spatial resolution, as shown in figure \ref{fig:colloc_by_l}. When $L$ increases, the high-resolution region $0 \le r < L$, where half of the collocation points are clustered, expands, which has a positive effect. However, it negatively impacts the resolution, especially in the high-resolution region, where $\Delta$ increases with $L$. Increasing $N = M+2$ may compensate for the loss in resolution. However, if $M$ is already at a practical limit due to the computing budget, expanding the high-resolution region by increasing $L$ should stop when $\Delta$ remains satisfactorily small. The requirement for satisfaction should be specific to the eigenmodes to be resolved, which will be discussed in each analysis section later. Similar discussions can be made in the opposite direction when decreasing $L$.

In spectral space, changing $L$ entirely replaces the complete basis function set. For instance, when $L=A$ and $L=B$, the spectral method can be constructed on either of two different complete basis sets, i.e., $\{ P_{A_{|m|}}^{m} ,\;P_{A_{|m|+1}}^{m} \; \cdots \}$ or $\{ P_{B_{|m|}}^{m} ,\;P_{B_{|m|+1}}^{m} ,\; \cdots \}$. Since orthogonality among the basis functions does not necessarily hold across the basis sets, an eigenmode found with $L=A$ can differ from that found with $L=B$. If $B$ differs from $A$ by an infinitesimal amount, our method makes it possible to find eigenmodes that continuously vary if they exist. This was thought to be hardly achievable via classic eigenvalue solvers due to discretisation \citep[cf.][p. 11]{Mao2011}. Once the numerical method's convergence is secured by sufficiently large $M$ and $N$, we explore such non-normal eigenmodes that vary continuously by fine-tuning $L$.

\subsection{Validation}
To confirm the numerical validity of our method, we compared some eigenvalues from the discrete branch of the spectra with those previously calculated by \citet{Mayer1992}. They also used a spectral collocation method but with Chebyshev polynomials as radial basis functions over an artificially truncated radial domain, rather than the mapped Legendre basis functions over an unbounded radial domain we use. For comparison, we linearly scaled the eigenvalues reported in \citet{Mayer1992} to match the $q$-vortex model used in our study because the azimuthal velocity component is scaled by $q$ in their study, whereas we adjust the axial velocity component.

We compared the most unstable eigenvalue calculations for the inviscid case $m=1$, $\kappa=0.5$, $q=-0.5$ (or equivalently $m=1$, $\kappa=-0.5$, $q=0.5$) and the viscous case $m=0$, $\kappa=0.5$, $q=1$, $\Rey=10^4$ in table~\ref{tab:valid}. We conducted the calculations using three different numbers of basis elements $M$ (20, 40, and 80) and three different map parameters $L$ (8, 4, and 2). Our results show that the trend towards convergence is apparent as $M$ increases and $L$ decreases. As we discuss in terms of the characteristic resolution parameter $\Delta$ defined in \eqref{resolution}, both parameters influence the numerical resolution. Increasing $M$ leads to an increase in the number of radial collocation points $N$, while decreasing $L$ improves spatial resolution by filling the inner high-resolution region ($0 \le r < L$) with more collocation points (see figure \ref{fig:colloc_by_l}). However, this comes at the expense of reducing the range of the high-resolution region and effectively shrinking the radial domain by placing the collocation point with the largest radius at $r_N = L \sqrt{(1+\zeta_N)/(1-\zeta_N)}$, which can lead to inaccuracies if any significant portion of the solution exists either in the outer low-resolution region or outside the effective limit. The convergence test of $\sigma^{\dagger}_{\text{viscous}}$ with $M = 20$ in table~\ref{tab:valid} partially demonstrates this concern. When we compare the eigenvalues computed with $L=4$ and $L=2$, the latter shows no clear improvement in convergence compared to the former, despite having a smaller $L$. Even small $L$ causes the eigenvalue's real part to move further away from the reference value of \citet{Mayer1992}. Therefore, we must keep in mind that blindly pursuing small $L$ does not guarantee better convergence, although using large $M$ is always favoured for numerical convergence.

\begin{table}
  \begin{center}
\def~{\hphantom{0}}
  \begin{tabular}{r || cc| c | c}
                      &   $M$  & $L$ & $\sigma^\dagger_{\text{inviscid}}$ & $\sigma^\dagger_{\text{viscous}}$     \\[5pt]
    Present study~    &   ~20~ & ~8~ & 0.37755989 + 0.112913723$\mathrm{i}$           & 0.00011969 + 0.01679606$\mathrm{i}$         \\
                      &   ~20~ & ~4~ & 0.40527381 + 0.099406043$\mathrm{i}$           & 0.00018939 + 0.01658207$\mathrm{i}$         \\
                      &   ~20~ & ~2~ & 0.40525621 + 0.099437298$\mathrm{i}$           & 0.00014902 + 0.01656308$\mathrm{i}$         \\[2pt]
                      &   ~40~ & ~8~ & 0.40522876 + 0.099370546$\mathrm{i}$           & 0.00017892 + 0.01632424$\mathrm{i}$         \\
                      &   ~40~ & ~4~ & 0.40525620 + 0.099437300$\mathrm{i}$           & 0.00018406 + 0.01640824$\mathrm{i}$         \\
                      &   ~40~ & ~2~ & 0.40525620 + 0.099437300$\mathrm{i}$           & 0.00018463 + 0.01640723$\mathrm{i}$         \\[2pt]
                      &   ~80~ & ~8~ & 0.40525620 + 0.099437300$\mathrm{i}$           & 0.00018478 + 0.01640740$\mathrm{i}$         \\
                      &   ~80~ & ~4~ & 0.40525620 + 0.099437300$\mathrm{i}$           & 0.00018469 + 0.01640717$\mathrm{i}$         \\
                      &   ~80~ & ~2~ & 0.40525620 + 0.099437300$\mathrm{i}$           & 0.00018469 + 0.01640717$\mathrm{i}$         \\[5pt]
    \citet{Mayer1992}~&   --   & --  & 0.40525620 + 0.099437300$\mathrm{i}$           & 0.00018469 + 0.01640717$\mathrm{i}$
  \end{tabular}
  \caption{Comparison of the eigenvalues associated with the most unstable mode (indicated with a superscript $\dagger$) for the inviscid case with $m=1$, $\kappa = 0.5$, $q=-0.5$ and for the viscous case with $m=0$, $\kappa = 0.5$, $q=1$, $\Rey = 10^4$. The table illustrates how the values change when we alter the map parameter $L$ and the number of radial mapped Legendre basis functions $M$. The last row displays the values obtained by \citet{Mayer1992}, who employed up to 200 radial Chebyshev basis functions. Their published eigenvalues were appropriately rescaled to fit the $q$-vortex model employed in our study. Our numerically computed eigenvalues tend towards a fixed point as we increase $M$ beyond 40. It should be noted that the size of the matrix eigenvalue problem system is $2M$ for our method and $3M$ for that of \citet{Mayer1992}. Thus, even when using the same $M$, our method is expected to require $(2/3)^3$ less work than theirs.}
  \label{tab:valid}
  \end{center}
\end{table}

The high-resolution range of the present method, represented by $L$, should not match the domain truncation radius in the method of \citet{Mayer1992}. Adjusting the high-resolution range through $L$ has no impact on the unbounded nature of the domain and can be customised essentially. However, altering the domain truncation radius fundamentally harms the unbounded nature of the domain and must be set to its maximum computing limit. On the other hand, we achieve the same accuracy as \citet{Mayer1992} with roughly three times smaller $M$, which supports the numerical efficiency of our method. Presumably, our method is around ten times more computationally efficient in solving matrix eigenvalue problems that scale as $O(M^3)$. We believe this is mainly because their simple algebraic mapping of Chebyshev collocation points \citep[see][p. 357]{Ash1995} clusters approximately one-third of the collocation points near the artificial outer radial boundary, where vortex motion is near zero and not important by assumption. Such collocation points do not significantly contribute to solving the problem, resulting in an inefficient use of numerical resources.

\begin{figure}
  % \vspace{0.1in}
  \centerline{\includegraphics[width=\textwidth,keepaspectratio]{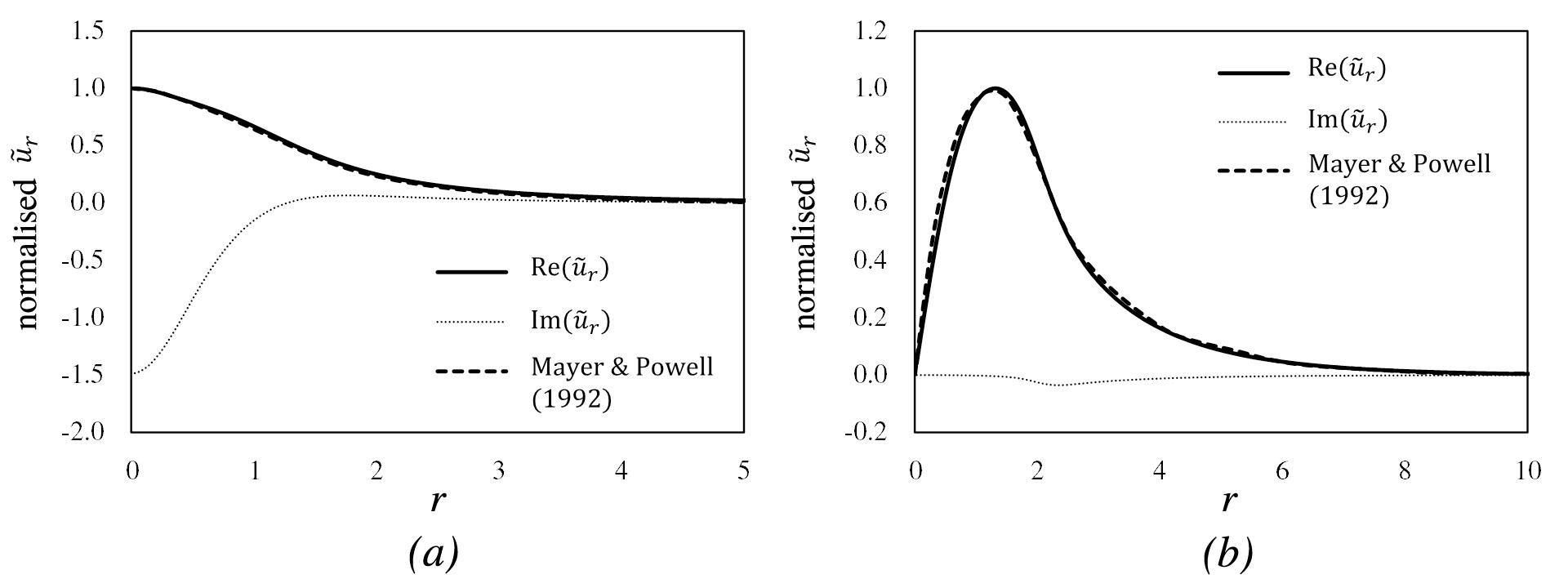}}
  \caption{A comparison of our numerical calculation with that of \citet{Mayer1992}. Shown is the radial velocity component of the most unstable eigenmode for the validation cases $(a)$ $(m, \kappa, q, \Rey) = (1, 0.5, -0.5, \infty)$ and $(b)$ $(m, \kappa, q, \Rey) = (0, 0.5, 1, 10^4)$, where the maximum of $\Real(\Tilde{u}_r)$ is normalised to unity. Numerical parameters are $M=80$ and $L=2$. Note that \citet{Mayer1992} only plotted the real parts of the eigenmodes.}
\label{fig:validation}
\end{figure}

Note that the eigenmodes shown here are regular and have no singularities, as depicted in figure~\ref{fig:validation}. Such regular eigenmodes are expanded by a finite number of radial basis elements that are already regular, and as shown in table~\ref{tab:valid}, their numerical results converge exponentially with increasing $M$. However, singular eigenmodes can only be expressed exactly when an infinite sum of mapped Legendre functions is taken \citep[see][]{Gottlieb1977}. Nonetheless, as stated in the preliminary remarks, we are essentially interested in physical eigenmodes, i.e., those without singularities and computed numerically with small spectral residual error. The current validation is strong enough to underpin this objective. Later in this paper (see \S \ref{viscouscriticallayereigenmodes}), we present some eigenmodes that have convincing signatures of viscous remnants after regularising the inviscid critical-layer singularities. These singularities become regularised but still nearly singular regions of local rapid oscillations. We can find the value of $M$ at which these eigenmodes are spatially resolved, even if it typically goes beyond 80. Also note that in this respect, we only peripherally examine their inviscid counterparts with the critical-layer singularities using our numerical method (see \S \ref{inviscidcriticallayereigenmodes}).

\section{Spectrum}\label{spectrum}
Solving an eigenvalue problem $\lambda \bm{x} = \mathcal{L} \bm{x}$ is often equivalent to finding the spectrum of the linear operator $\mathcal{L}$, denoted $\sigma ( \mathcal{L} )$. A number of previous studies that investigated a linearised version of the Navier-Stokes equations, epitomised by the Orr-Sommerfeld equation, have already adopted the term ``spectra'' \citep{Grosch1978, Jacobs1998} to account for eigenmodes of the linearised equations. In our study, we also employ this concept to characterise eigenmode families found in the linear analysis of the $q$-vortex. We first state the definition of the spectrum for the reader's convenience.

\begin{definition}
    Given that a bounded linear operator $\mathcal{L}$ operates on a Banach space $\mathcal{X}$ over $\mathbb{C}$, $\sigma(\mathcal{L})$ consists of all scalars $\lambda \in \mathbb{C}$ such that the operator $(\mathcal{L} - \lambda)$ is not bijective and thus $(\mathcal{L} - \lambda)^{-1}$ is not well-defined.
\end{definition}

If a complex scalar $\lambda$ is an eigenvalue of $\mathcal{L}$, then it belongs to $\sigma (\mathcal{L})$; however, the inverse statement is generally not true. This is because, by definition, the spectrum of $\mathcal{L}$ includes not only a type of $\lambda$ that makes $(\mathcal{L} - \lambda)$ non-injective but also another type of $\lambda$ by which $(\mathcal{L} - \lambda)$ is injective but not surjective. The former ensures the presence of a non-trivial eigenmode in $\mathcal{X}$, which therefore comprises the set of ordinary eigenvalues, while the latter does not. However, if $(\mathcal{L} - \lambda)$ has a dense range, $\lambda$ can be an \textit{approximate} eigenvalue in the sense that there exists an infinite sequence $(\bm{e}_j \in \mathcal{X}\setminus\left\{\bm{0}\right\})$ for which
\begin{equation}
    \lim_{j\rightarrow\infty}{\left\lVert \mathcal{L} \bm{e}_j - \lambda \bm{e}_j \right\rVert} = 0.
    \label{limiteig}
\end{equation}
In our method, $\bm{e}_j$ and $\mathcal{X}$ can be taken as a mapped Legendre series of the first $j$ basis elements in \eqref{finiteexpand} and the Hilbert space, respectively. Even if the sequence limit $\bm{e}_\infty$ does not belong to $\mathcal{X}$, it can still be regarded as an eigenmode solution in a \textit{rigged} manner, by permitting discontinuities, singular derivatives, or non-normalisabilities (i.e., rigged Hilbert space). In the literature related to fluid dynamics, both ordinary and approximate cases are considered as eigenvalues. They are classified either as discrete in the complex $\sigma$-plane, or as continuous in association with the eigenmodes possessing singularities. Despite their singular behaviour, understanding eigenmodes associated with continuous spectra may be important because they contribute to a complete basis for expressing an arbitrary perturbation \citep{Case1960, Fabre2006, Roy2014}.

\begin{figure}
  % \vspace{0.1in}
  \centerline{\includegraphics[width=\textwidth,keepaspectratio]{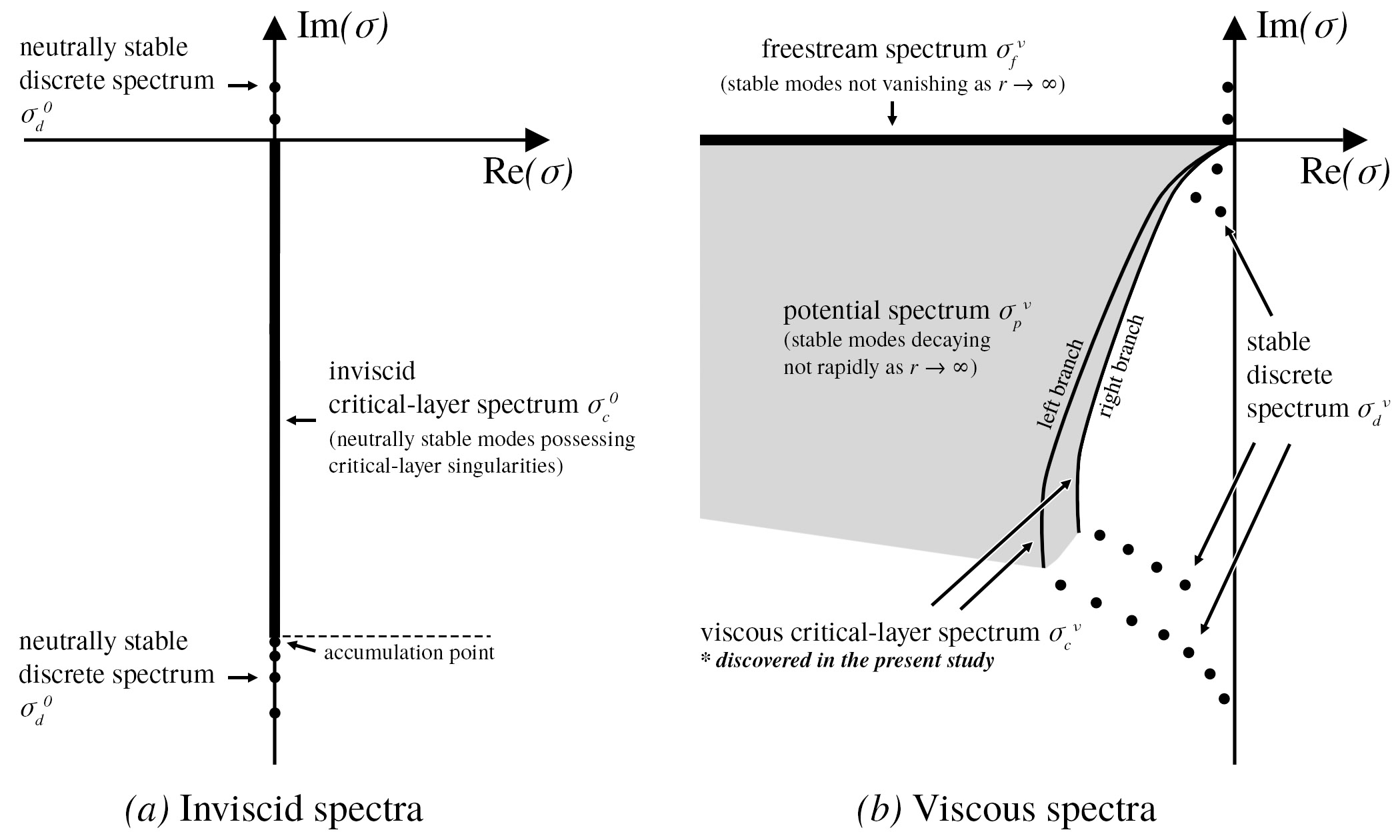}}
  \caption{Schematic diagrams of the spectra of the eigenvalues of a $q$-vortex of $(a)$ $\mathcal{L}_{m\kappa}^{0}$ for inviscid problems where $\nu \equiv 0$ \citep[see][]{Mayer1992,Heaton2007,Gallay2020} and $(b)$ $\mathcal{L}_{m\kappa}^{\nu}$ for viscous problems with finite $\Rey$, including $\nu \rightarrow 0^+$ \citep[see][]{Fabre2006, Mao2011}. Each schematic exhibits a set of eigenvalues where $m$ and $\kappa$ are fixed. The cases illustrated here assume $m > 0$. These spectra are shown here because they are representative, but they do not embrace all of the different families of spectra. The labels attached here are used throughout the main body of the text. Note that figures of the true numerical spectra computed by us, rather than schematics, follow in \S \ref{inviscidlinearanalysis} and \S \ref{viscouslinearanalysis}, and that the viscous critical-layer spectrum, consisting of \textit{two} distinct curves in $(b)$, were discovered via the present numerical analysis and were not previously identified.
}
\label{fig:spectrum_scheme}
\end{figure}

In figure \ref{fig:spectrum_scheme}, schematic diagrams of the spectra in relation to the $q$-vortices are presented. These illustrations assume that $m$ is positive. The exact spectra differ depending on the values of $m$, $\kappa$, $q$, $\Rey$, and the symmetries which are explained next. Some families of the spectra are not displayed because they are not within the main scope of this study. For instance, in the inviscid spectra, the unstable discrete spectrum and its symmetric stable counterpart frequently appear for some $m$, $\kappa$, and $q$. However, they vanish as $q$ becomes sufficiently large (e.g., $|q| > 2.31$) \citep[see][]{Heaton2007}. For the Lamb-Oseen vortex where $q \rightarrow \infty$, it was analytically proven that all of the eigenvalues are located on the imaginary axis, irrespective of $m$ and $\kappa$, indicating that all eigenmodes must be neutrally stable \citep[see][]{Gallay2020}.

There are three notable space-time symmetries in this eigenvalue problem. First, because the linearised equations admit real solutions for the velocity/pressure eigenmodes, regardless of the values of $q$ and the viscosity (including the case $\nu \equiv 0$), if $(\tilde{u}_r, \tilde{u}_\phi, \tilde{u}_z, \tilde{p})$ and $\sigma$ are an eigenmode and eigenvalue with wavenumbers $(m, \kappa)$, then  $(\tilde{u}_r^{*}, \tilde{u}_\phi^{*}, \tilde{u}_z^{*}, \tilde{p}^{*})$ is also an eigenmode with eigenvalue $\sigma^{*}$ and with $(-m, -\kappa)$.
Next, for the inviscid case, with any value of $q$, the linearised equations are time-reversible, and as a consequence if $(\tilde{u}_r, \tilde{u}_{\phi}, \tilde{u}_z, \tilde{p})$ and $\sigma$ are a velocity/pressure eigenmode and eigenvalue with wavenumbers ($m, \kappa)$,  then $(\tilde{u}_r^{*}, -\tilde{u}_{\phi}^{*}, -\tilde{u}_z^{*}, -\tilde{p}^{*})$ is also an eigenmode with eigenvalue $-\sigma^{*}$ and with the same $(m, \kappa)$. This symmetry makes the spectra symmetric about the imaginary axis in the left panel of figure \ref{fig:spectrum_scheme} but not in the right panel. 
Third, for the inviscid case with any value of $q$, we could combine the two symmetries above and obtain the fact that if $(\tilde{u}_r, \tilde{u}_{\phi}, \tilde{u}_z, \tilde{p})$ and $\sigma$ are a velocity/pressure eigenmode and eigenvalue with wavenumber ($m, \kappa)$,  then $(\tilde{u}_r, -\tilde{u}_{\phi}, -\tilde{u}_z, -\tilde{p})$ is also an eigenmode with eigenvalue $-\sigma$ with wavenumbers $(-m, -\kappa)$. 

In particular, for the inviscid case with $q \rightarrow \infty$ (i.e., with  $\bar{U}_z = 0$), the linearised equations are also invariant under $z \rightarrow -z$. In this case, if $(\tilde{u}_r, \tilde{u}_{\phi}, \tilde{u}_z, \tilde{p})$ and $\sigma$ are a velocity/pressure eigenmode and eigenvalue with wavenumbers ($m, \kappa)$,  then $(\tilde{u}_r, \tilde{u}_{\phi}, -\tilde{u}_z, \tilde{p})$ is also an eigenmode with eigenvalue $\sigma$ and with wavenumbers $(m, -\kappa)$. This symmetry can be combined with either or both of the two earlier listed symmetries to produce additional, but not independent symmetries; for example, if $(\tilde{u}_r, \tilde{u}_{\phi}, \tilde{u}_z, \tilde{p})$ and $\sigma$ are a velocity/pressure eigenmode and eigenvalue with wavenumbers ($m, \kappa)$,  then $(\tilde{u}_r^{*}, -\tilde{u}_{\phi}^{*}, \tilde{u}_z^{*}, -\tilde{p}^{*})$ is also an eigenmode with eigenvalue $-\sigma^{*}$ and with $(m, -\kappa)$.

Based on the two-dimensional Orr-Sommerfeld equation, \citet{Lin1961} argued that the spectra of eigenmodes of viscous flows are discrete. However, for unbounded viscous flows, \citet[pp. 156-157]{Drazin2004} stated that this is incorrect, and there is a continuous spectrum associated with eigenmodes that vary sinusoidally in the far field instead of vanishing. The presence of continuous spectra associated with the $q$-vortices due to spatial unboundedness was also discussed by \citet{Fabre2006} and \citet{Mao2011}. One example of the continuous spectrum is the viscous \textit{freestream} spectrum, named by \citet{Mao2011} and denoted $\sigma_f^{\nu}$ here, which is located on the left half of the real axis in the complex $\sigma$-plane in figure~\ref{fig:spectrum_scheme}(b). However, the eigenmodes in this spectrum persist rather than go to zero as $r\rightarrow\infty$. As stated in \S \ref{sec:prelim}, we are only interested in eigenmodes that we classify as physical. We have defined eigenmodes in which the velocity and vorticity do not decay harmonically at radial infinity as non-physical. Since our numerical method was specifically designed not to deal with such non-physical eigenmodes, we do not discuss them further in this paper and clarify that our method is not the tool for those who wish to investigate $\sigma_f^\nu$. We remark that \citet{Bolle2020} argued that the viscous freestream spectrum is rather an ``artefact'' of the mathematical model of an unbounded domain. With the exception of the viscous \textit{freestream} eigenmodes, our numerical method is capable of computing the families of eigenvalues and eigenmodes indicated in figure \ref{fig:spectrum_scheme}.

For the inviscid and viscous discrete spectra, denoted $\sigma_d^{0}$ and $\sigma_d^{\nu}$, respectively, the unstable eigenmodes of the $q$-vortices with finite $q$ have been extensively studied \citep{Stewartson1983, Mayer1992}, particularly for small $q$ \citep{Lessen1974, Heaton2007}. However, it is unclear whether these instabilities would be significant for aeronautical applications that are known to have large $q~( \approx 4)$ \citep[see][pp. 258-259]{fabre2004}. As the discrete spectra and related instabilities, which have been well-studied, are not the main focus of the present study, the unstable branches in $\sigma_d^{0}$ and $\sigma_d^{\nu}$, which may be detectable for small $q$ and large $\Rey$, are omitted in figure \ref{fig:spectrum_scheme}.

\begin{figure}
  \vspace{0.1in}
  \centerline{\includegraphics[width=\textwidth,keepaspectratio]{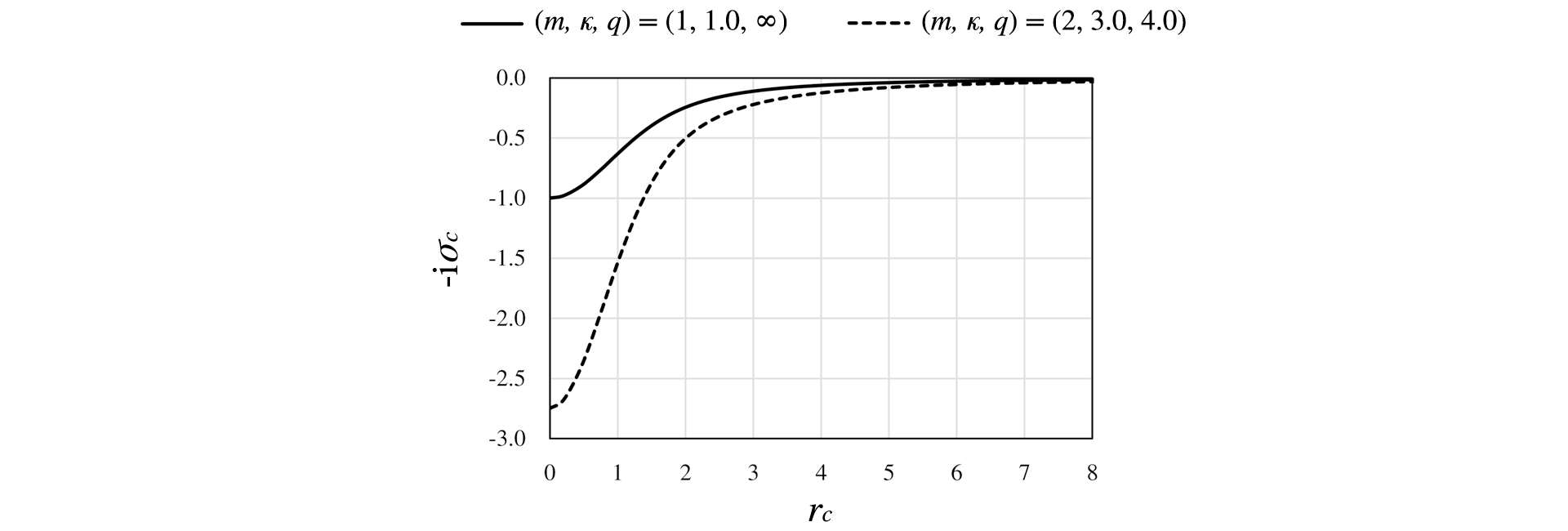}}
  \caption{Critical-layer singularity radial location $r_c$ versus critical layer eigenvalue $\sigma_c$ with fixed $m$, $\kappa$ and $q$. See \eqref{sigma_r_c_qvort} and \eqref{sigma_r_c_lamb}. The two illustrated cases where $(m, \kappa, q) = (1, 1.0, \infty)$ and $(m, \kappa, q) = (2, 3.0, 4.0)$ are investigated in later analyses.
  }
\label{fig:sigma_r_c}
\end{figure}

Instead, we pay attention to the eigenmodes associated with the inviscid critical-layer spectrum, denoted $\sigma_c^{0}$, which has been known to be related to further transient growth of wake vortices \citep{HEATON2007b, Mao2012}. For the inviscid $q$-vortex $\sigma_c^{0}$ is determined as a subset of $\sigma(\mathcal{L}_{m \kappa}^{0})$, which is
\begin{equation}
    \sigma_c^{0} = \left\{ \sigma_c \in \mathrm{i}\mathbb{R} \; \bigg| \; \exists r_c \in (0, \infty) ~~ -\mathrm{i} \sigma_c + \frac{m(1-e^{-r_c^2})}{r_c^2} + \frac{\kappa e^{-r_c^2}}{q}=0 \right\} \subset \sigma \left( \mathcal{L}_{m\kappa}^{0} \right).
    \label{inviscidspectrum}
\end{equation}
When $q \rightarrow \infty$, \eqref{inviscidspectrum} reduces to the expression given in \citet{Gallay2020}, which applies to the Lamb-Oseen vortex case. Considering the fact that $\sigma_c^{0}$ is due to an inviscid singularity \citep{LeDizes2004}, we deduce the expression in \eqref{inviscidspectrum} through the following steps. The singularity can be straightforwardly identified by further reducing the governing equations, as shown in \citet[p. 94]{Mayer1992}, originally done by \citet{Howard1962}. Breaking the eigenvalue problem form in \eqref{continuity_linpert} and \eqref{momentum_linpert} and performing further reduction, we obtain the following second-order differential equation:
\begin{equation}
    \gamma^2 \frac{d}{dr}\left(\frac{r}{\kappa^2 r^2 + m^2} \frac{d(r\Tilde{u}_r )}{dr}\right) - \left( \gamma^2 + a\gamma + b \right) \Tilde{u}_r = 0 ,
\end{equation}
where
\begin{equation}
    \gamma \equiv -\mathrm{i} \sigma + \frac{m \overline{U}_\phi (r)}{r} + \kappa \overline{U}_z (r),
\end{equation}
\begin{equation}
    a \equiv r \frac{d}{dr} \left[ \frac{r}{\kappa^2 r^2 + m^2} \left(\frac{d\gamma}{dr} + \frac{2m \overline{U}_\phi (r)}{r^2} \right) \right],
\end{equation}
\begin{equation}
    b \equiv \frac{2\kappa m \overline{U}_\phi (r)}{\kappa^2 r^2 + m^2} \left( \frac{d \overline{U}_z}{dr} - \frac{\kappa}{m} \frac{d(r \overline{U}_\phi)}{dr} \right).
\end{equation}
The equation becomes singular when $\gamma = 0$, which is feasible when there exist $\sigma_c \in \mathrm{i} \mathbb{R}$ and $r_c \in (0, \infty)$ such that
\begin{equation}
    -\mathrm{i} \sigma_c + \frac{m \overline{U}_\phi (r_c)}{r_c} + \kappa \overline{U}_z (r_c) = 0,
    \label{critpoint}
\end{equation}
or equivalently,
\begin{equation}
    \Real (\sigma_c) = 0 , ~~~ \Imag (\sigma_c) = - \frac{m \overline{U}_\phi (r_c)}{r_c} - \kappa \overline{U}_z (r_c).
    \label{critpoint-2}
\end{equation}
Substituting the $q$-vortex velocity profile into \eqref{critpoint} shows the relationship between the imaginary eigenvalue $\sigma_c^{0}$ and the radial location $r_c$ of the critical layer:
\begin{equation}
    \sigma_c = -\mathrm{i} \left[ \frac{m(1-e^{-r_c^2})}{r_c^2} + \frac{\kappa e^{-r_c^2}}{q} \right]
    \label{sigma_r_c_qvort},
\end{equation}
and for the Lamb-Oseen vortex with $1/q \equiv 0$
\begin{equation}
    \sigma_c = -\mathrm{i} \left[ \frac{m(1-e^{-r_c^2})}{r_c^2} \right].
    \label{sigma_r_c_lamb}
\end{equation}
For every eigenmode associated with $\sigma_c$, it must contain at least one singularity at $r = r_c$, which is what we have been referring to as a critical-layer singularity. As a result, the continuum of eigenvalues on the imaginary axis forms $\sigma_c^{0}$, as depicted in figure \ref{fig:spectrum_scheme}$(a)$. For the $q$-vortices with positive $m$, $\kappa$, and $q$ (including $q\rightarrow \infty$), which we will consider in later analyses, the supremum of $-\mathrm{i}\sigma_c$ is $0$ (as $r_c \rightarrow \infty$) and the infimum of $-\mathrm{i}\sigma_c$ is $-m - \kappa /q$ (as $r_c \rightarrow 0$). Also in this case, there is a one-to-one correspondence between $\sigma_c$ and $r_c$ as $m (1 -e^{-r^2})/ r^2 + (\kappa/q) e^{-r^2}$ is monotonic with respect to $r$ (see figure~\ref{fig:sigma_r_c}).

On the other hand, viscosity regularises the critical-layer singularities of the eigenmodes of $q$-vortices. It is of physical importance to identify how viscosity transforms inviscid spectra, such as $\sigma_c^{0}$, into a subset of the viscous spectra $\sigma(\mathcal{L}_{m\kappa}^{\nu})$ and to determine which branches of $\sigma_c^{0}$ vanish and what new eigenmodes are created. According to \citet{Heaton2007}, for non-zero viscosity, $\sigma_c^{0}$ is replaced by a large number of closely packed discrete eigenmodes, but a detailed explanation was not given. Numerical observations by \citet{Bolle2020} identified randomly scattered eigenvalues in the shaded region in figure \ref{fig:spectrum_scheme}$(b)$, suggesting that they are the viscous remnants of $\sigma_c^{0}$. \citet{Mao2011}, who earlier discovered this region, named it the \textit{potential} spectrum, denoted $\sigma_p^{\nu}$, and suggested that it could be continuous based on the shape of the surrounding pseudospectra. The ($\varepsilon$-)pseudospectrum is defined as follows \citep{trefethen_spectra_2005}.

\begin{definition}
    Let $R(z;\mathcal{L}) \equiv (\mathcal{L} - z)^{-1}$ be the resolvent of $\mathcal{L}$ at $z \in \mathbb{C} \setminus \sigma (\mathcal{L}) $. For $\varepsilon > 0$, the $\varepsilon$-pseudospectrum, denoted $\sigma_{\varepsilon} (\mathcal{L})$, is the set
    \begin{equation}
        \sigma_{\varepsilon}(\mathcal{L}) \equiv \bigg\{ z \in \mathbb{C} \; \bigg| \; \left\lVert R(z;\mathcal{L}) \right\rVert > \frac{1}{\varepsilon} \bigg\}.
    \end{equation}
\end{definition}
\noindent Note that the lower bound of the resolvent norm is determined by the inequality
\begin{equation}
    \left\lVert R(z;\mathcal{L}) \right\rVert \ge \sup_{\mu \in \sigma (\mathcal{L})} \frac{1}{\left| z - \mu  \right|},
    \label{resolvdist}
\end{equation}
where equality holds if the resolvent is normal \citep[pp. 9-10]{Bolle2020}. For discrete eigenvalues, when $\varepsilon$ is sufficiently small, the $\varepsilon$-pseudospectrum is formed by an open disk that surrounds the eigenvalue. However, when it comes to continuous spectra, \citet{Mao2011} pointed out that as $\varepsilon$ approaches zero, the $\varepsilon$-pseudospectrum tends to cover the entire region in the complex $\sigma$-plane that is equivalent to $\sigma_p^{\nu}$, as shown in figure~\ref{fig:spectrum_scheme}$(b)$. They proposed that this region comprises entirely of the viscous continuous spectra together with $\sigma_f^{\nu}$, which is located on the negative real axis. Such an asymptotic topology of pseudospectra implies the presence of continuous spectra in this region.

Although this argument appears reasonable, it requires careful examination for the following reasons. Firstly, as we numerically solve the eigenvalue problem, solutions that do not exhibit convergence may result from spurious modes due to discretisation. While randomly scattered eigenvalues may be true examples of eigenmodes within the continuous spectrum, they can also be spurious eigenmodes created by the disretised approximation of $\mathcal{L}_{m\kappa}^{\nu}$. Secondly, describing the pseudospectra of $\mathcal{L}_{m\kappa}^{\nu}$ as proximity to the spectrum is valid only if $R(z;\mathcal{L}_{m\kappa})$ is normal and the equality in \eqref{resolvdist} holds. According to \citet{Bolle2020}, the resolvent is selectively non-normal in a frequency band where $\sigma_p^{\nu}$ is located, meaning that $R(z;\mathcal{L}_{m\kappa})$ can take a large value even if $z$ is not actually close to $\sigma(\mathcal{L}_{m\kappa})$. Lastly, for the sake of rigour, the shape of the potential spectrum, as depicted in the schematic in figure \ref{fig:spectrum_scheme}$(b)$, should be considered suggestive. This is because, to the best of our knowledge, its presence has only been numerically proposed in the discretised problem with increasing $M$ (i.e., $\mathsfbi{L}_{m\kappa}^{\nu}$), but has not been analytically verified in the original problem (i.e., $\mathcal{L}_{m\kappa}^{\nu}$). It should be noted that in the present study, we premise the analytic presence of the potential spectrum as depicted in figure \ref{fig:spectrum_scheme}$(b)$, so that numerical eigenvalues found on the $\varepsilon$-pseudospectrum of $\mathsfbi{L}_{m\kappa}^{\nu}$ in the limit of $\varepsilon \rightarrow 0$ with a sufficiently large value of $M$ can be considered the discretised representation of this analytic entity, and therefore non-spurious.

Although $\sigma_p^{\nu}$ is known to be associated with stable eigenmodes that decay to zero as $r\rightarrow\infty$, their decay rates in $r$ have been reported to be much slower than the exponential decay rates of the discrete eigenmodes \citep[][]{Mao2011}. In the following section, we will show that the decay behaviours of the inviscid critical-layer eigenmodes are comparable to those of the discrete eigenmodes. Therefore, we cast doubt on whether $\sigma_p^{\nu}$ accurately represents the viscous remnants of $\sigma_c^{0}$ that result from the viscous regularisation of the critical layers. If there exist spectra associated with eigenmodes that possess not only regularised critical-layer structures due to viscosity but also exhibit radial decay behaviours similar to those seen in the inviscid critical-layer eigenmodes, it would be accurate to refer to them as the \textit{true} viscous remnants of $\sigma_c^{0}$. We propose to distinguish these spectra and call them the ``viscous critical-layer spectrum,'' denoted $\sigma_{c}^{\nu}$. Using the present numerical method, we will demonstrate that $\sigma_c^{\nu}$ is formed by two distinct curves near the right end of $\sigma_p^{\nu}$, as depicted in figure \ref{fig:spectrum_scheme}$(b)$.

\section{Inviscid linear analysis}\label{inviscidlinearanalysis}
The eigenvalue problem $\sigma \left[ \mathbb{P}_{m\kappa} ( \Tilde{\bm{u}} ) \right] = \mathcal{L}_{m\kappa}^{0} \left[ \mathbb{P}_{m\kappa} ( \Tilde{\bm{u}} ) \right]$ is analysed by finding the spectra of the discretised operator $\mathsfbi{L}_{m\kappa}^{0}$ and their associated eigenmodes. Since the number of spatially resolved \textit{discrete} eigenmodes is typically far less than $M$ due to the spatial resolution limit, the majority of numerical eigenmodes should be associated with the \textit{continuous} critical-layer spectrum $\sigma_c^{0}$. Although $\sigma_c^{0}$ is associated with neutrally stable eigenmodes, its numerical counterpart often creates a ``cloud'' of incorrect eigenvalues clustered around the true location of $\sigma_c^{0}$, as observed by \citet{Mayer1992, fabre2004, Heaton2007}. However, the previous studies that observed this incorrect spectrum paid less attention to its correction, which is our major interest, as they were primarily interested in discrete unstable modes that can be resolved out of (and thus are sufficiently far from) the cloud. When discrete unstable eigenmodes are present for small $q$, the most unstable one prevails in the linear instability of the $q$-vortex. Therefore, the presence of these incorrect eigenmodes may not be problematic.

On the other hand, for large $q$ (typically, $|q|>1.5$ according to \citet{Lessen1974}, or $|q|>2.31$ according to \citet{Heaton2007}, depending on the parameter values of $m$ and $\kappa$) where the inviscid $q$-vortex is linearly neutrally stable and the eigenmodes are located on $\mathrm{i} \mathbb{R}$ of the complex $\sigma$-plane. Although the flow is analytically neutrally stable, incorrect eigenmodes may appear in association with eigenvalues clustered around the imaginary axis, leading to the incorrect conclusion that the flow is linearly unstable because some of the eigenvalues lie in the right half of the complex $\sigma$-plane ($\Real (\sigma) > 0$). We focus our attention on the analysis of large or infinite $q$ cases as any unstable eigenmodes occurring in the analysis are incorrect. In what follows, we demonstrate that these incorrect eigenmodes are \textit{under-resolved} eigenmodes of the inviscid critical-layer eigenmodes and can be corrected by adjusting the numerical parameters so that they correctly exhibit their neutrally stable nature ($\Real (\sigma) = 0$) in our numerical analysis.

\subsection{Numerical spectra and eigenmodes}
\begin{figure}
  % \vspace{0.1in}
  \centerline{\includegraphics[width=\textwidth,keepaspectratio]{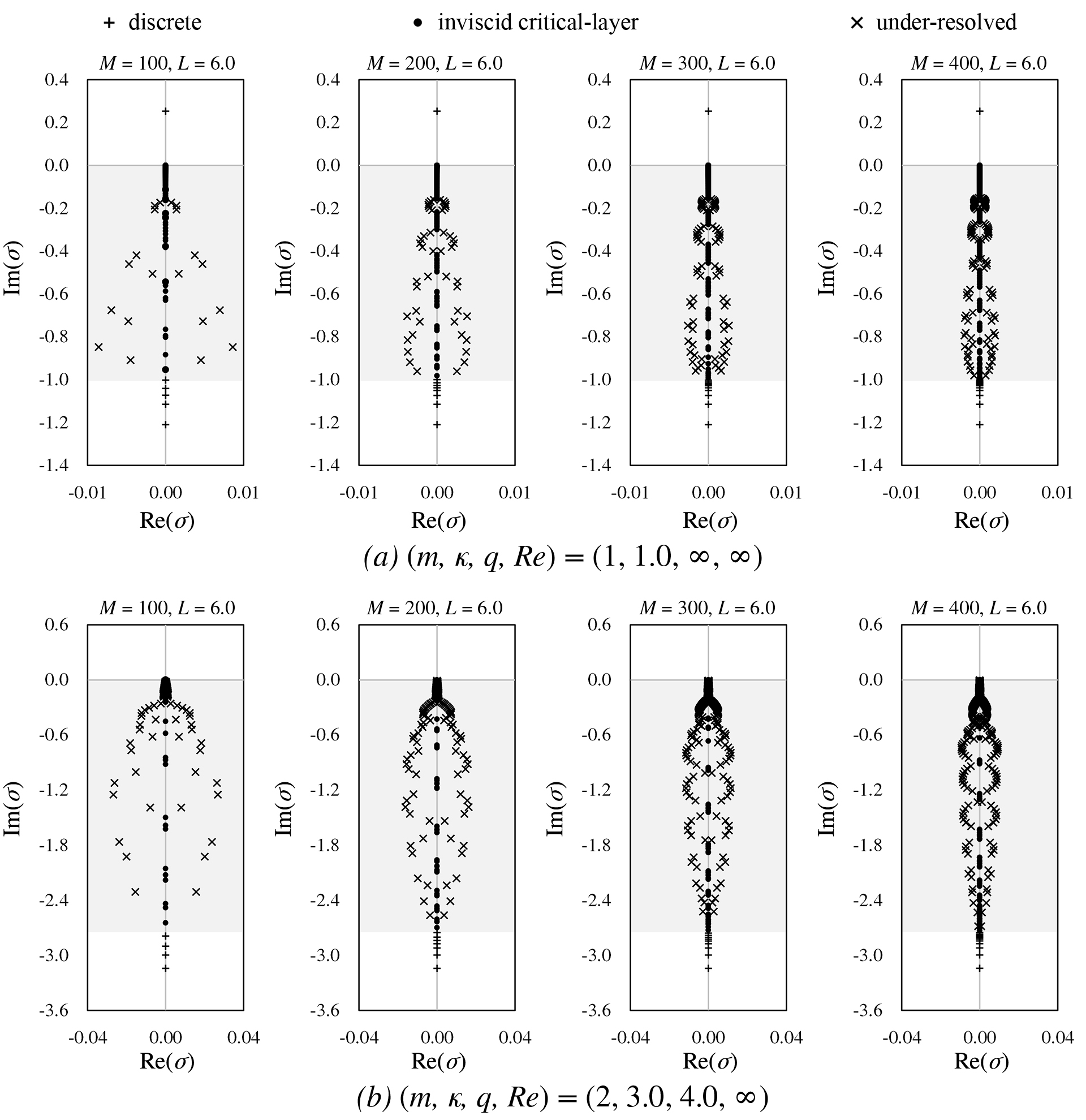}}
  \caption{Numerical spectra computed with zero viscosity $(a)$ for the Lamb-Oseen vortex ($q \rightarrow \infty$) in $(m, \kappa) = (1, \; 1.0)$ and $(b)$ for the strong swirling Batchelor vortex ($q = 4.0)$ in $(m, \kappa) = (2, \; 3.0)$ with respect to $M=100,~200,~300$ and $400$. $L$ is fixed at $6.0$ and $N=M+2$. A shaded band in each plot indicates the non-normal region where $\sigma_c^{0}$ appears. The larger $M$ we use, the closer the numerical spectra is to their true shape (see figure \ref{fig:spectrum_scheme}$(a)$). However, with sufficiently large values of $M$ and with appropriately tuned values of $L$, the under-resolved can be corrected, making all eigenvalues lie on the imaginary axis; see figure~\ref{fig:inviscid_vary_l}.}
\label{fig:inviscid_vary_m}
\end{figure}

In figure \ref{fig:inviscid_vary_m}, we present the eigenvalues of two inviscid vortices: the Lamb-Oseen vortex with $(m, \kappa, q) = (1, 1.0, \infty)$ and the strong swirling Batchelor vortex with $(m, \kappa, q) = (2, 3.0, 4.0)$. By comparing these two vortices, we demonstrate their common properties and extract features that can be generalised to vortices with large $q$ and moderate $m$ and $\kappa$ of order unity, which are thought to be relevant for practical aeronautical applications \citep[see][pp. 258-259]{fabre2004}. To observe the effect of the numerical parameter $M$, we computed each vortex in four ways: with $M = 100,~ 200,~300$, and $400$. Analytically, every eigenvalue is expected to lie on $\mathrm{i}\mathbb{R}$. The shaded area in each plot is the non-normal region of the spectra, indicating the frequency band that includes the analytic range of $\sigma_c^{0}$.

Clearly, all these \textit{numerical} spectra contain some eigenvalues that are incorrect (i.e, not on the imaginary axis). We can observe three families of numerical eigenvalues. A discrete family ($+$) corresponds to $\sigma_d^{0}$, where the eigenvalues are discrete and located outside the shaded area. An inviscid critical-layer family ($\bullet$) corresponds to $\sigma_c^{0}$. Its eigenvalues lie on the imaginary axis, are within the shaded area, and the number of them increases as $M$ increases. Finally, a family of under-resolved eigenvalues ($\times$), which, had they been spatially well-resolved, would have been eigenvalues belonging to $\sigma_c^{0}$ and lie on the imaginary axis. Instead, these eigenvalues lie off the imaginary axis and within the shaded area. These under-resolved eigenvalues are characterised by non-zero real parts with absolute values typically greater than $10^{-10}$ as a result of numerical discretisation errors. The eigenvalues form clouds of structures that are symmetric about the imaginary axis. The cloud structures are due to insufficient spatial resolution, and the absolute values of the real parts of the eigenvalues tend to increase as the value of $q$ decreases. As $M$ increases, the absolute values of the real parts of the eigenvalues tend to decrease, and the cloud of eigenvalues gets ``squeezed'' to the imaginary axis, which is similar to the ``squeeze'' observed by \citet{Mayer1992} when they increased the number of Chebyshev basis elements in their spectral method calculation.

\subsubsection{Discrete eigenmodes}\label{invisciddiscreteeigenmodes}
\begin{figure}
  % \vspace{0.1in}
  \centerline{\includegraphics[width=\textwidth,keepaspectratio]{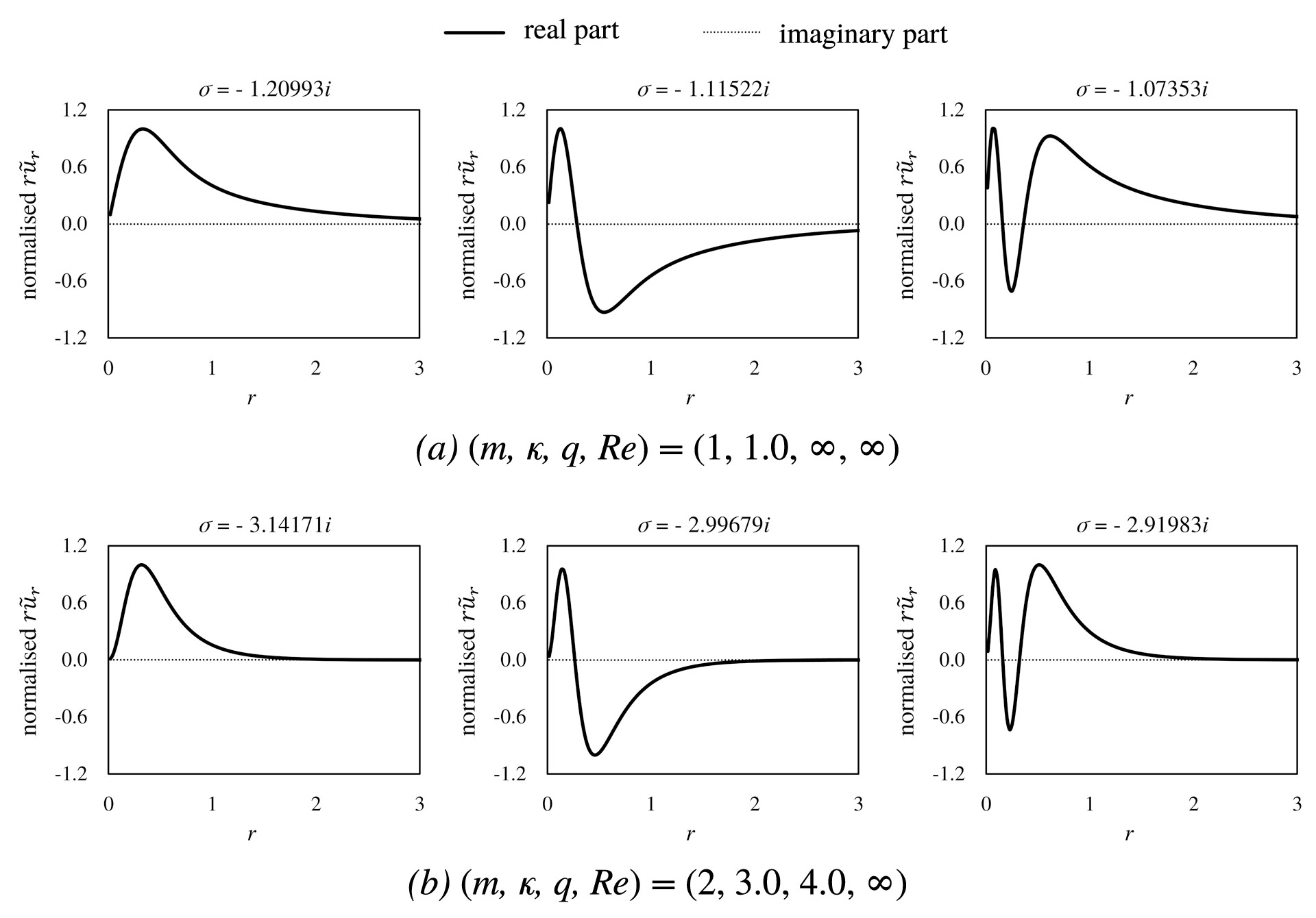}}
  \caption{Radial velocity profiles of the inviscid discrete eigenmodes associated with three largest $|\Imag (\sigma)|$ $(a)$ for the Lamb-Oseen vortex ($q \rightarrow \infty$) in $(m, \kappa) = (1, \; 1.0)$ and $(b)$ for the strong swirling Batchelor vortex ($q = 4.0)$ in $(m, \kappa) = (2, \; 3.0)$. The maximum of $\Real(r \Tilde{u}_r)$ is normalised to unity. $M=400$ and $L=6.0$ are used. The number of ``wiggles'' in and around the vortex core distinguishes each discrete eigenmode. Note that, for the eigenmodes that are neutrally stable, the phase of the eigenmodes can be chosen such that the radial velocity components are made to be either real or pure imaginary for all $r$.}
\label{fig:inviscid_discrete}
\end{figure}

Although $\sigma_d^{0}$ and the discrete eigenmodes are not the main focus of this paper, it is worthwhile to confirm their convergence properties. Figure \ref{fig:inviscid_vary_m} shows that the discrete eigenmodes associated with eigenvalues away from the accumulation points \citep[see][pp. 14-16]{Gallay2020} (i.e., intersections of the imaginary axis with the lower boundary of the shaded regions in figure \ref{fig:inviscid_vary_m}) are spatially resolved for $M \geq 100$, $L=6$, and $N=M+2$. For these values of $L$, $M$, and $N$, each eigenvalue approaches a fixed point as $M$ increases. The discrete eigenmodes are distinguishable from each other by their radial structures and, in particular, by the number of ``wiggles'' (intervals between two neighboring zeroes) as a function of radius. Typically, the eigenmodes with eigenvalues farthest from the accumulation points have the fewest wiggles, as shown in figure \ref{fig:inviscid_discrete}. The discrete eigenmodes have an increasing number of wiggles as the eigenvalue approaches the accumulation point, forming a countably infinite, linearly independent set in the eigenspace of $\mathcal{L}_{m\kappa}^{0}$.

The eigenmodes with discrete eigenvalues and $\Imag(\sigma) / m > 0$ in
figure \ref{fig:inviscid_vary_m} were referred to as ``countergrade'' by \citet{Fabre2006}. They appear to exist only for eigenmodes with specific values of $m$, including $m = \pm 1$ \citep[see][]{Gallay2020}. However, we remark that these eigenmodes are also legitimate solutions to the problem and can be spatially resolved using our numerical method, just like those shown in figure \ref{fig:inviscid_discrete}. They are also expected to be crucial for triad-resonant interactions among the eigenmodes and will be actively considered in further instability studies.

The numerically computed eigenmodes correspond to the eigenvectors of the $2M \times 2M$ matrix $\mathsfbi{L}_{m\kappa}^{0}$, which implies that the maximum number of numerical eigenmodes that can be obtained is $2M$ under double-precision arithmetic. The number of discrete eigenmodes that our numerical solver can find increases with an increase in $M$. For instance, in the case of a strong swirling Batchelor vortex illustrated in figure \ref{fig:inviscid_vary_m}$(b)$, the number of discrete eigenmodes (i.e., in the $\sigma_d^{0}$ spectrum) is $4$, $7$, $9$, and $11$ with respect to $M=100$, $200$, $300$, and $400$, respectively. This behaviour is expected because a finer spatial resolution is required to resolve more wiggles in the eigenmode structure. If $n$ wiggles exist in the vortex core region $(r \le 1.122)$, whose non-dimensionalised scale is of order unity, the necessary spatial resolution to resolve all the wiggles is $O(1/n)$. As $\Delta = 2L/(M+2) \sim O(1/M)$ in our analysis, the proportionality of $n$ to $M$ is verified. The implication of this scaling is that the number of discrete eigenmodes accounts for only a small portion of the total number of numerical eigenmodes computed, and the vast majority are associated with the non-regular, continuous part of the spectrum, $\sigma_{c}^{0}$.

\begin{figure}
  % \vspace{0.1in}
  \centerline{\includegraphics[width=\textwidth,keepaspectratio]{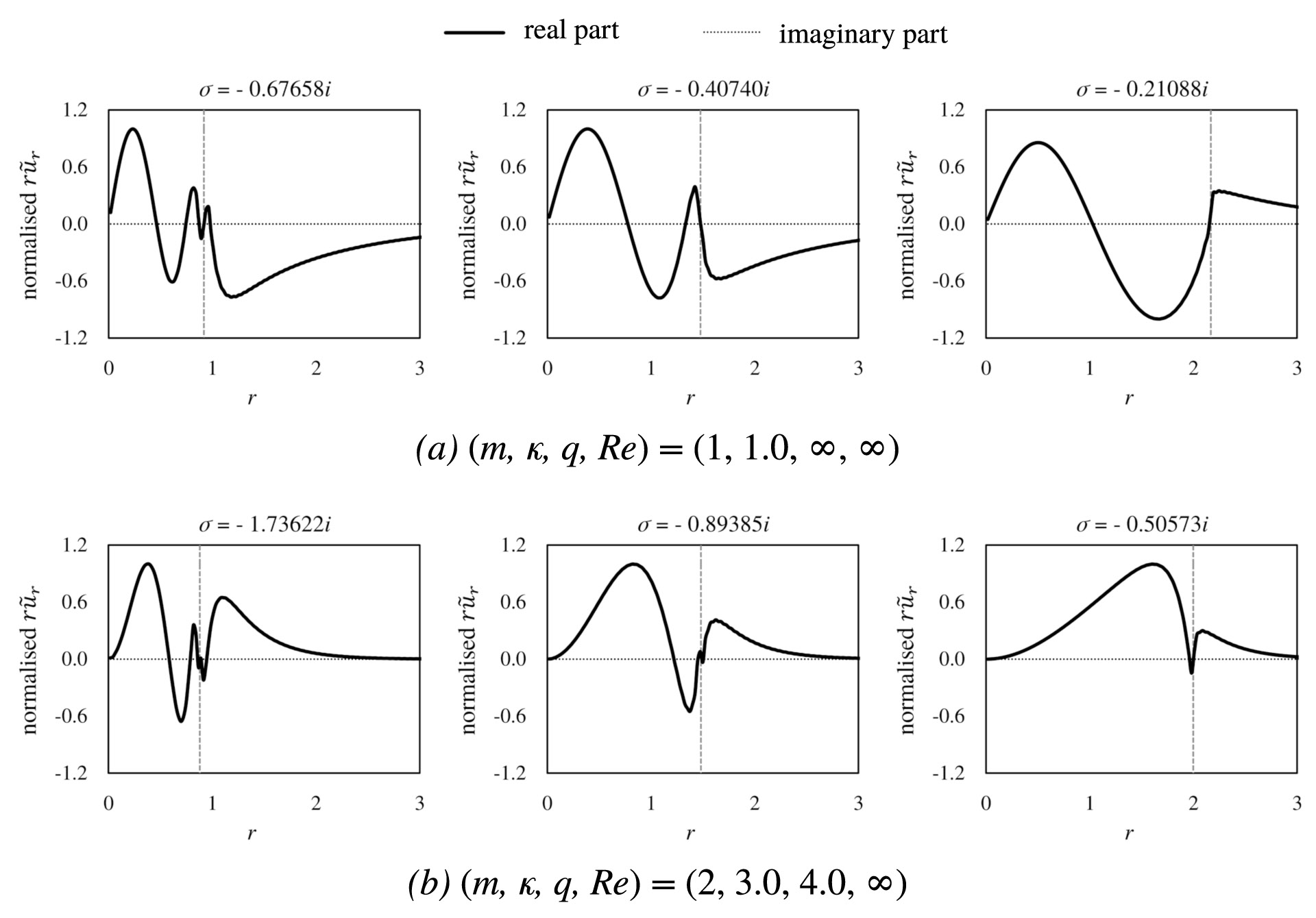}}
  \caption{
  Radial velocity profiles of three inviscid, critical-layer eigenmodes $(a)$ for the Lamb-Oseen vortex ($q \rightarrow \infty$) in $(m, \kappa) = (1, \; 1.0)$ and $(b)$ for the strong swirling Batchelor vortex ($q = 4.0)$ in $(m, \kappa) = (2, \; 3.0)$. The maximum of the real part of $r \Tilde{u}_r$ is normalised to unity. $M=400$, $N=M+2$, and $L=6.0$ are used. For each eigenmode, the vertical dashed line indicates the critical layer location $r_c$ determined by \eqref{sigma_r_c_qvort}. Note that all of the radial components of the velocity can be made to be real-valued for all $r$ by a proper choice of phase as they are neutrally stable as well as discrete ones.
  }
\label{fig:inviscid_critical}
\end{figure}

\subsubsection{Inviscid critical-layer eigenmodes}\label{inviscidcriticallayereigenmodes}
We emphasise that our essential interest lies in eigenmodes with small, but non-zero viscosity. This ensures that the eigenmodes can be physical and do not have difficult-to-compute singularities. Nevertheless, it is still intriguing to compute the eigenmodes with $\nu \equiv 0$, which are numerically (not physically) regularisable by the spatial discretisation. By selecting a suitably large value of $M$ and an appropriate value for the mapping parameter $L$ (see \S \ref{sec:correction}), we can resolve the spatial structure of the inviscid eigenmode outside the critical-layer singularity neighbourhood well. In addition, the numerical error in the eigenvalue, caused by the slow decay of spectral coefficients or the Gibbs phenomenon around the critical-layer singularity, can be kept adequately small and the eigenvalues correctly lie on the imaginary axis.

Figure \ref{fig:inviscid_critical} shows some critical-layer eigenmodes, which were numerically obtained with $M=400$. The real parts of the eigenvalues are zero, and the velocity components are either real or purely imaginary for all $r$, with a suitable phase choice. Typically, $r_c$ increases as $|\sigma|$ decreases along the critical-layer spectrum. The singular behaviour of abrupt slope change commonly occurs at the critical layer singularity, as predicted analytically by \eqref{sigma_r_c_qvort}. As stated in \S \ref{sec:prelim}, we cannot claim that they are perfectly resolved due to the presence of the singularity and the continuous nature of their associated spectrum. However, our focus is not on their exact convergence but rather on their well-behaved spatial structure outside the neighbourhood of the singularity, achieved by using a large $M$, along with purely imaginary eigenvalues that conform to analytic expectations. We use this information later to study the spatial correspondence of eigenmodes with non-zero viscosity to determine which viscous eigenmodes are of physical relevance.

For $r < r_c$, the radial velocity components of the inviscid critical-layer eigenmodes oscillate in $r$, and the number of oscillations decreases as the value of $r_c$ increases (or equivalently, as $|\sigma|$ decreases). Consequently, when $r_c > r_{cc}$ for some value $r_{cc}$, there is no longer one full oscillation. In our numerical investigation, we found that for the Lamb-Oseen vortex with $(m,\kappa) = (1,1.0)$, $r_{cc}$ equals $2.2$, which corresponds to $\sigma = -0.21\mathrm{i}$. We believe that our numerically found value of $r_{cc}$ approximately coincides with the theoretical threshold of $r=2.124$, at which the analytic solutions obtained by the Frobenius method change form regarding the roots of the indicial equation \citep[see][p. 20 and p. 50]{Gallay2020}. For $r > r_c$, the radial velocity components of the critical-layer eigenmodes are not oscillatory, and the amplitudes of $r\tilde{u}_r$ achieve the local maximum or minimum values close to $r=r_c$, before decreasing monotonically as rapidly as those of the discrete eigenmodes, as shown in figure~\ref{fig:inviscid_discrete}.

\subsubsection{Under-resolved eigenmodes}
The under-resolved eigenmodes, which, if resolved, would be part of the spectrum with $\sigma_c^{0}$, have eigenvalues in the complex $\sigma$-plane on either side of the imaginary axis in the shaded region in figure~\ref{fig:inviscid_vary_m}. The eigenvalues come in pairs, with one unstable and one stable eigenmode. The reflection symmetry with respect to the imaginary axis is due to the fact that the analytic operator $\mathcal{L}_{m\kappa}^{0}$ is time-reversible \citep[cf.][p. 10]{Bolle2020}. Therefore, the eigenmode $(\tilde{u}_{r}, \tilde{u}_{\phi}, \tilde{u}_{z}, \tilde{p})$ with eigenvalue $\sigma_s$ corresponds to the eigenmode $(\tilde{u}_{r}^{*}, -\tilde{u}_{\phi}^{*}, -\tilde{u}_{z}^{*}, -\tilde{p}^{*})$ with eigenvalue $-\sigma_s^*$.

\begin{figure}
  % \vspace{0.1in}
  \centerline{\includegraphics[width=\textwidth,keepaspectratio]{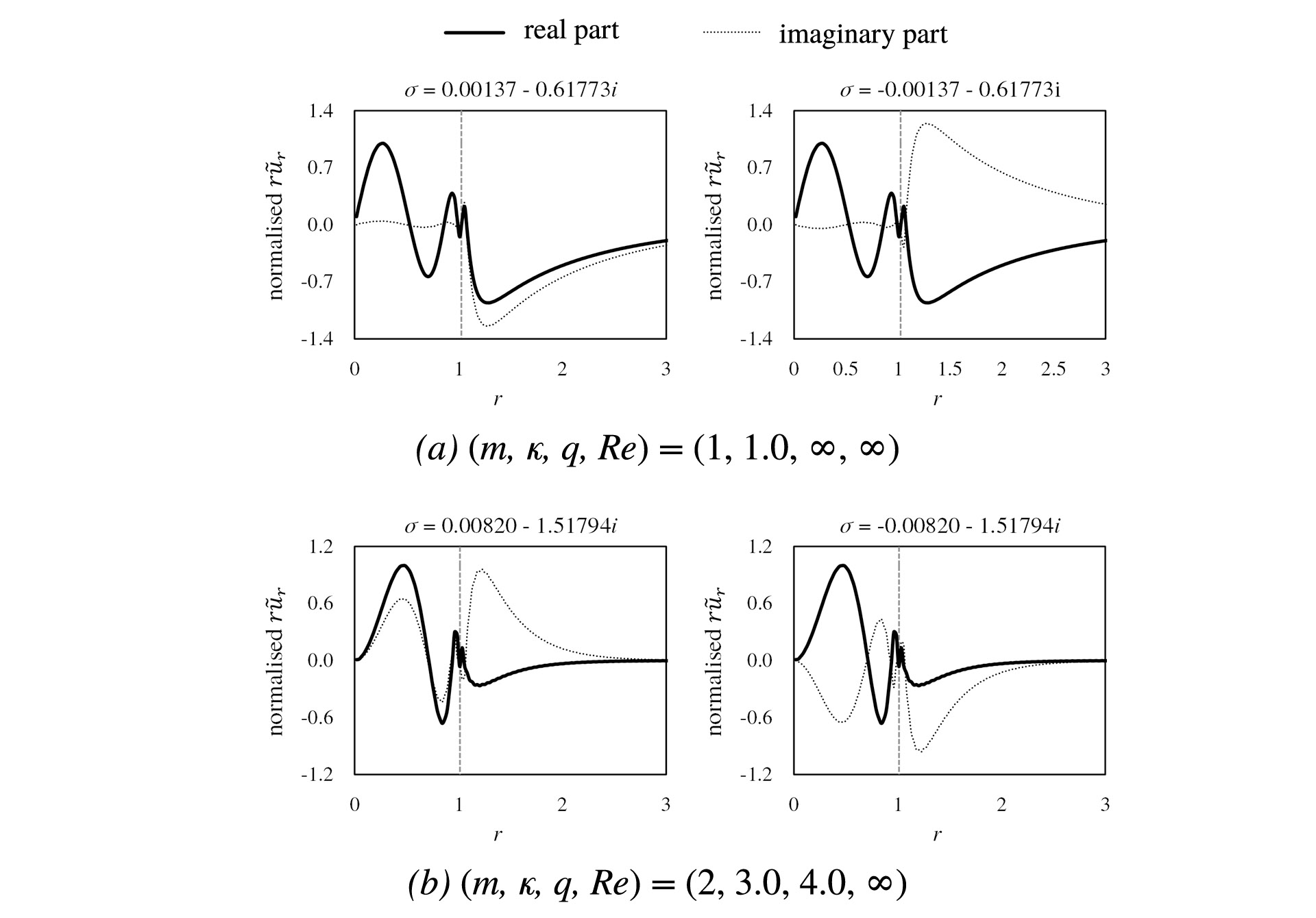}}
  \caption{Radial velocity profiles of two inviscid under-resolved eigenmodes whose eigenvalues are symmetric about the imaginary axis $(a)$ for the Lamb-Oseen vortex ($q \rightarrow \infty$) in $(m, \kappa) = (1, \; 1.0)$ and $(b)$ for the strong swirling Batchelor vortex ($q = 4.0)$ in $(m, \kappa) = (2, \; 3.0)$. The maximum of the real part of $r \Tilde{u}_r$ is normalised to unity. $M=400$ and $L=6.0$ are used. For each eigenmode, an abrupt slope change occurs at the vertical dashed line at the critical layer location, $r =r_c$ (which is determined from \eqref{sigma_r_c_qvort} by ignoring the real part of the eigenvalue), indicating that they will become correct critical-layer eigenmodes given more resolution.
  }
\label{fig:inviscid_spurious}
\end{figure}

Some examples of under-resolved eigenmodes are shown in figure~\ref{fig:inviscid_spurious}. These eigenmodes are qualitatively incorrect because (1) unlike the eigenmodes in figure~\ref{fig:inviscid_discrete} and figure~\ref{fig:inviscid_critical}, there is no choice of phase that makes their radial components real for all $r$, and more importantly, because (2) we know that their eigenvalues should be purely imaginary when $q$ is sufficiently large, and they are not. However, these eigenmodes appear to exhibit no other distinguishing properties, except for the two properties listed above, from the inviscid critical-layer eigenmodes in figure~\ref{fig:inviscid_critical}. It should be noted that they have been called ``\textit{spurious}'' in previous numerical studies \citep[see][]{Mayer1992, Heaton2007}, of which the usage was similar to our clarification given in \S \ref{sec:prelim}. However, instead of following convention, we propose naming these numerical eigenmodes ``\textit{under-resolved}'' eigenmodes of the continuous part of the inviscid spectrum. In this way, we put more emphasis on the fact that adjusting the numerical parameters can ``correct'' these eigenmodes so that neither of the two key properties listed above applies.

By examining the spatial structure of the under-resolved eigenmodes, we can detect sudden changes in slope at the critical-layer singularity point at $r = r_c$. The value of $r_c$ is obtained by setting the imaginary part of either of the eigenvalues $\Imag(\sigma_s)$ to $\sigma_c$ in \eqref{sigma_r_c_qvort}. The break in slope confirms that the under-resolved eigenmodes originate from $\sigma_c^{0}$ and indicates that they have lost their neutrally stable property due to numerical errors at the critical-layer singularity.

Correcting the under-resolved eigenmodes is crucial, not only for correctly evaluating $\sigma_c^{0}$ but also for the following reasons. Despite their invalid origin, half of the under-resolved eigenmodes in $\Real(\sigma) > 0$ erroneously suggest that the wake vortex is linearly unstable. In the future, we plan to use the computed velocity eigenmodes from the present numerical method to initialise an initial-value code that solves the full nonlinear equations of motion given by \eqref{continuity_pert} and \eqref{momentum_pert}. Inappropriately computed eigenmodes that grow erroneously, rather than remain neutrally stable, are likely to corrupt these calculations.

\subsection{Correction of the under-resolved eigenmodes}\label{sec:correction}
An intriguing question is whether the under-resolved eigenmodes tend towards something as $M$ increases. What is the potential outcome of such convergence? In the beginning of this section, it was argued that the real part of eigenvalues remains at zero (i.e., all eigenmodes are neutrally stable) when $q$ is sufficiently large. In figure~\ref{fig:inviscid_vary_m}, this can be observed as the ``squeeze'' of the eigenvalue cluster towards the imaginary axis. However, we have also indicated that the imaginary part of eigenvalues may not converge to a fixed point, instead continuing to evolve along the imaginary axis. Therefore, instead of concentrating on the convergence of individual under-resolved eigenmodes to a fixed point, it is more pragmatic to aim to ``correct'' the set of eigenmodes as a whole, that is, to restore their neutrally stable nature. The ``correction'' means that we comprehensively treat the entire set of eigenmodes as a single entity, which complies with the usage of this term in this section up to this point.

To ``correct'' the under-resolved eigenmodes, We first consider increasing $M$ to its largest possible value within the available computing resources. However, increasing $M$ is generally undesirable because it always comes at a steep computational expense; the cost of finding the eigenmodes is proportional to $(2M)^3$. Instead, we may consider dealing with the mapping parameter $L$, where the novelty and usefulness of our method come from. $L$ controls the spatial resolution locally as a function of $r$. As seen from the resolution parameter $\Delta$ in \eqref{resolution}, $L$ controls the spatial resolution by providing more resolution near the radial origin (i.e., $0 \leq r < L$). It is important to note that changing or tuning $L$ does not affect the cost of computation.

\begin{figure}
  % \vspace{0.1in}
  \centerline{\includegraphics[width=\textwidth,keepaspectratio]{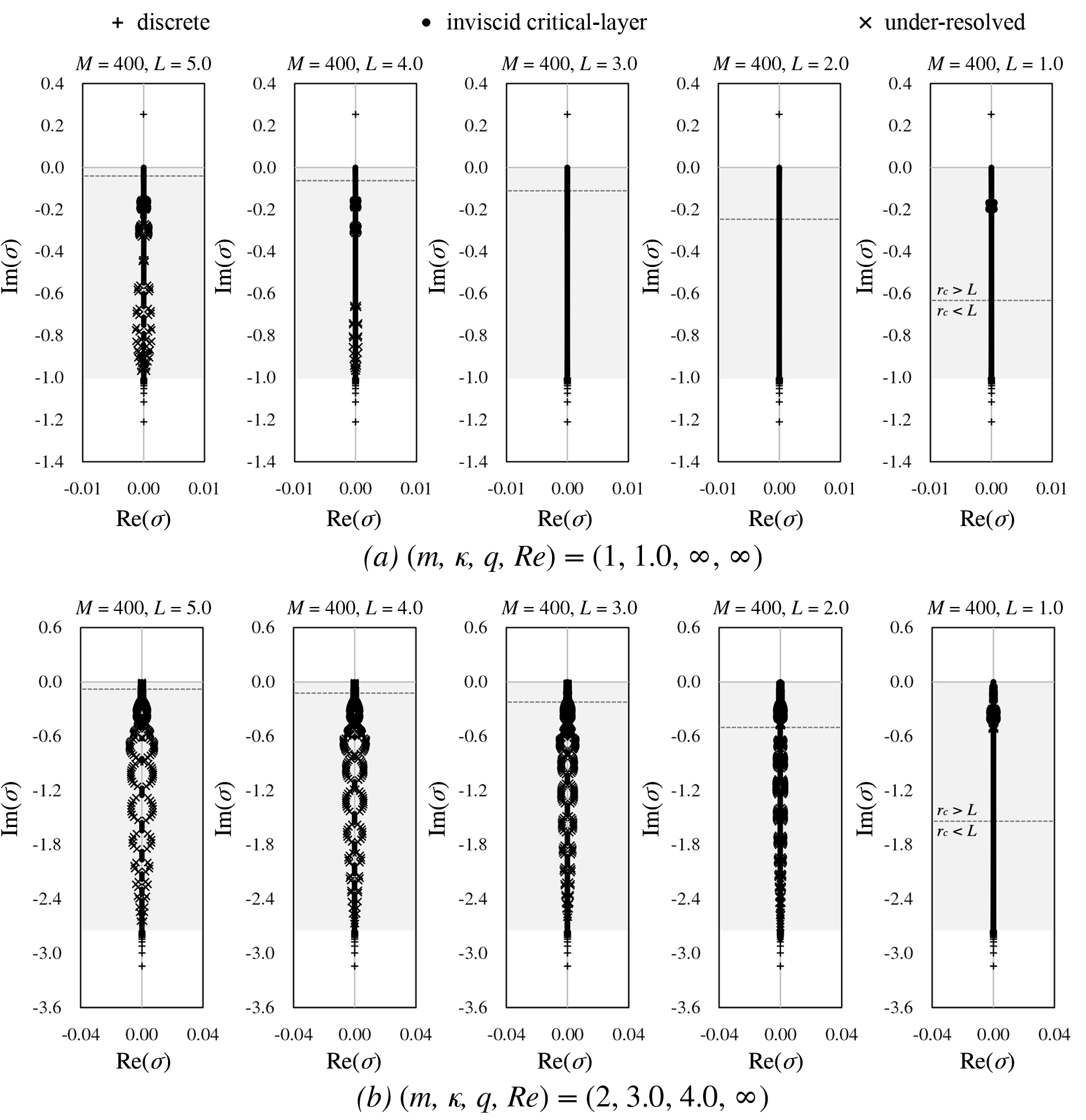}}
  \caption{Numerical spectra computed at zero viscosity $(a)$ for the Lamb-Oseen vortex ($q\rightarrow \infty$) in $(m,\kappa)=(1,~1.0)$ and $(b)$ for the strong swirling Batchelor vortex ($q=4.0$) in $(m,\kappa)=(2,~3.0)$ with respect to $L= 5.0,~4.0,~3.0,~2.0$ and $1.0$. $M$ is fixed at $400$ and $N=M+2$. n each plot, a shaded band indicates the non-normal region in which $\sigma_c^{0}$ appears, and a horizontal dashed line represents the threshold used to determine if the critical layer $r=r_c$ is located within the high-resolution region $0\le r <L$. It should be noted that there is a one-to-one correspondence between a critical-layer eigenvalue $\sigma$ and a critical-layer radius $r_c$, as seen in \eqref{sigma_r_c_qvort}. Furthermore, $r_c$ approaches zero at the bottom of the shaded band, $\Imag(\sigma) = m + \kappa /q$, and monotonically increases towards infinity as $|\sigma|$ becomes smaller. By tuning $L$, under-resolved eigenmodes can be corrected without requiring additional computing resources.
  }
\label{fig:inviscid_vary_l}
\end{figure}

For a fixed $M$, with $N=M+2$, figure \ref{fig:inviscid_vary_l} shows five numerical eigenvalue spectra for two prescribed cases with different values of $L$, varying from $1.0$ to $5.0$. Overall, decreasing $L$ brings the numerical spectra closer to the imaginary axis. In particular, some values of $L$ enable complete resolution of $\sigma_c^{0}$ on the imaginary axis, which cannot be achieved by increasing $M$ within a modest computing budget. However, decreasing $L$ does not always shrink the clouds of eigenvalues closer to the imaginary axis. We separate the numerically computed eigenvalues and eigenmodes into two categories: those with the critical layer $r=r_c$ located in the high-resolution region $0 \le r < L$, and those where $r_c$ is in the low-resolution region $r \ge L$. Figure \ref{fig:inviscid_vary_l} indicates this separation with horizontal dashed lines. For the former, the cloud structures vanish as $L$ decreases, and $\sigma_c^0$ is correctly resolved. In contrast, for the latter, the cloud structures persist or even recur if $L$ is too small, resulting from excessive concentration of collocation points solely around the center. Once $\sigma_c^{0}$ is satisfactorily resolved, adjusting $L$ should stop to keep the portion of $\sigma_c^{0}$ resolved in the high-resolution region as large as possible. For instance, in figure \ref{fig:inviscid_vary_l}, we propose setting $L$ between $3.0$ and $4.0$ for the Lamb-Oseen vortex case and between $1.0$ and $2.0$ for the Batchelor vortex case.

To provide a detailed explanation of what we have seen, we must revisit the differences in the way $M$ and $L$ operate in the current numerical method, as stated in \S \ref{sec:numericalparameters}. One of the roles of $L$ is to serve as a tuning parameter for spatial resolution in physical space, whereas $M$ determines the number of basis elements used in spectral space. Increasing $M$ allows us to handle eigenmodes with more complex shapes, such as (nearly) singular functions, which often have more wiggling and are thus more numerically sensitive. $M$ has only an indirect effect on spatial resolution through $N$, which is required to be greater than or equal to $M$. On the other hand, the critical-layer singularity is essentially a phenomenon that occurs in physical space. Although using more spectral basis elements relates to improving spatial resolution because we set $N = M+2$, the main contribution to dealing with the critical-layer singularity with minimal errors comes from the latter. Therefore, it can be more effective to use $L$ to directly control resolution and suppress the emergence of under-resolved eigenmodes, rather than using $M$. It is worth noting that increasing $N$ to very large values while keeping $M$ constant can also reduce the number of under-resolved eigenvalues to some extent. This observation supports that high spatial resolution is crucial for suppressing under-resolved eigenmodes.

If one aims to correct the under-resolved eigenmodes and obtain $\sigma_c^{0}$ using the present numerical method, the following steps are suggested to properly set up the numerical parameters. Assuming that $M$ is already at the practical maximum due to finite computing budget, and $N$ follows $M+2$:
\begin{enumerate}
    \item Start with an arbitrarily chosen value of $L$ and gradually decrease it if under-resolved eigenmodes exist, until they vanish in the high-resolution region $0 \le r < L$. This step improves spatial resolution, helping to identify the critical-layer singularity with less numerical error despite the discretisation.
    \item If there are no eigenvalue clouds around the imaginary axis, increase $L$ as long as they do not appear in the numerical spectra. This step expands the high-resolution region where the critical-layer singularity can be accurately treated.
\end{enumerate}

\subsection{Pairing in the inviscid critical-layer spectrum}\label{sec:pairing}

\begin{figure}
  % \vspace{0.1in}
  \centerline{\includegraphics[width=\textwidth,keepaspectratio]{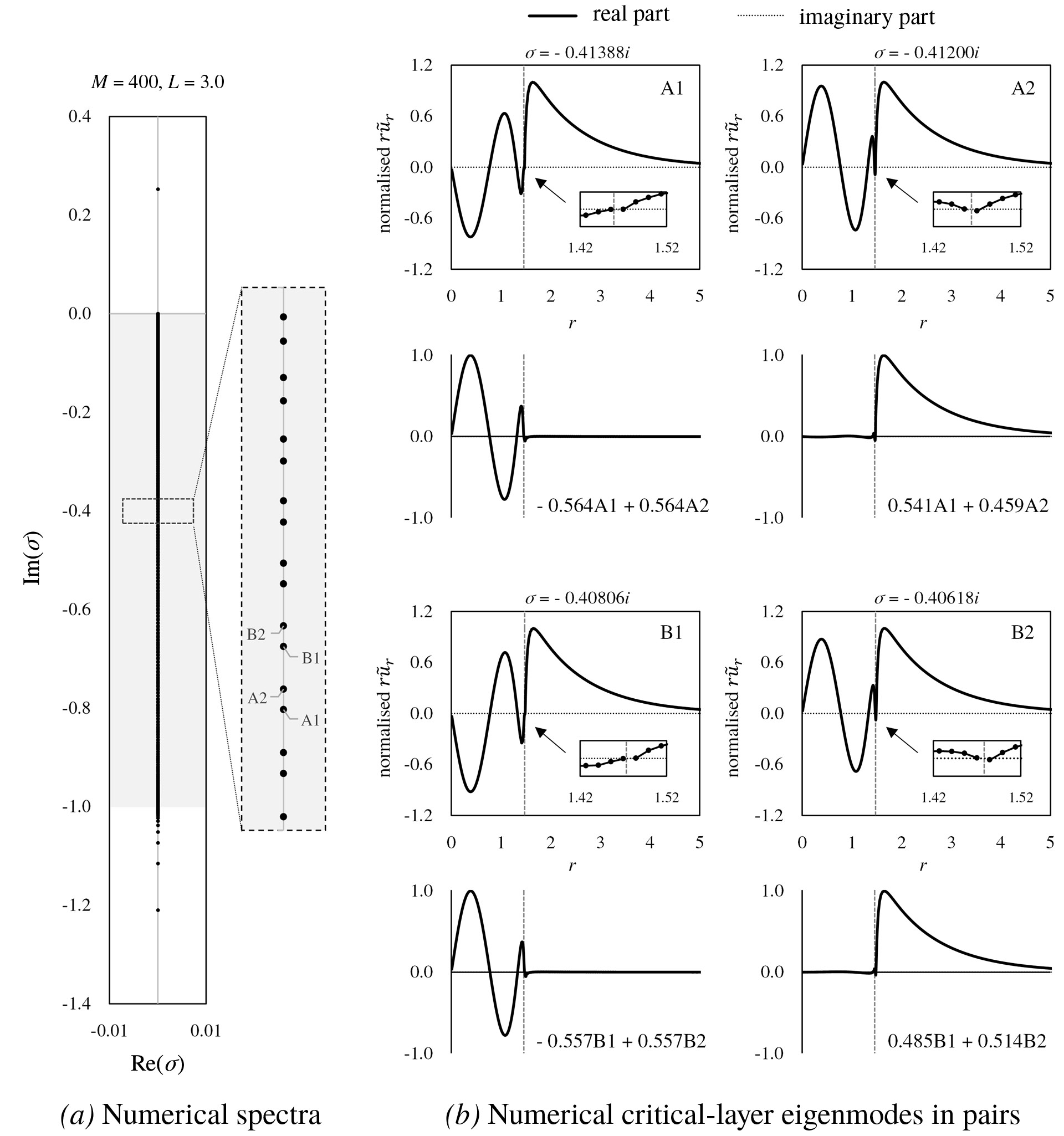}}
  \caption{$(a)$ Numerical inviscid spectra with no under-resolved eigenmodes for the Lamb-Oseen vortex ($q\rightarrow\infty$) in $(m,\kappa)=(1,~1.0)$ along with a magnified part exhibiting the pairing phenomenon, and $(b)$ four radial velocity profiles of the critical-layer eigenmodes from two neighboring pairs, labelled as A1/2 and B1/2. Here, $M=400$, $L=3.0$, and $N=M+2$. Note the similarity in structure within each pair, and the change in the critical layer location (marked by vertical dashed lines) by one collocation point between these neighboring pairs. This pairing phenomenon stems from the singular degeneracy in $\sigma_c^{0}$. The linear combination of the pair constructs two independent solutions that are singular at the same critical-layer location and are nearly zero on either $(0,r_c)$ or $(r_c, \infty)$.}
\label{fig:inviscid_pairing}
\end{figure}

In the numerical spectrum of $\sigma_c^{0}$, we observe that numerical eigenvalues tend to appear in pairs. This pairing phenomenon is demonstrated in the left panel of figure \ref{fig:inviscid_pairing}, which illustrates the Lamb-Oseen vortex case with $(m,\kappa) = (1,~1.0)$ that we computed with $M=400$ and $L=3.0$ (see figure \ref{fig:inviscid_vary_l}$(a)$). We argue that the pairing in our numerical results arises from a degeneracy resulting from the critical-layer singularity. We refer to \citet[][pp. 19-21]{Gallay2020}, who used the Frobenius method to construct analytic solutions to the problem under the assumption of non-zero $m$ and $\kappa$ with $q \rightarrow \infty$. They showed that if a critical-layer singularity occurs at $r=r_c$, there exists a unique solution with scalar multiplication that is only non-zero on $(0, r_c)$ and another one that is only non-zero on $(r_c, \infty)$. Here we call them inner and outer solutions, respectively. For both the inner and outer solutions, the radial velocity components can be made real-valued by an appropriate choice of phase, since their degenerate eigenvalue is purely imaginary. These two solutions are independent of each other, and their linear combination should be the general form of an inviscid critical-layer eigenmode that is singular at $r=r_c$.

We can observe these analytic characteristics in our numerically computed pairs. In the right panel of figure \ref{fig:inviscid_pairing}, we present the $r \tilde{u}_r$ profiles of the critical-layer eigenmodes from two neighboring pairs. In each pair, the velocity profiles have an abrupt change in slope across an interval between two collocation points, whose location matches the critical-layer singularity radius calculated by \eqref{sigma_r_c_qvort}. The difference in $r_c$ among the neighboring pairs corresponds to the collocation interval, indicating their continuous emergence. Furthermore, by linearly combining these paired eigenmodes, we can construct the inner and outer solutions as derived analytically, each of which is approximately zero on $(0,r_c)$ or $(r_c, \infty)$. Although their eigenvalues are slightly different, we believe that it is due to the numerical error resulting from the spatial discretisation, which slightly breaks the degeneracy. This error decreases with increasing $M$.

Strictly speaking, the discussion made here is limited to infinite $q$, because the analytic results found in \citet[][]{Gallay2020} were verified in the Lamb-Oseen vortex case, and we can only compare this case. Nonetheless, we remark that we have numerically observed this same pairing phenomenon with finite $q$ (e.g., $q=4.0$). We conjecture that the pairing phenomenon exists for values of $q$ where the eigenmodes are all neutrally stable.

\section{Viscous linear analysis}\label{viscouslinearanalysis}
We numerically examine the viscous eigenvalue problem $\sigma \left[ \mathbb{P}_{m\kappa} ( \Tilde{\bm{u}} ) \right] = \mathcal{L}_{m\kappa}^{\nu} \left[ \mathbb{P}_{m\kappa} ( \Tilde{\bm{u}} ) \right]$ by studying the spectra of the discretised operator $\mathsfbi{L}_{m\kappa}^{\nu}$ and their associated eigenmodes. Due to viscous regularisation, the viscous eigenmodes do not exhibit critical-layer singularities. Instead, at or near the locations where the inviscid critical layer would have been, the viscous eigenmodes have thin layers characterised by large amplitudes and small-scale oscillations, with widths proportional to $\Rey^{-1/3}$ \citep{Maslowe1986, LeDizes2004}. Note that the $\Rey^{-1/3}$ law is a well-established analytic principle, similar to the $\Rey^{-1/2}$ law for the laminar viscous boundary layer thickness. Several classic textbooks have already provided an in-depth description of this principle \citep[see][]{lin1955theory, Drazin2004}. 

The families of viscous eigenmodes are not just small corrections to the inviscid eigenmodes; the addition of the viscous term, despite being small for $\Rey^{-1}$, serves as a \textit{singular} perturbation \citep{Lin1961}. This is because it increases the spatial order of the set of equations that govern the eigenmodes. Therefore, the linear stability features of wake vortices from vanishing viscosity can differ from the purely inviscid instability characteristics \citep[see][p. 258]{fabre2004}. It is well known that exactly inviscid flows where $\nu \equiv 0$ often behave quite differently from high Reynolds number flows where $\nu \rightarrow 0^+$. In particular, not only do the locations of the eigenvalues in the complex $\sigma$-plane change, but new families can also be created. One example is the freestream spectrum $\sigma_f^\nu$ shown in figure \ref{fig:spectrum_scheme}. This spectrum consists of non-normalisable eigenmodes that do not vanish as $r \rightarrow \infty$ and is mathematically derivable. However, its non-physical behaviour at radial infinity renders it unsuitable for computation using our method. Otherwise, all other families that we depicted in figure \ref{fig:spectrum_scheme} are in the scope of the analysis. 

\citet{Mao2011} and \citet{Bolle2020} reported that the inviscid critical-layer spectrum $\sigma_c^{0}$ changes with viscosity and spreads to an area on the left half-plane of the complex $\sigma$-plane, which they called the potential spectrum $\sigma_p^{\nu}$. In this section, we demonstrate that our numerical method can produce randomly scattered eigenvalues, which represent $\sigma_p^{\nu}$ numerically as reported by previous authors, and investigate their spatial characteristics. Also, we identify and describe the viscous critical-layer eigenmodes associated with the spectrum $\sigma_c^{\nu}$ (see figure~\ref{fig:spectrum_scheme}), which, to the best of our knowledge, have not been distinguished before.

\subsection{Numerical spectra and eigenmodes}
\begin{figure}
  % \vspace{0.1in}
  \centerline{\includegraphics[width=\textwidth,keepaspectratio]{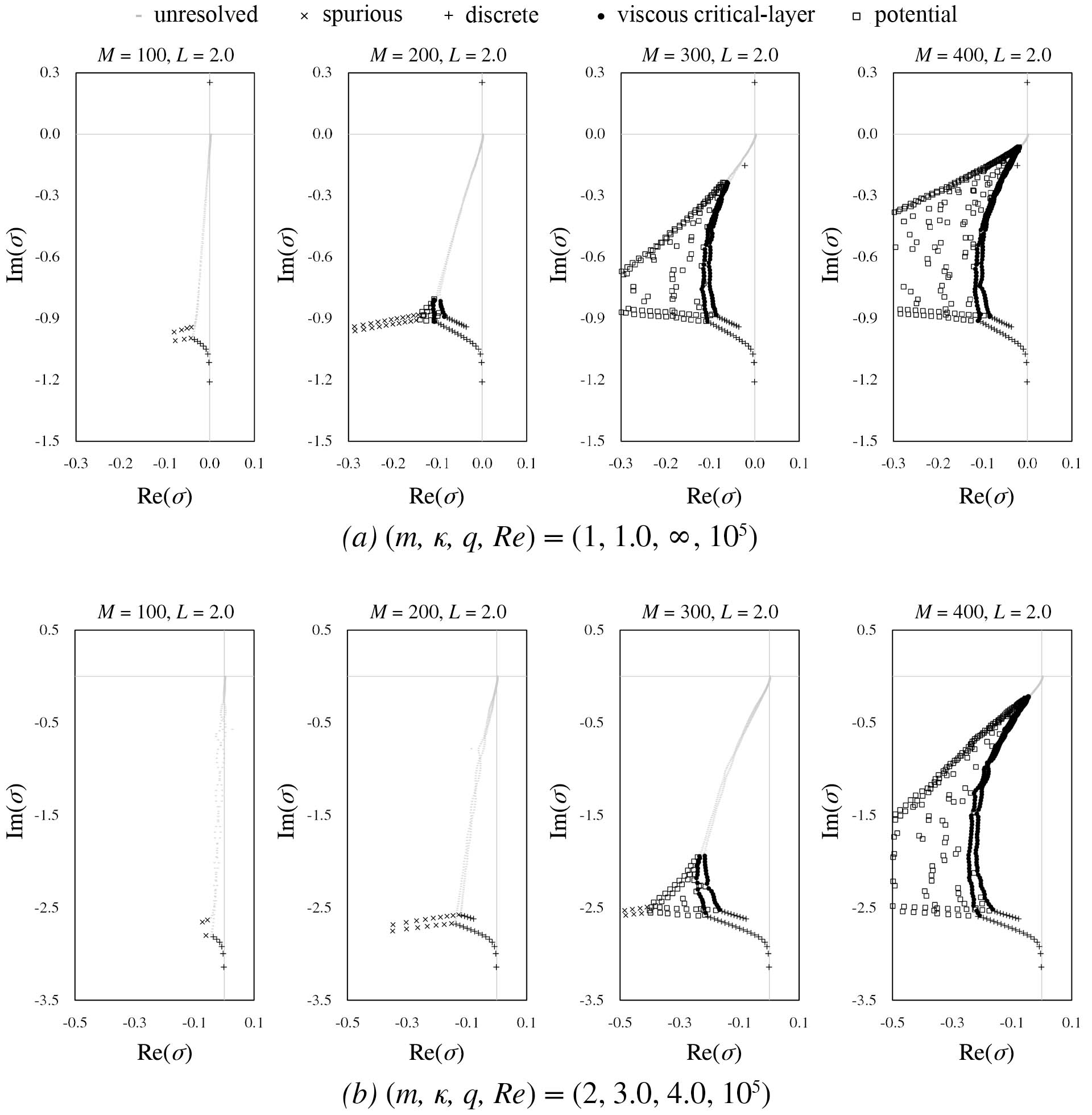}}
  \caption{Numerical viscous spectra at $\Rey = 10^5$ $(a)$ for the Lamb-Oseen vortex ($q\rightarrow\infty$) in $(m,\kappa)=(1,1.0)$ and $(b)$ for the strong swirling Batchelor vortex ($q=4.0$) in $(m,\kappa) = (2,3.0)$ with respect to $M = 100,~200,~300$ and $400$. $L$ is fixed at $2.0$ and $N=M+2$. Larger $M$ enables more portion of the spectra to be resolved. Near the right boundary of the potential spectrum there are  two distinct branches of the viscous critical-layer spectrum. See movie~1 for animation.}
\label{fig:viscous_vary_M}
\end{figure}

The Lamb-Oseen vortex with $(m, \kappa, q) = (1, 1.0, \infty)$ and the strong swirling Batchelor vortex with $(m, \kappa, q) = (2, 3.0, 4.0)$ are analysed at $\Rey = 10^5$. As with the prior inviscid analysis, our aim is to identify common characteristics in the linear vortex dynamics with viscosity for large $q$ and moderate $m$ and $\kappa$. In the current analysis, however, we have a specific focus on the physical relevance of each eigensolution. To observe how the viscous numerical spectra converge, we compute four numerical spectra for each case, with different values of $M$ ranging from $100$ to $400$. The spectra are presented in figure \ref{fig:viscous_vary_M}, or movie 1. Movies of the spectrum animations are provided with the supplementary material available at [JFM]. Based on these spectra, we classify the numerical eigenmodes into five families: unresolved ($\color{lightgray}\bm{-}$), discrete ($+$), spurious ($\times$), potential ($\Box$), and viscous critical-layer ($\bullet$). Note that with increasing numerical resolution, an eigenmode in the unresolved family will always evolve into an eigenmode in one of the other four families.

The eigenvalues in the discrete spectrum, $\sigma_d^{\nu}$, converge to fixed points with increasing $M$. These eigenvalues populate two distinct discrete branches shown near the bottom of the panels in figure \ref{fig:viscous_vary_M}, in addition to a few locations outside of these branches. The eigenmodes in the discrete spectrum were known previously and were studied by other authors. For example, according to \citet{Fabre2006}, the lower branch was designated as the \textsf{C}-family, while the upper one was labelled the \textsf{V}-family. 

With viscosity, none of the eigenmodes have critical-layer singularities as they are regularised, and no eigenvalues lie exactly along the imaginary axis. In the non-normal region, the spectrum of eigenmodes, $\sigma_p^{\nu}$, that we have labelled as potential, occupies an area in the complex $\sigma$-plane that stretches out towards $\Real(\sigma)\rightarrow-\infty$ as $M$ increases. However, unlike the region shown in the schematic in figure \ref{fig:spectrum_scheme}, there is a gap between the upper bound of this numerical spectrum and the real axis on which the freestream spectrum $\sigma_f^{\nu}$ is located. The reason is that we force solutions to vanish at radial infinity due to the decaying nature of the spectral basis elements. Therefore, the gap should be considered a peculiar product of our numerical method that excludes solutions with decay rates that are too slow in $r$, such as those with velocity decaying slower than $O(r^{-1})$ in $r$ and $\phi$, or $O(r^{-2})$ in $z$ in the far field; see \eqref{decayrigor} in \S \ref{sec:tpdecomp}.

The fact that the numerically computed eigenvalues in $\sigma_p^{\nu}$ shift towards the left side of the complex $\sigma$-plane as $M$ increases coincides with \citet{Mao2011}. Additionally, the number of potential eigenmodes increases with increasing $M$. Up to the largest value of $M$ that we have explored, the numerical eigenvalues in $\sigma_p^{\nu}$ tend to emerge randomly. This random scattering can be understood as the spectrum's extremely high sensitivity to numerical errors even in the order of machine precision (see \S \ref{pseudospectral}).

On the other hand, we observe a moving branch of numerical eigenvalues attached to the left end of $\sigma_p^{\nu}$. They also never converge with respect to $M$, and the values of their $|\Real(\sigma)|$ increase rapidly. We explicitly label them as spurious because of their absolutely irregular spatial characteristics, as shown later in \S \ref{viscousspuriouseigenmodes}. Although they are not removable, we can pull them away by setting $M$ to a large value.

Last but not least, we report the presence of two new distinct branches of viscous critical-layer eigenvalues, which are seen on the right side of the area containing the potential eigenvalues. The two branches of these eigenvalues, $\sigma_{c}^{\nu}$, as depicted in figure \ref{fig:spectrum_scheme}, converge to distinct loci. We distinguish $\sigma_{c}^{\nu}$ from $\sigma_{p}^{\nu}$ because of their unique bifurcating shape. Furthermore, this is the only part of the spectra in the non-normal region that approach fixed points at finite $M$, along with the discrete spectrum. As the name suggests, we argue that their associated eigenmodes are not only non-spurious but also physical, since they are the \textit{true} viscous remnants of the inviscid critical-layer eigenmodes, as explained with details in \S \ref{viscouscriticallayereigenmodes}.

\subsubsection{Discrete eigenmodes}\label{viscousdiscreteeigenmodes}
\begin{figure}
  % \vspace{0.1in}
  \centerline{\includegraphics[width=\textwidth,keepaspectratio]{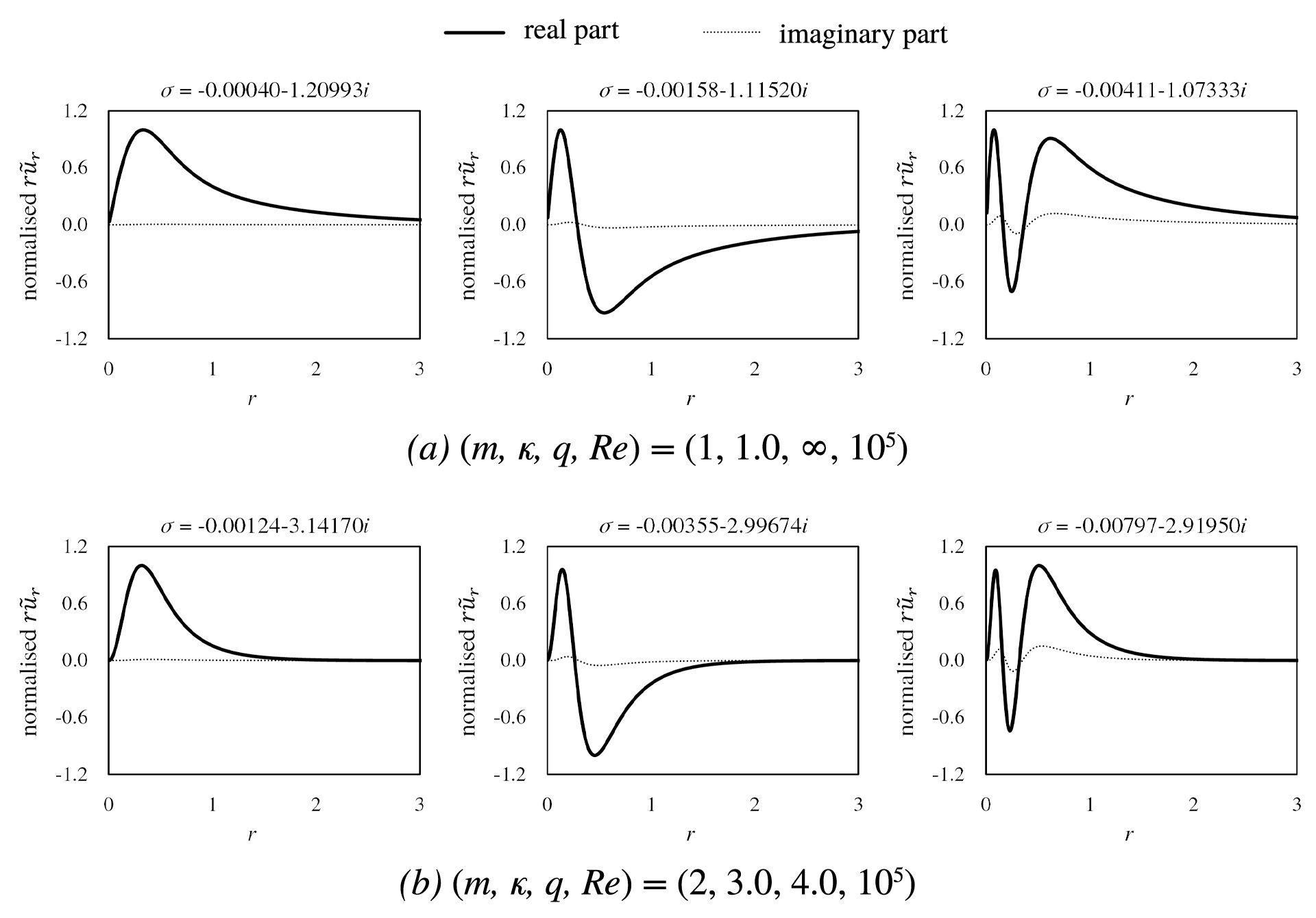}}
  \caption{Radial velocity profiles of the viscous discrete eigenmodes associated with three smallest $\Imag (\sigma)$ $(a)$ for the Lamb-Oseen vortex ($q \rightarrow \infty$) in $(m, \kappa) = (1, \; 1.0)$ and $(b)$ for the strong swirling Batchelor vortex ($q = 4.0)$ in $(m, \kappa) = (2, \; 3.0)$. The maximum of $\Real(r \Tilde{u}_r)$ is normalised to unity. $M=400$ and $L=2.0$ are used. Compare with the inviscid counterparts in figure \ref{fig:inviscid_discrete}, and notice that viscosity only marginally affects these eigenmodes.}
\label{fig:viscous_discrete}
\end{figure}

Figure \ref{fig:viscous_discrete} presents three viscous discrete eigenmodes with respect to each base flow, whose spatial structures are inherited from the inviscid discrete eigenmodes  displayed in figure \ref{fig:inviscid_discrete}. They remain non-singular throughout the whole radial domain. Viscosity only marginally affects the spatial structures of these eigenmodes compared to their inviscid counterparts, making the velocity components have slightly non-zero imaginary parts due to viscous perturbation. The number of wiggles in the eigenmodes still determines their spatial characteristics. Moreover, those eigenmodes with more wiggles near $r=0$ are more stable over time, i.e., $|\Real(\sigma)|$ increases. This phenomenon is physically justifiable since the spatial gradient of velocity components becomes steeper when the spacing between the wiggles is reduced, and viscous diffusion should, therefore, be more intensive. These eigenmodes are physical as they are regular, well-resolved solutions to the linearised Navier-Stokes equations on the $q$-vortex. They are typically characterised by modest wiggles that are spatially resolved, have rapid monotonous decay in $r$, and clearly correspond to the inviscid discrete eigenmodes associated with $\sigma_{d}^{0}$.

\subsubsection{Spurious eigenmodes}\label{viscousspuriouseigenmodes}
\begin{figure}
  % \vspace{0.1in}
  \centerline{\includegraphics[width=\textwidth,keepaspectratio]{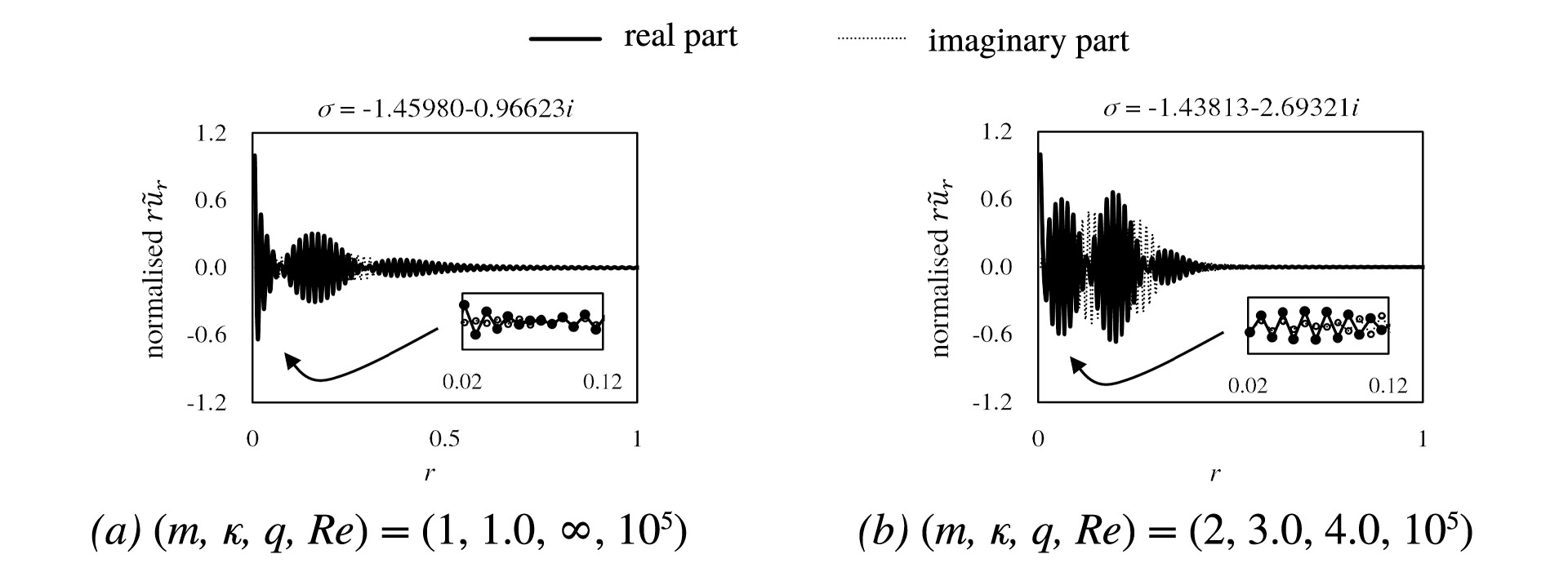}}
  \caption{Radial velocity profiles of a representative viscous spurious eigenmode $(a)$ for the Lamb-Oseen vortex ($q \rightarrow \infty$) in $(m, \kappa) = (1, \; 1.0)$ and $(b)$ for the strong swirling Batchelor vortex ($q = 4.0)$ in $(m, \kappa) = (2, \; 3.0)$. The maximum of $\Real(r \Tilde{u}_r)$ is normalised to unity. $M=400$ and $L=2.0$ are used. Non-trivial and irregularly fast oscillations with alternating sign at every collocation point, as shown in each inset for magnification, manifest that they are spurious.}
\label{fig:viscous_spurious}
\end{figure}

Two numerically computed eigenmodes that represent the viscous spurious eigenmodes are shown in figure \ref{fig:viscous_spurious}. We have not observed any signs of convergence up to $M=400$. These eigenmodes are not spatially resolved, as evidenced by irregularly fast oscillations that alternate at every collocation point. It is apparent that they are neither analytically nor physically meaningful. Therefore, we will not perform an in-depth analysis of them.

\subsubsection{Potential eigenmodes}\label{viscouspotentialeigenmodes}
\begin{figure}
  % \vspace{0.1in}
  \centerline{\includegraphics[width=\textwidth,keepaspectratio]{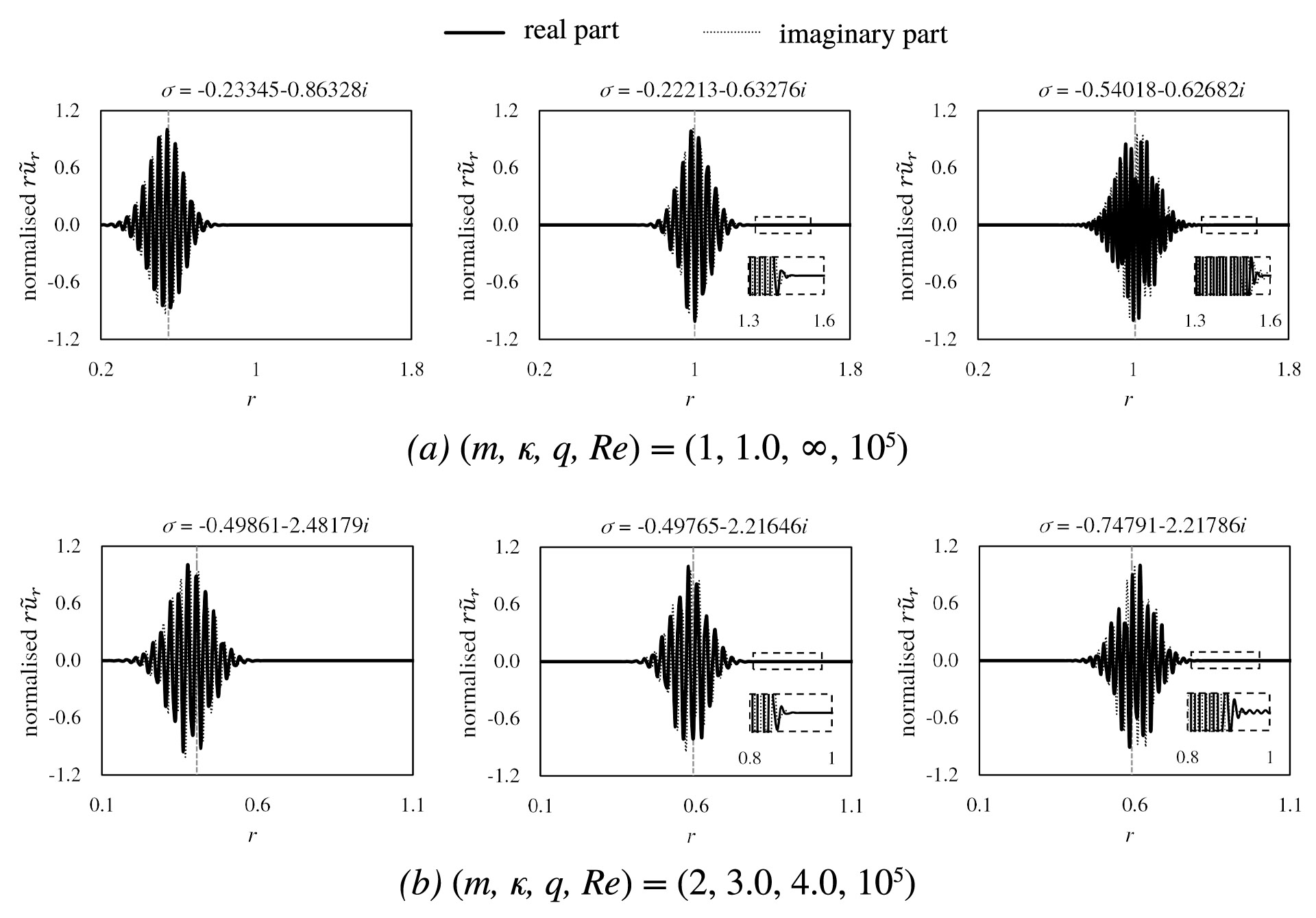}}
  \caption{Radial velocity profiles of three viscous potential eigenmodes $(a)$ for the Lamb-Oseen vortex ($q \rightarrow \infty$) in $(m, \kappa) = (1, \; 1.0)$ and $(b)$ for the strong swirling Batchelor vortex ($q = 4.0)$ in $(m, \kappa) = (2, \; 3.0)$. The maximum of $\Real(r \Tilde{u}_r)$ is normalised to unity with the use of $M=400$ and $L=2.0$. The first and middle two potential eigenmodes exhibit similar $\Real(\sigma)$, and their number of major oscillations is comparable. The middle and last two eigenmodes have similar $\Imag(\sigma)$, and their major oscillatory positions are similar. Each vertical dashed line indicates the critical layer location $r_c$, which is estimated by setting each $\Imag(\sigma)$ to $\sigma_c$ in \eqref{sigma_r_c_qvort}. Each inset within a dashed box reveals small-amplitude wiggles where $r\Tilde{u}_{r} \sim O(10^{-5})$ that persist at large $r$ even when the amplitude seems to be nearly zero, indicating their slow radial decay rates.}
\label{fig:viscous_potential}
\end{figure}

Next, we examine the numerical eigenmodes associated with $\sigma_p^{\nu}$, or the potential eigenmodes. If we look at the randomly scattered eigenvalues while increasing $M$, it is possible to observe common spatial characteristics of these eigenmodes that are spatially resolved with a sufficiently large value of $M$, unlike the \textit{spurious} family mentioned above. Figure \ref{fig:viscous_potential} presents the three representative potential eigenmodes for each vortex case, using $M=400$. We note that we have selected eigenmodes whose smallest wiggle is captured with more than two collocation points to ensure that we validly discuss their common spatial features. These eigenmodes are characterised by excessive wiggles, resulting in slow radial decay rates \citep[cf.][]{Mao2011}. They exhibit generally faster decay rates in time (i.e., larger $|\Real(\sigma)|$) than the discrete ones, as more wiggles demand steeper spatial gradients vulnerable to viscous diffusion.

The potential eigenmodes have wiggles that are usually concentrated roughly near the inviscid critical-layer singularity locations, which is estimated by setting their $\Imag(\sigma)$ to $\sigma_c$ in \eqref{sigma_r_c_qvort}. In other words, as noted by \citet{Mao2011}, they take the form of ``wavepackets,'' whose major oscillatory components are localised both in physical and spectral spaces. The correspondence between these ``wavepackets'' and the inviscid critical-layer singularity locations leads us to posit that the potential eigenmodes originate from the viscous regularisation of the critical layers. From a mathematical standpoint, the introduction of the viscous term serves only to ensure their regularisation and does not impose any restrictions on their appearance following regularisation, such as thickness and wave amplitude. This may explain why potential eigenmodes exhibit various wavepacket widths at different locations.

In figure \ref{fig:viscous_potential}, the first and second eigenmodes have similar decay rates in time, i.e., $\Real(\sigma_1) \simeq \Real(\sigma_2)$, which relates to the fact that they also have a similar number of wiggles at their major oscillatory positions. On the other hand, the second and third eigenmodes have similar wave frequencies, i.e., $\Imag(\sigma_2) \simeq \Imag(\sigma_3)$, which means that their major oscillatory locations are close. As the number of wiggles increases, $|\Real(\sigma)|$ becomes large, and the major oscillatory structure extends to a wide range in $r$. This extension likely contributes to the retardation of radial decay rates, as the wiggles remain at large radii in small scales (see the insets in figure \ref{fig:viscous_potential}).

There are several noteworthy factors that should be pointed out regarding the spatial characteristics of these eigenmodes. Although they appear physical, they make it difficult to believe that they are the \textit{true} viscous remnants of the inviscid critical-layer eigenmodes. First, potential eigenmodes' wavepackets can have varying widths even at the same $\Rey$, indicating the absence of a clear scaling relationship between wavepacket widths and the important physical parameter $\Rey$. Second, it is challenging to identify a clear spatial similarity to the inviscid critical-layer eigenmodes. The typical radial decaying behaviour of the viscous eigenmodes appears slow and oscillatory, as shown in figure \ref{fig:viscous_potential}, in contrast to the inviscid critical-layer eigenmodes that exhibit monotonically rapid radial decay (see figure \ref{fig:inviscid_critical}). We postulate that the rapid radial decaying behaviour in $\sigma_c^{0}$ must be sustained for its \textit{true} viscous remnants since the viscous regularisation effect should be highly localised around the critical-layer singularity. Therefore, a subsequent question should arise as to which other eigenmodes in the non-normal region can be considered the \textit{true} viscous remnants of $\sigma_c^{0}$. As the name suggests, we claim that the viscous critical-layer eigenmodes associated with $\sigma_c^{\nu}$ offer the answer, which we set forth in the following.

\subsubsection{Viscous critical-layer eigenmodes}\label{viscouscriticallayereigenmodes}
\begin{figure}
  % \vspace{0.1in}
  \centerline{\includegraphics[width=\textwidth,keepaspectratio]{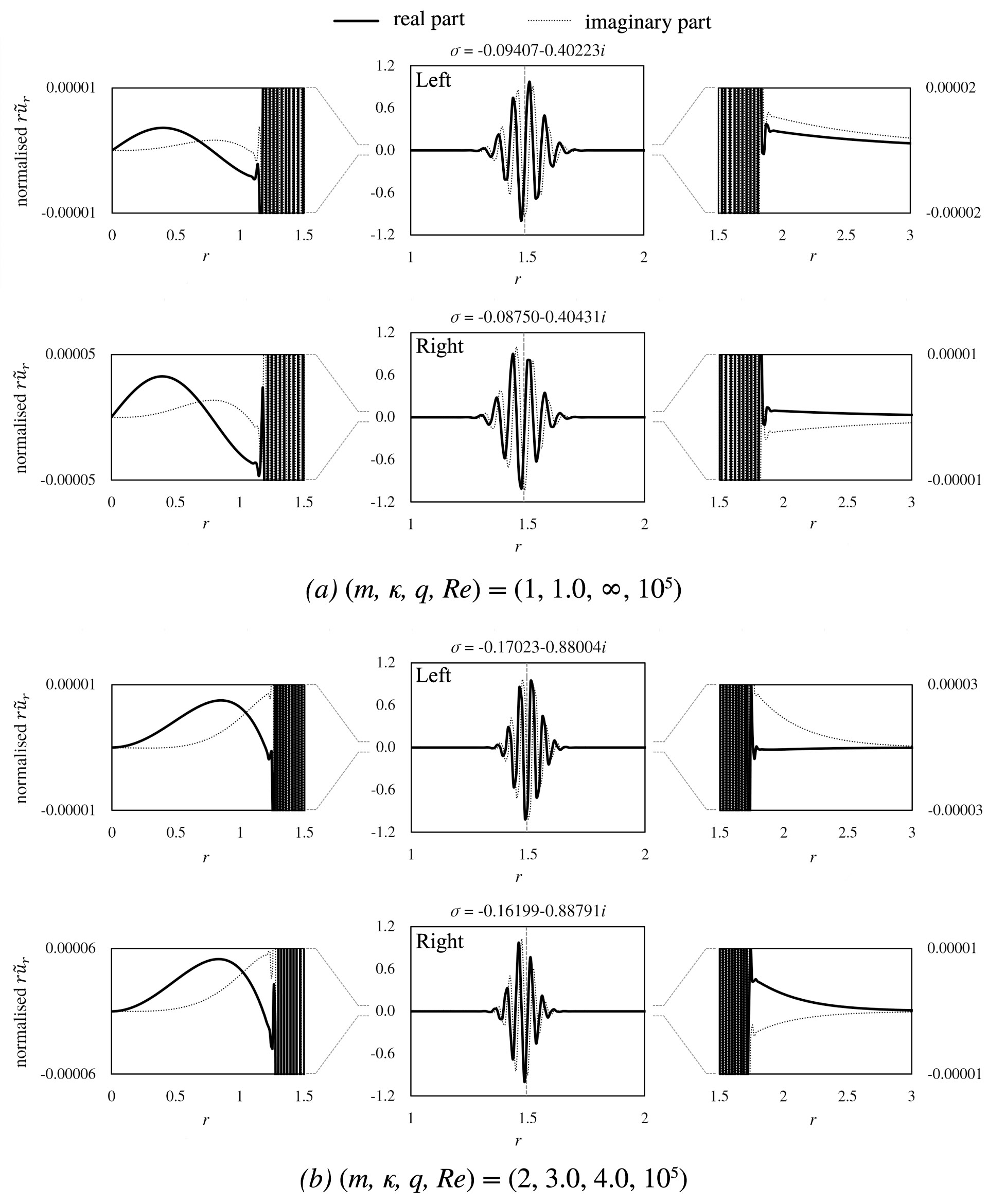}}
  \caption{Two viscous critical-layer eigenmodes with nearly identical $\Imag(\sigma)$. $(a)$ Radial component of the velocity eigenmode of the Lamb-Oseen vortex ($q \rightarrow \infty$) with $(m, \kappa) = (1, \; 1.0)$ and $(b)$ of the Batchelor vortex ($q = 4.0)$ with $(m, \kappa) = (2, \; 3.0)$. The maximum of $\Real(r \Tilde{u}_r)$ is normalised to unity. $M=400$ and $L=2.0$ are used. Each vertical dashed line indicates the location of the viscous critical layer estimated by setting $\Imag(\sigma)$ equal to $\sigma_c$ in \eqref{sigma_r_c_qvort}. Those locations are nearly equal to the centroid of the magnitude of $r \tilde{u}_r$. Due to the similarity of the shape of small-amplitude structures in the right and left columns, where $r \Tilde{u}_r \sim O(10^{-5})$, to the inviscid critical-layer eigenmodes (compare them with the middle column panels in figure~\ref{fig:inviscid_critical}$(a)$ and $(b)$, respectively), we hypothesise that these nearly degenerate viscous critical-layer eigenmodes are the viscous analogues of the inviscid two-fold degenerate critical-layer eigenmodes.}
\label{fig:viscous_critical}
\end{figure}

Figure \ref{fig:viscous_critical}$(a)$ shows two viscous critical-layer eigenmodes of a Lamb-Oseen vortex with values of $\Imag(\sigma_{c}^{\nu})$ that are within 6\% of each other. Due to the closeness of their eigenvalues and their similar appearances, we believe that they evolved from a pair of degenerate inviscid critical layer eigenmodes. The eigenvalue of the eigenmode in the upper row of figure \ref{fig:viscous_critical}$(a)$ is in the left branch, while the lower row is in the right branch of $\sigma_c^{\nu}$ in figure ~\ref{fig:spectrum_scheme} and figure~\ref{fig:viscous_vary_M}. We believe that the regions of large amplitude oscillations shown in the middle column of figure~\ref{fig:viscous_critical}$(a)$ are the \textit{true} remnants of the inviscid critical layers.

The central locations of the critical layers in the middle column of figure~\ref{fig:viscous_critical}$(a)$, which we define as the centroid of the magnitude of $r\tilde{u}_r$, are nearly equal to the inviscid critical-layer singularity locations $r_c$, as estimated by setting $\Imag(\sigma_c^{\nu})$ to $\sigma_c$ in \eqref{sigma_r_c_qvort}. An important qualitative difference from the potential eigenmodes in figure~\ref{fig:viscous_potential} and the critical-layer eigenmodes is that in the radial regions outside the large amplitude oscillations, the viscous critical-layer eigenmodes decay monotonically, while the decay of the potential eigenmodes is highly oscillatory. Figure \ref{fig:viscous_critical}$(b)$ shows two eigenmodes of the Batchelor vortex, which have similar eigenvalues (differing by only 6\%). Their properties are similar to those of the eigenmode of the Lamb-Oseen vortex.

These numerical eigenmodes and eigenvalues exhibit good convergence with increasing $M$ and are spatially resolved. For physical relevance, it is worthwhile to investigate their structures outside the remnant critical layers. By normalising the oscillation amplitude in the remnant critical layer to be of order unity, we can identify small-scale perturbation structures outside the critical layer of $O(10^{-5})$ or less. We note the similarity in shape of these small-scale perturbations to the inviscid critical-layer eigenmodes of similar $\Imag(\sigma)$ (see the middle column of panels in figure~\ref{fig:inviscid_critical}), where each part in $(0, r_c)$ and $(r_c, \infty)$ appears to be a scalar multiple of each side of the inviscid solutions (see \S \ref{sec:pairing}). This is one indication that the viscous critical-layer eigenmodes are truly inherited from the inviscid critical-layer eigenmodes. Note that viscosity has a profound influence on the structure of the eigenmode at radial locations inside the remnant critical layer, where it locally regularises the critical layer's singularity, but viscosity has only marginal impact on the eigenmode at radial locations outside the remnant critical layer. Therefore, we expect the inviscid critical-layer eigenmodes (in figure~\ref{fig:inviscid_critical}) and the viscous critical-layer eigenmodes (in figure~\ref{fig:viscous_critical}) to look similar in the regions outside the critical layer.

As can be seen in the viscous spectra, the decay rates in time of the viscous critical-layer eigenmodes are comparable to those of the viscous discrete eigenmodes, indicating that they can last for a relatively long time against viscous diffusion. Moreover, when comparing an eigenmode in the left branch with another in the right branch of $\sigma_c^{\nu}$, no notable structural difference is observed between them. This observation is further supported by the fact that these eigenmodes lie in the non-normal region. A more detailed analysis of the viscous critical-layer eigenmodes is presented later in this paper, dealing with $L$ and including the viscous remnant critical layers conforming to the $\Rey^{-1/3}$ scaling law (see \S \ref{sec:viscousresolution}) and the continuity of the viscous critical-layer spectrum $\sigma_c^{\nu}$ (see \S \ref{sec:viscouscontinuity}).

If $\sigma_c^{\nu}$ is the truly regularised descendant of $\sigma_c^{0}$ with the correct critical layer thickness, an important question that remains to be answered is how the spectrum of a single straight line bifurcates into two distinct branches. This bifurcation is physically meaningful because the separation of the branches, or equivalently the difference in $\Real(\sigma)$, is significantly larger compared to the extent of purely numerical error at the same level of $M$, such as the eigenvalue difference found in the pairing phenomenon in $\sigma_c^{0}$ (see figure~\ref{fig:inviscid_pairing}). Recall that there exist numerous singular, degenerate eigenmodes associated with the same eigenvalue due to the critical-layer singularities in $\sigma_c^{0}$. We can infer that the viscous effect perturbs these two-fold degenerate singular eigenmodes and splits them into two regularised eigenmodes with marginally different eigenvalues. Hence, we expect that the emergence of $\sigma_c^{\nu}$ in two bifurcating curvy branches is not accidental but explicable by means of perturbation theory dealing with two-fold degeneracy \citep[pp. 300-305]{sakurai_modern_2021}.

It is worth discussing why $\sigma_c^{\nu}$ was not distinguished by previous researchers. When we compare our numerical method with that of \citet{Mao2011} or \citet{Bolle2020}, we see that they truncated the radial domain at a large but finite $r$ and applied a homogeneous boundary condition there. In contrast, our method essentially involves the entire radial domain $0 \leq r < \infty$, and each basis function $P_{L_m}^n(r)$ obeys the boundary conditions that we want to apply. As a result, our truncated spectral sums, expanded by $P_{L_m}^n(r)$ as the basis elements of the Galerkin method, implicitly and exactly impose the boundary conditions on the solutions, regardless of the value of $M$ used. However, the boundary condition at $r \rightarrow \infty$ is only approximately satisfied by the others. Considering the sensitivity of the numerical spectra to numerical errors (see \S \ref{pseudospectral}), the truncation is likely to impede the numerical convergence of $\sigma_c^{\nu}$ because an approximate far-field radial boundary condition introduces errors. For instance, in the numerical spectrum plot provided by \citet[][p. 8]{Mao2011}, we can see \textit{faint} traces of the two bifurcating branches at the location of $\sigma_c^{\nu}$ found in our results. Nonetheless, the results were substantially disturbed with respect to the radial truncation as well as the number of spectral elements, and the authors could not distinguish them from $\sigma_p^{\nu}$.

\subsection{Optimal choice of $L$ to resolve the viscous critical layers}\label{sec:viscousresolution}
One of our goals is to accurately compute the viscous critical-layer modes. Clearly, we should use the largest $M$ (with $N \equiv M+2$) that our computational budget allows, which in this analysis is $M=400$. We are interested in finding all of the viscous critical layer eigenmodes and eigenvalues, not just one, nor are we interested in finding them one-at-a-time. Unlike previous studies that looked at individual eigemmodes and stretched the radial domain locally around the location of that eigenmode's critical layer to maximise the resolution there \citep[e.g.,][]{LeDizes2005}, our numerical method is designed for a fixed $\Rey$, $m$, and $\kappa$ to compute the entire radial domain for all of the eigenmodes, regardless of the locations of their critical layers, using the same radial collocation points.

\begin{figure}
  % \vspace{0.1in}
  \centerline{\includegraphics[width=\textwidth,keepaspectratio]{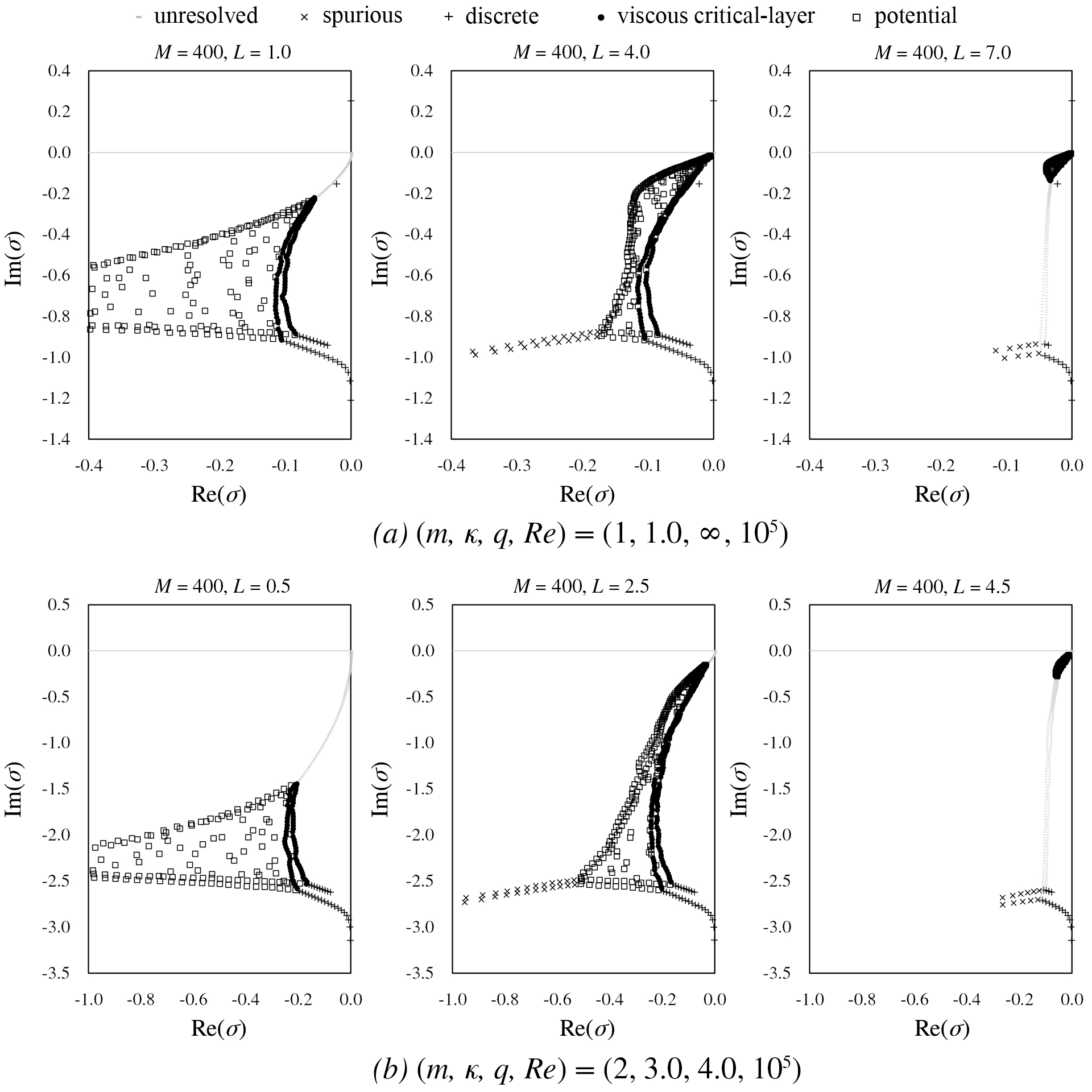}}
  \caption{Changes of numerical viscous spectra $(a)$ for the Lamb-Oseen vortex ($q\rightarrow\infty$) in $(m,\kappa)=(1,1.0)$ and $(b)$ for the strong swirling Batchelor vortex ($q=4.0$) in $(m,\kappa) = (2,3.0)$ with respect to three different $L$ values. For animation, see movie 2. $M$ is fixed at $400$ and $N=M+2$. If we aim to optimally resolve the critical-layer spectrum, we should appropriately tune $L$ to find a balance between (left) the expansion of the high-resolution region $0 \le r < L$, and (right) the deterioration of the overall resolution represented by $\Delta \sim O(L)$. The middle one shows the optimal $L$, denoted $L_{\mathrm{opt}}$, which minimises the emergence of the numerical potential spectrum. Thus, most numerical eigenvalues in the non-normal region belong to the viscous critical-layer eigenvalues. See movie~2 for animation.}
\label{fig:viscous_vary_L}
\end{figure}

Choosing a small value of $L$ is advantageous because the spatial resolution of our method is $\Delta = L/(M+2)$ (see \eqref{resolution}), and we need to have $\Delta$ smaller than the critical-layer thickness to resolve it. However, only half of the collocation points lie in the vast range between $L$ and infinity, so eigenmodes with critical-layer radii with $r_c > L$ will have few collocation points (if any) within their critical layers and therefore be spatially under-resolved. The optimal value of $L$, denoted $L_{\mathrm{opt}}$, must be a ``Goldilocks'' value: not too big or too small. Figure~\ref{fig:viscous_vary_L} demonstrates another reason why $L_{\mathrm{opt}}$ is a ``Goldilocks'' value. The figure displays the eigenvalues in the imaginary plane for $\Rey = 10^5$ with three different values of $L$. The left column of the figure represents a scenario where $L$ is small, and viscous critical-layer eigenmodes with small $r_c$ (and therefore large $|\sigma_c^{\nu}|$; see \eqref{sigma_r_c_qvort} - \eqref{sigma_r_c_lamb}) are spatially well-resolved. However, eigenmodes with small values of $|\sigma_c^{\nu}|$ and large $r_c$ are not adequately resolved. The right column of the figure shows the case with a large $L$. In this case, only eigenmodes with small $|\sigma_c^{\nu}|$ and large $r_c$ are well-resolved.

Nevertheless, figure~\ref{fig:viscous_vary_L} reveals another reason for the ``Goldilocks'' behaviour. The panels in the left column exhibit a clear separation between the potential eigenvalues $\sigma_p^{\nu}$ and the two new branches of viscous critical-layer eigenvalues $\sigma_c^{\nu}$. As $L$ increases, the potential eigenvalues shift towards the right in the complex plane (middle column). When $L$ becomes sufficiently large, the potential eigenvalues become intertwined with those of the viscous critical-layer eigenmodes (right column), and the latter set of eigenmodes are no longer well-resolved spatially. 

Upon detailed examination of the viscous critical-layer eigenmodes of the Lamb-Oseen vortex with $(m, \kappa) = (1, 1.0)$ and the Batchelor vortex with $(q = 4.0)$ and $(m, \kappa) = (2, 3.0)$, with $M = 400$ and $\Rey = 10^5$, we found that $\Delta = L/(M+2)$ is just small enough to resolve the viscous critical-layer thicknesses when $L = 4.0$ and $2.5$, respectively. Figure~\ref{fig:viscous_vary_L} illustrates that these values of $L$ also represent the maximum values where the eigenvalues of the potential eigenmodes remain distinct from those of the viscous critical-layer eigenmodes. Thus, we believe that these values of $L$ are the ``Goldilocks'' values: large enough to maximise the region $0 \le r_c < L$, providing a sufficient number of collocation points to resolve the eigenmodes, and small enough that $\Delta = L/(M+2)$ adequately resolves the critical-layer thicknesses. Our procedure for determining the optimal value $L_{\mathrm{opt}}$ is similar to how we found the optimal $L$ for resolving the inviscid critical-layer eigenvalues $\sigma_c^{0}$ in $\S$~\ref{sec:correction}:
\begin{enumerate}
    \item Start with $L$ of order unity (i.e., the core radius of the unperturbed aircraft wake vortex), and increase $L$ to expand the high-resolution region $0 \le r_c < L$.
    \item Stop increasing $L$ just before the spatial resolution is so poor that the $\sigma_p^{\nu}$ and $\sigma_c^{\nu}$ eigenvalues intertwine as shown in the middle panels of figure \ref{fig:viscous_vary_L}.
\end{enumerate}

\subsection{Use of $L_{\mathrm{opt}}$ to find the scaling behavior of the critical layer thickness with $\Rey$}
\begin{figure}
  % \vspace{0.1in}
  \centerline{\includegraphics[width=\textwidth,keepaspectratio]{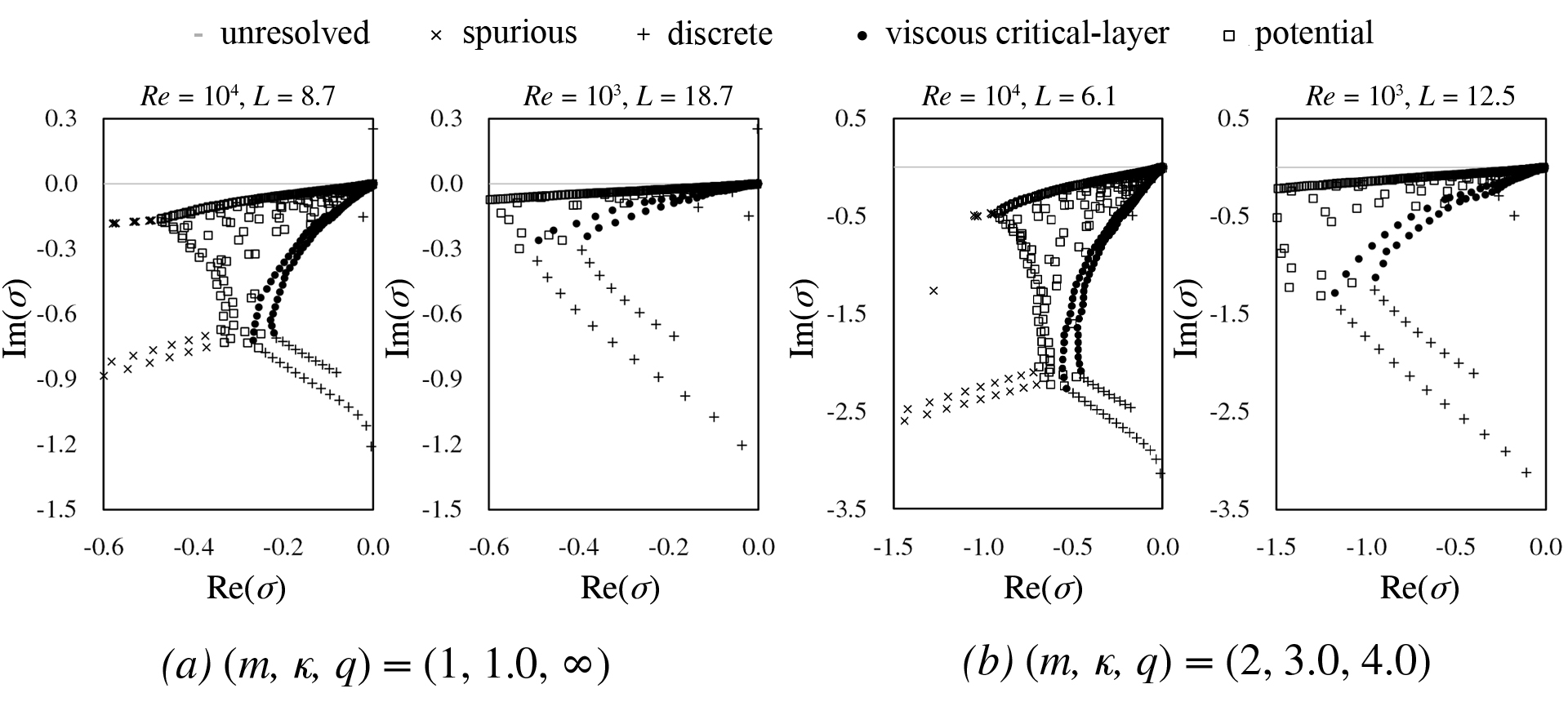}}
  \caption{Numerical viscous spectra with $L_{\mathrm{opt}}$ at $\Rey = 10^4$ and $10^3$ $(a)$ for the Lamb-Oseen vortex ($q\rightarrow\infty$) in $(m,\kappa)=(1,1.0)$ and $(b)$ for the strong swirling Batchelor vortex ($q=4.0$) in $(m,\kappa) = (2,3.0)$. $M$ is fixed at $400$ and $N=M+2$.}
\label{fig:viscous_resolution}
\end{figure}

\begin{figure}
  % \vspace{0.1in}
  \centerline{\includegraphics[width=\textwidth,keepaspectratio]{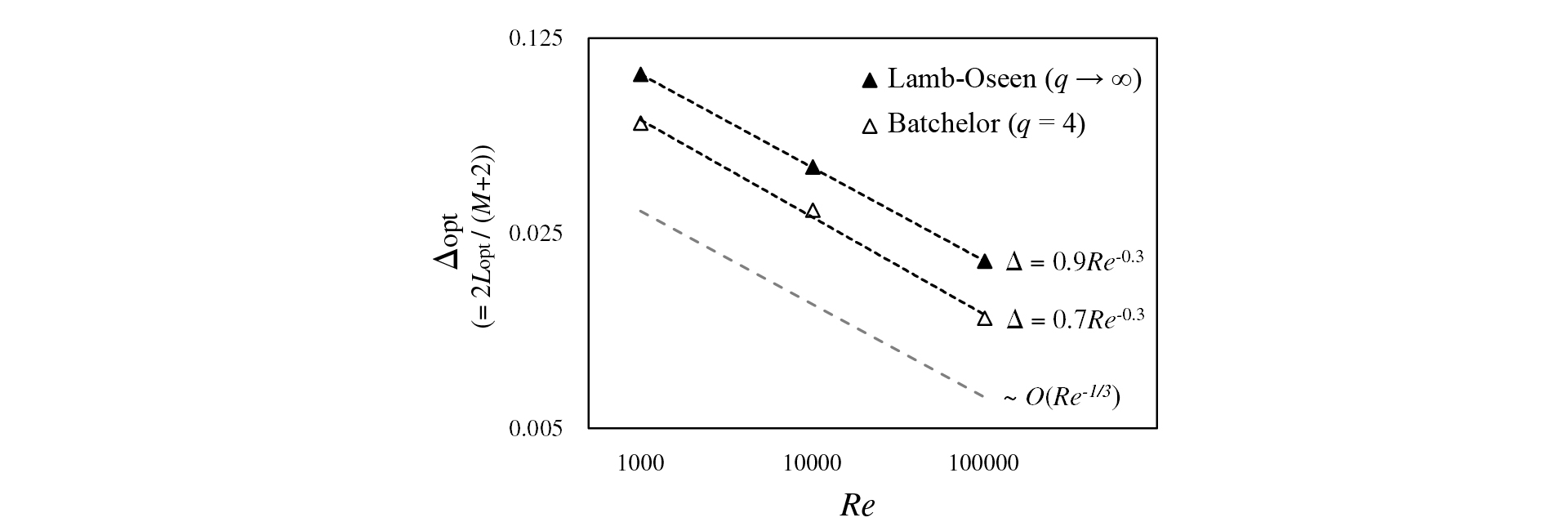}}
  \caption{The optimal numerical resolution $\Delta_{\mathrm{opt}} \equiv 2L_{\mathrm{opt}}/(M+2)$, at the fixed $M=400$, to resolve the critical-layer spectrum with respect to $\Rey$. The trend indicates $\Delta_{\mathrm{opt}} \propto \Rey^{-1/3}$. The presented cases of $\Rey = 10^3$, $10^4$ and $10^5$ for each vortex can be found in figure~\ref{fig:viscous_vary_L} and figure~\ref{fig:viscous_resolution}.
  }
\label{fig:viscous_optimal}
\end{figure}

We hypothesise that the values of $L$ at which the potential and viscous critical-layer eigenvalues intermingle in figure~\ref{fig:viscous_vary_L} and where $L/(M+2)$ just barely resolves the critical layer thickness are the same for all $\Rey$. We believe that the loss of numerical spatial resolution causes the two families of eigenvalues to become non-distinct from one another. To partially test this hypothesis, we calculate $L_{\mathrm{opt}}$ using the two-step procedure mentioned above, using the data in figure~\ref{fig:viscous_vary_L} and figure~\ref{fig:viscous_resolution}. We then assume that $\Delta_{\mathrm{opt}} \equiv L_{\mathrm{opt}}/(M+2)$ represents the critical layer thickness. Plotting $\Delta_{\mathrm{opt}}$ as a function of $\Rey$ in figure~\ref{fig:viscous_optimal} demonstrates that the critical layer thickness (if our hypothesis is correct) scales approximately as $\Rey^{-1/3}$. This scaling agrees with previous analyses using asymptotic expansions \citep{Maslowe1986, LeDizes2004}. 

\subsection{Continuity in the viscous critical-layer spectrum}\label{sec:viscouscontinuity}

\subsubsection{Pseudospectral analysis}\label{pseudospectral}
\begin{figure}
  % \vspace{0.1in}
  \centerline{\includegraphics[width=\textwidth,keepaspectratio]{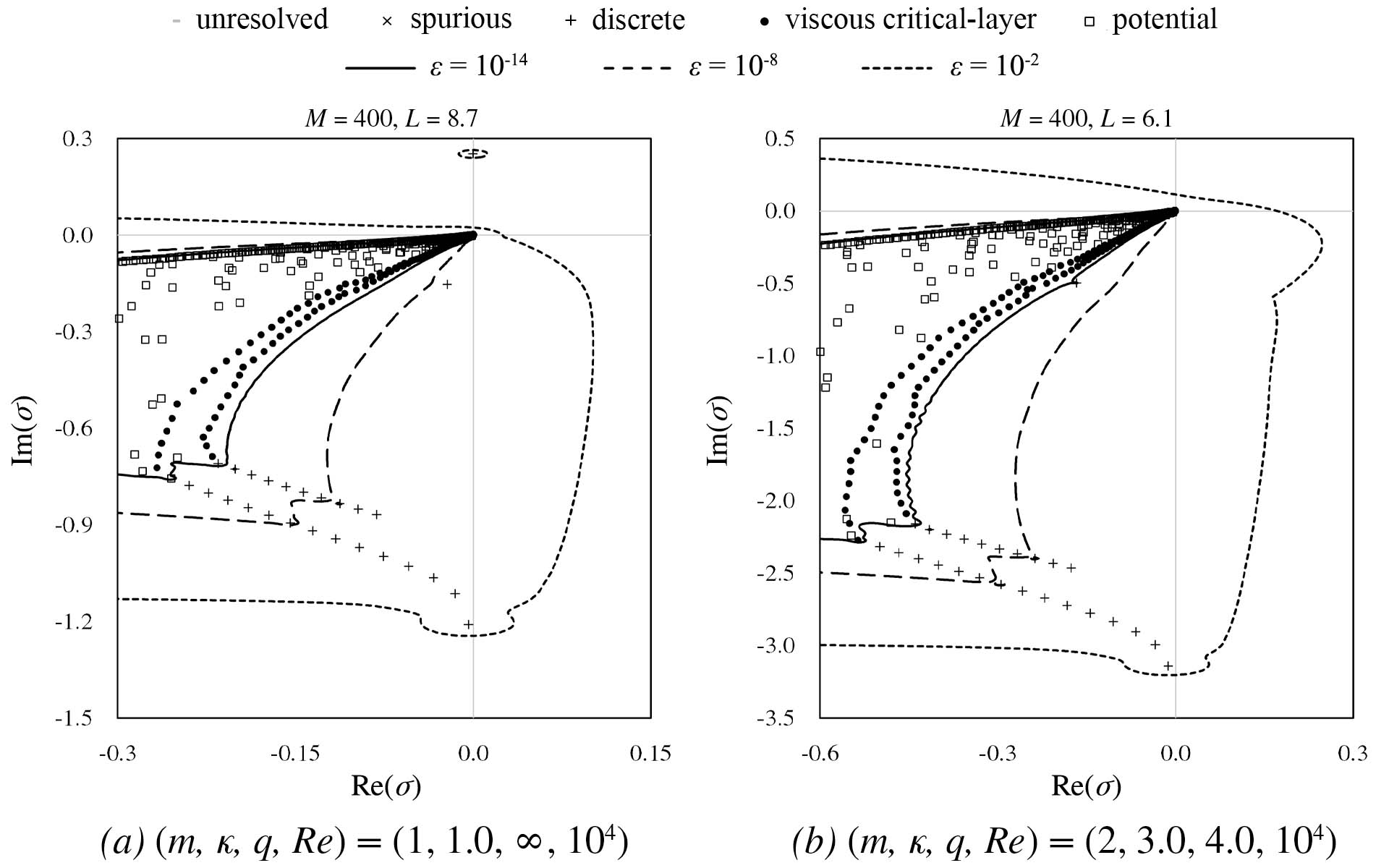}}
  \caption{$\varepsilon$-pseudospectrum bounds of $\varepsilon = 10^{-14},~ 10^{-8}$ and $10^{-2}$ with respecto to $\mathsfbi{L}_{m\kappa}$ at $\Rey=10^4$ $(a)$ for the Lamb-Oseen vortex ($q\rightarrow\infty$) in $(m,\kappa) = (1,1.0)$ and $(b)$ for the strong swirling Batchelor vortex ($q=4.0$) in $(m,\kappa) = (2,3.0)$. To construct the matrix, we use $M=400$ and $N=M+2$. $L$ is optimally chosen. We can infer from their formation which part of the spectra is continuous and how big the maximum transient growth is.}
\label{fig:viscous_pseudospectra}
\end{figure}

Finding the pseudospectra of the viscous operator $\mathsfbi{L}_{m\kappa}$, we can obtain evidence that the spectra $\sigma_p^{\nu}$ and $\sigma_c^{\nu}$ fill the continuous region in the complex $\sigma$-plane, as depicted in the schematic in figure \ref{fig:spectrum_scheme}. According to \citet{Mao2011}, the $\varepsilon$-pseudospectra around the potential and critical-layer eigenvalues seem to enclose the entire area when $\varepsilon$ is small, as shown in figure \ref{fig:viscous_pseudospectra}. In addition, we present the $\varepsilon$-pseudospectrum with $\varepsilon$ as small as $10^{-14}$, which is much smaller than the values used by \citet{Mao2011} or \citet{Bolle2020}. Therefore, we believe that our observation provides strong empirical support for the continuity of the non-normal region that we have numerically resolved.

Furthermore, based on the alternative statement of the pseudospectra given by Trefethen (2005), any point in the $\varepsilon$-pseudospectra of $\mathsfbi{L}_{m\kappa}$ can be on the spectrum of $\mathsfbi{L}_{m\kappa} + \mathsfbi{E}$ for some small disturbance $\mathsfbi{E}$ where $\left\lVert \mathsfbi{E} \right\rVert < \varepsilon$. Since $\varepsilon = 10^{-14}$ is almost comparable to the double-precision machine arithmetic used in modern computing, one possible explanation for the random scattering of the numerical eigenvalues in the numerical representation of $\sigma_{p}^{\nu}$ is that they are perturbed by machine-dependent precision errors serving as $\mathsfbi{E}$.

As an aside, we observe that the $\varepsilon$-pseudospectrum of $\varepsilon = 10^{-2}$ protrudes into the right half-plane of the complex $\sigma$-plane, as shown in figure \ref{fig:viscous_pseudospectra}. It is well-known that the supremum of the real parts of $\sigma \in \sigma_{\varepsilon}(\mathsfbi{L}_{m\kappa})$, denoted $\alpha_{\varepsilon}$ and referred to as the $\varepsilon$-pseudospectral abscissa \citep{trefethen_spectra_2005}, is relevant to the lower bound of the maximum transient growth of the stable system with an arbitrary initial state. of $\bm{x} = \bm{x}_0$ where $\left \lVert \bm{x}_0 \right \rVert = 1$,
\begin{equation}
    \frac{\partial \bm{x}}{\partial t} = \mathsfbi{L}_{m\kappa} \bm{x}.
\end{equation}
The supremum of $\alpha_{\varepsilon} / \varepsilon$ in $\varepsilon > 0$ determines the lower bound of the maximum transient growth of the system \citep{APKARIAN2020}, or 

\begin{equation}
     \sup_{t \ge 0}\left \lVert e^{\mathsfbi{L}_{m\kappa}t} \right \rVert \; \ge \; \sup_{\varepsilon > 0} \frac{\alpha_{\varepsilon (\mathsfbi{L}_{m\kappa})}}{\varepsilon} .
\end{equation}
The fact that the $\varepsilon$-pseudospectral abscissa of $\varepsilon = 10^{-2}$ occurs in the frequency band coinciding with the critical-layer spectrum implies the significance of this spectrum in regards to the transient vortex growth, which needs more investigation in further studies.

\begin{figure}
  % \vspace{0.1in}
  \centerline{\includegraphics[width=\textwidth,keepaspectratio]{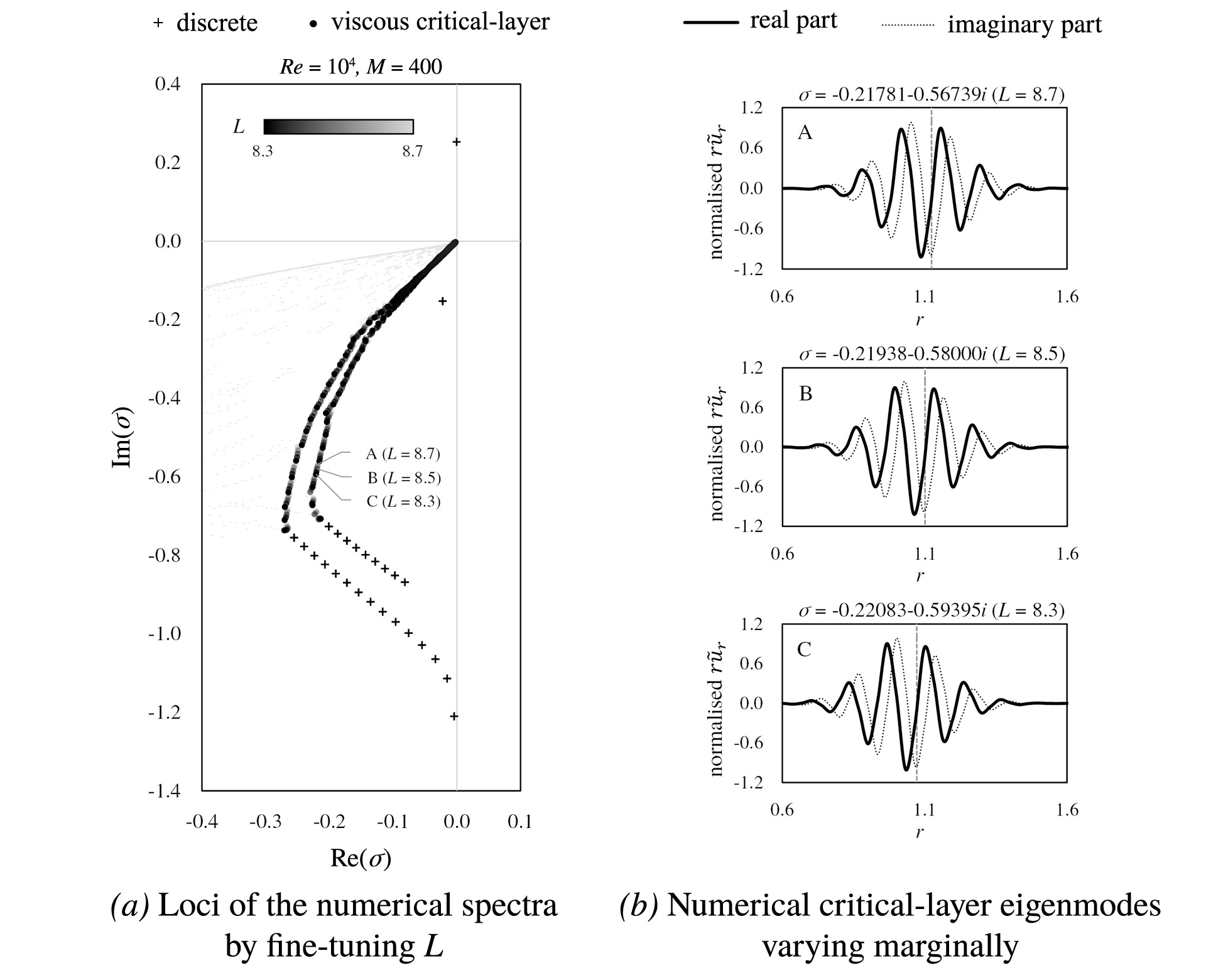}}
  \caption{$(a)$ Loci of the numerical spectra for the Lamb-Oseen vortex ($q\rightarrow\infty$) in $(m,\kappa)= (1,1.0)$ obtained by fine-tuning $L$ from $8.3$ to $8.7$, where $\Rey = 10^4$, and $(b)$ three viscous critical-layer eigenmodes that marginally vary, all of which are obtained from different $L$. Unlike the discrete spectrum that does not change with respect to $L$, the critical-layer spectrum is continuously filled by numerical eigenvalues associated with valid critical-layer eigenmodes.}
\label{fig:viscous_continuity}
\end{figure}

\subsubsection{Loci of the numerical spectra}
One issue with pseudospectra is that they cannot provide non-normal eigenmodes corresponding to each eigenvalue point in the continuum. Instead, pseudomodes can be constructed in association with pseudospectra as an approximation of the eigenmodes, which were introduced and described by \citet{trefethen_spectra_2005}. Unfortunately, pseudomodes do not necessarily satisfy the exact governing equations and boundary conditions \citep[see][p. 11]{Mao2011}.

In our numerical method, it is possible to find critical-layer eigenmodes whose spatial structures continuously vary by fine-tuning $L$. Recalling the role of $L$ (see \S\ref{numericalmethod}), we know that it changes the entire $P_{L_m}^n(r)$ in the basis function set. If we replace $L$ and solve the eigenvalue problem again, we can expect the eigenmodes generated from a new $L$ not necessarily to be identical to the eigenmodes generated from an old $L$. Moreover, if this parametric change occurs in parts of the spectra where numerical convergence with respect to $M$ is ensured, including $\sigma_d^{\nu}$ and $\sigma_{c}^{\nu}$, the loci of them with respect to $L$ should genuinely reflect the analytic spectra.

Based on the idea described above, we create the loci of the numerical spectra with respect to $L$ for the Lamb-Oseen vortex case, where $(m,~\kappa,~q) = (1,~1.0,~\infty)$ at $\Rey = 10^4$ in figure \ref{fig:viscous_continuity}$(a)$. To draw the loci, the viscous eigenvalue problem is solved multiple times with fine-tuning $L$ from $8.3$ to $8.7$ with $M = 400$, where both $\sigma_d^{\nu}$ and $\sigma_c^{\nu}$ are found to be well-resolved. The other parts of the spectra, including $\sigma_p^{\nu}$, are excluded due to no clear convergence with respect to $M$. That being said, we note that the loci of $\sigma_p^{\nu}$ with varying $L$ sweep over the shaded area depicted in the schematic in figure \ref{fig:spectrum_scheme}.

As for $\sigma_d^{\nu}$, its locus is completely invariant against changes in $L$. It makes sense because there is no chance to find an intermediate form of two discrete eigenmodes. The locus of $\sigma_d^{\nu}$ remaining discrete rather strengthens our method's robustness for any $L$. On the contrary, the locus of $\sigma_c^{\nu}$ is notably different from that of $\sigma_d^{\nu}$; as $L$ changes, the eigenvalue points on two branches of $\sigma_c^{\nu}$ also move and eventually fill in two distinct curves as in figure \ref{fig:spectrum_scheme}. In figure \ref{fig:viscous_continuity}$(b)$, it can be confirmed that the critical-layer eigenmodes with slightly different eigenvalues, having only a marginal difference in their spatial structures, are obtained from varying $L$. By comparing the two loci of $\sigma_d^{\nu}$ and $\sigma_c^{\nu}$, we can conclude the continuity of the critical-layer spectrum.

For reference, we report the polynomial fitting results up to $6^{th}$ order of some loci of $\sigma_c^{\nu}$ among what we have explored. In the case $(m, \kappa, q) = (1, 1.0, \infty)$ at $\Rey = 10^4$, the left and right branches of $\sigma_c^{\nu}$ in the complex $\sigma$-plane are fitted as
\begin{equation}
\begin{aligned}
    \sigma_r = ~& - (1.905 \times 10^1) \cdot \sigma_i^6 
             - (4.562 \times 10^1) \cdot \sigma_i^5
             - (4.138 \times 10^1) \cdot \sigma_i^4 \\ &
             - (1.741 \times 10^1) \cdot \sigma_i^3 
             - (2.761 \times 10^0) \cdot \sigma_i^2 
             + (5.348 \times 10^{-1}) \cdot \sigma_i,
\end{aligned}
\end{equation}
and
\begin{equation}
\begin{aligned}
    \sigma_r = ~& - (4.682 \times 10^0) \cdot \sigma_i^6 
             - (8.233 \times 10^0) \cdot \sigma_i^5
             - (6.816 \times 10^0) \cdot \sigma_i^4 \\ &
             - (3.243 \times 10^0) \cdot \sigma_i^3 
             - (2.636 \times 10^{-1}) \cdot \sigma_i^2 
             + (6.108 \times 10^{-1}) \cdot \sigma_i,
\end{aligned}
\end{equation}
where $\sigma_r$ and $\sigma_i$ indicate the real and imaginary parts of $\sigma$, respectively. 
In the case $(m, \kappa, q) = (2, 3.0, 4.0)$ at $\Rey = 10^4$, the left and right branches of $\sigma_c^{\nu}$ are fitted as
\begin{equation}
\begin{aligned}
    \sigma_r = ~& + (1.071 \times 10^{-2}) \cdot \sigma_i^6 
             + (2.553 \times 10^{-2}) \cdot \sigma_i^5
             - (9.022 \times 10^{-2}) \cdot \sigma_i^4 \\ &
             - (3.417 \times 10^{-1}) \cdot \sigma_i^3 
             - (1.906 \times 10^{-1}) \cdot \sigma_i^2 
             + (4.764 \times 10^{-1}) \cdot \sigma_i ,
\end{aligned}
\end{equation}
and
\begin{equation}
\begin{aligned}
    \sigma_r = ~& + (7.098 \times 10^{-2}) \cdot \sigma_i^6 
             + (4.052 \times 10^{-1}) \cdot \sigma_i^5
             + (8.111 \times 10^{-1}) \cdot \sigma_i^4 \\ &
             + (6.323 \times 10^{-1}) \cdot \sigma_i^3 
             + (2.508 \times 10^{-1}) \cdot \sigma_i^2 
             + (4.833 \times 10^{-1}) \cdot \sigma_i.
\end{aligned}
\end{equation}
We will work on the analytic formulation of $\sigma_c^{\nu}$ to better understand the bifurcation in future studies. These fitting forms will be considered for comparison and validation.

\section{Conclusion}\label{conclusion}
In this study, we proposed a numerical method that is capable of computing eigenmodes and eigenvalues for linear stability analyses of aircraft wake vortices with high time-efficiency and accuracy compared to previous studies. Also, we established a means of unambiguously verifying whether the numerically computed eigenmodes and eigenvalues are physical, spatially resolved, or spurious.

We developed a numerical method for the linear stability analysis of aircraft wake vortices, and applied this method to the $q$-vortex model, which is a non-dimensional vortex model that portrays the Lamb-Oseen or Batchelor vortices, used as the base vortex profile. Our numerical method employs algebraically mapped associated Legendre functions $P_{L_m}^n(r)$, defined in \eqref{maplegfun}, as Galerkin basis functions for the spectral expansion of functions in a radially unbounded domain. We found these basis functions to be suitable as they capture the correct boundary conditions, including analyticity at the origin and rapid decay in the far field. By applying the poloidal-toroidal decomposition to the linearised governing equations, we reduced the problem size for computation while preserving the spatial order of the equations. Furthermore, we believe that our numerical method is preferable for linear analyses of vortex dynamics for the following reasons.

\begin{enumerate}
    \item Our method, the mapped Legendre spectral collocation method, converts the original vortex stability problem into a standard matrix eigenvalue problem of toroidal and poloidal streamfunctions. In comparison to other methods that lead to a generalised matrix eigenvalue problem of primitive variables, our method effectively reduces the number of state variables of the problem from four to two, and the number of matrices constructed for eigenvalue computation from two to one.
    \item Our method does not require extra treatments for analyticity and boundary conditions in a radially unbounded domain. The use of toroidal and poloidal streamfunctions expanded by $P_{L_n}^m(r)$ guarantees that computed linear perturbation velocity fields are analytic at $r=0$ and decay to zero as $r \rightarrow \infty$. This prevents artificial interference in the problem, such as truncation of the radial domain and imposition of artificial boundary conditions at the point of truncation, which likely cause unnecessary numerical errors.
    \item Our method allocates collocation points properly around the vortex core, ensuring that half of them remain within the high-resolution region of $0 \leq r < L$ while the other half contribute to sustaining the domain's unboundedness, where $L$ is the map parameter of associated Legendre functions. In comparison to the numerical method proposed by \citet{Mayer1992}, our method requires about three times fewer radial basis elements, which is expected to result in roughly ten times greater efficiency in terms of computing time. Moreover, $L$ offers an additional degree of computational freedom, enabling us to adjust the spatial resolution without requiring extra computing resources to match the smallest radial length scale to be resolved.
\end{enumerate}

We numerically computed eigenmodes and eigenvalue spectra with azimuthal and axial wavenumbers of order unity for strong swirling $q$-vortices, and classified these eigenmodes and eigenvalue spectra into different families based on the criteria outlined in \S \ref{sec:prelim}, which determine whether they are physical, spatially resolved, or spurious. Some family, such as the \textit{freestream} family which do not decay at radial infinity, were beyond the scope of our analysis as we considered such non-vanishing solutions to be non-physical. For this reason, our method only calculates solutions that decay to zero. Our main focus was on physical eigenmodes that exist in the real world, i.e., those that can destabilise an aircraft wake vortex, with greater emphasis on critical layers. In this regard, we identified the following important families of eigenmodes and eigenvalue spectra, some of which we believe we distinguished for the first time.

\begin{enumerate}
    \item \textit{Discrete  family} (see \S \ref{invisciddiscreteeigenmodes} and \S \ref{viscousdiscreteeigenmodes}). They consist of entirely regular solutions to the linearised governing equations. Each of their eigenvalues is discrete, and approaches a fixed point as the number of spectral basis elements $M$ increases. The eigenmodes are characterised by ``wiggles'' around the vortex core, and monotonically rapid decay in the $r$ direction. All spatially resolved eigenmodes with small but finite viscosity are found to have their respective inviscid counterparts, exhibiting only marginal changes in their spatial structures. Without doubt, this family are physical.
    \item \textit{Inviscid critical-layer family} (see \S\ref{inviscidcriticallayereigenmodes}). The analytic presence of their spectrum on the imaginary axis arises from mathematical point singularities, which are given in \eqref{sigma_r_c_qvort}. Although the eigenmodes possess a critical-layer singularity, our numerical method yields well-behaved spatial structures outside the neighbourhood of the singularity when using a sufficiently large value of $M$. These structures are crucial for identifying the remnants of this family after adding small viscosity. However, their singular nature often causes the eigenmodes to be under-resolved, i.e., to have incorrect eigenvalues out of the imaginary axis, leading to a misjudgement of the wake vortex's linear instability. Adjusting the map parameter $L$ can help correct these errors so that the numerical spectrum reflects its analytic ground-truth (see \S \ref{sec:correction}). In the corrected inviscid critical-layer spectrum, eigenvalues tend to emerge in pairs. This phenomenon is understood as a marginal separation caused by numerical errors of two singular degenerate critical-layer eigensolutions, whose exact eigenvalues are supposed to be the same (see \S \ref{sec:pairing}).
    \item \textit{Potential family} (see \S \ref{viscouspotentialeigenmodes}), which were first proposed by \citet{Mao2011}. \citet[][p. 17]{Bolle2020} suggested this family be the viscous remnants of the inviscid critical-layer spectrum. The spectrum is supposed to fill continuously a portion of the left half of the complex eigenvalue plane, as depicted in the schematic in figure~\ref{fig:spectrum_scheme}. Its discretised representation can be found in our method through an area with randomly scattered numerical eigenvalues that keeps stretching out to the left as $M$ increases. We cannot establish the convergence of a particular eigenvalue to a fixed point due to the continuous nature of the spectrum. The random scattering makes it impossible to find a clear correspondence between the eigenvalue computed with $M+1$ basis elements and another computed with $M$ basis elements. Nevertheless, the eigenmodes are spatially resolved enough to identify their common spatial characteristics. They are typified by local rapid oscillations (``wavepackets'') around the corresponding critical-layer radius, estimated by setting the imaginary part of their respective eigenvalues to \eqref{sigma_r_c_qvort}. This implies that they stem from the viscous regularisation of the inviscid critical layers. Considering their uninteresting near-zero region outside the respective ``wavepackets'' together, we deem these eigenmodes to be physical. Nonetheless, the fact that their ``wavepackets'' can have varying widths even at the same $\Rey$ raises concern about the absence of a scaling relationship between wavepacket widths and $\Rey$. Moreover, their slow and oscillatory decaying behaviour does not resemble the inviscid critical-layer eigenmodes' rapid and monotonous decaying behaviour (see \S \ref{inviscidcriticallayereigenmodes}). Unlike the suggestion by \citet[][]{Bolle2020}, we argue that they do not represent the \textit{true} viscous remnants of the inviscid critical-layer family. The \textit{true} viscous remnants mean that they not only originate from the viscous regularisation but also exhibit spatial similarity to the inviscid critical-layer eigenmodes, in compliance with the $\Rey^{-1/3}$ scaling law for critical layers.
    \item \textit{Viscous critical-layer family} (see \S \ref{viscouscriticallayereigenmodes}), which are believed to be distinguished for the first time. As the name suggests, we argue that this family is the \textit{true} viscous remnants of the inviscid critical-layer spectrum. The spectrum of this family is identified near the right end of the potential spectrum as two distinct continuous curves. It shows good numerical convergence with respect to $M$, and their continuous loci are confirmed by fine-tuning $L$ (see \S \ref{sec:viscouscontinuity}). When spatially resolved, these eigenmodes exhibit thin and distinct local rapid oscillations at the inviscid critical-layer singularity radius as estimated above. This implies their origination from the viscous regularisation of the inviscid critical layers, as with the potential family. However, unlike the potential family, they are not only considered physical but also thought of as the \textit{true} viscous remnants of the inviscid critical-layer spectrum for the following reasons. First, the similarity in spatial structure to the corresponding inviscid critical-layer eigenmode is noticeable in the regions outside the critical layer. Second, the optimal resolution required to numerically compute the viscous critical-layer eigenmodes as many as possible is defined (see \S \ref{sec:viscousresolution}), providing a measure of the numerical resolution necessary to resolve the viscous critical-layer family overall. This optimal numerical resolution is found to be scaled in the order of $\Rey^{-1/3}$.
\end{enumerate}

The bifurcation of the viscous critical-layer spectrum has remained an unanswered question yet, and will be analytically examined based on our conjecture that viscosity breaks the singular degeneracies, which are numerically shown as the pairing phenomenon in the inviscid critical-layer spectrum. As the current study is limited to linear stability analyses, we plan to investigate the nonlinear or non-normal dynamics of the eigenmodes in the future. This investigation will include the triad-resonant instability among the degenerate eigenmodes and the transient growth with respect to the critical-layer eigenmodes. Moreover, we expect to use well-resolved eigenmodes, computed from the current method, as initial conditions for an initial-value problem solving the full, nonlinear governing equations of vortex motion.

\backsection[Supplementary movies]{\label{SupMat}Supplementary movies are available at [JFM].}

\backsection[Acknowledgements]{We would like to thank Jinge Wang (University of California, Berkeley) for providing discussions with respect to aircraft wake vortex instability and numerical method development.}

\backsection[Declaration of interests]{The authors report no conflict of interest.}

\backsection[Funding]{This research received no specific grant from any funding agency, commercial or not-for-profit sectors.}

\appendix
\section{Differential operators}\label{app}
\noindent For a $r -$dependent scalar function $f(r)$, the gradient and the Laplacian are 
\begin{equation}
    \bm{\nabla}_{m \kappa} f \equiv \frac{df}{dr}\bm{\hat{e}}_r + \frac{\mathrm{i}m}{r} f \bm{\hat{e}}_\phi+ \mathrm{i} \kappa f \bm{\hat{e}}_z,
    \label{mk_grad}
\end{equation}
\begin{equation}
    \nabla^2_{m \kappa} f \equiv \frac{1}{r} \frac{d}{dr} \left( r \frac{df}{dr} \right) - \frac{m^2}{r^2} f - \kappa^2 f.
    \label{mk_laplace}
\end{equation}

\noindent For a $r -$dependent vector field $\bm{F}(r)\equiv F_r (r) \bm{\hat{e}}_r + F_\phi (r) \bm{\hat{e}}_\phi + F_z (r) \bm{\hat{e}}_z $, the divergence, the curl and the vector Laplacian are
\begin{equation}
    \bm{\nabla}_{m \kappa} \cdot \bm{F} \equiv \frac{d F_r }{dr} + \frac{F_r}{r} + \frac{\mathrm{i}m}{r} F_\phi + \mathrm{i}\kappa F_z,
    \label{mk_div}
\end{equation}
\begin{equation}
\begin{aligned}
    \bm{\nabla}_{m \kappa} \times \bm{F} \equiv & \left(\frac{\mathrm{i}m}{r}F_z - \mathrm{i} \kappa F_\phi \right)\bm{\hat{e}}_r 
     + \left( \mathrm{i} \kappa F_r - \frac{dF_z}{dr} \right) \bm{\hat{e}}_\phi
    \\ & + \left( \frac{dF_\phi}{dr} + \frac{F_\phi}{r} - \frac{\mathrm{i}m}{r} F_r \right) \bm{\hat{e}}_z,
    \label{mk_curl}
\end{aligned}
\end{equation}
\begin{equation}
\begin{aligned}
    \nabla^2_{m \kappa} \bm{F} \equiv & \left( \nabla_{m\kappa}^2 F_r - \frac{F_r}{r^2} - \frac{2\mathrm{i}m}{r^2}F_\phi \right)\bm{\hat{e}}_r 
     + \left( \nabla_{m\kappa}^2 F_\phi - \frac{F_\phi}{r^2} + \frac{2\mathrm{i}m}{r^2}F_r \right) \bm{\hat{e}}_\phi 
    \\[5pt] & + \left( \nabla_{m\kappa}^2 F_z \right) \bm{\hat{e}}_z.
    \label{mk_veclaplace}
\end{aligned}
\end{equation}

\section{Analyticity at the origin}\label{app:b}
In literature studying swirling flows in a radially unbounded domain with respect to the perturbation with azimuthal wavenumber $m$ and axial wavenumber $\kappa$, i.e., $\bm{u}'= \Tilde{\bm{u}}(r;m,\kappa) e^{\mathrm{i}(m\phi + \kappa z)+\sigma t}$ and $p'= \Tilde{p}(r;m,\kappa) e^{\mathrm{i}(m\phi + \kappa z)+\sigma t}$, the boundary conditions in terms of primitive variables $(\tilde{u}_r, \tilde{u}_\phi, \tilde{u}_z, \tilde{p})$ have been typically expressed as
\begin{equation}
    \begin{cases}
    \Tilde{u}_r = \Tilde{u}_\phi = 0, ~~ \Tilde{u}_z ~\text{and}~\Tilde{p}~\text{finite}  & \text{for } m = 0 \\ 
    \dfrac{d \Tilde{u}_r}{dr} = \Tilde{u}_r + m \Tilde{u}_\phi = \Tilde{u}_z = \Tilde{p} = 0 & \text{for } |m| = 1 \\
    \Tilde{u}_r = \Tilde{u}_\phi = \Tilde{u}_z = \Tilde{p} = 0 & \text{for } |m| > 1
    \end{cases}
    ~~\text{at } r=0, ~~ \Tilde{\bm{u}}, ~\Tilde{p} \rightarrow 0 ~~ \text{as } r \rightarrow \infty.
    \label{bcs}
\end{equation}
These conditions were first suggested by \citet{Batchelor1962} and the detailed derivation can be found in \citet[][pp. 339-342]{Ash1995}. Our numerical method naturally complies with the far-field condition as all spectral basis elements, $P_{L_n}^{m}(r)$, are designed to vanish at radial infinity. Additionally, our method's handling of velocity functions at the origin not only meets the centerline condition given above, but also leads to a more accurate function behaviour. This is verified in the following.

The derivation of the centerline condition begins with
\begin{equation}
    \lim_{r \rightarrow 0} \frac{\partial \bm{u}'}{\partial \phi} = 0,
    \label{coordsing_1}
\end{equation}
to remove the coordinate singularity at $r=0$, ensuring smoothness. As the pressure term is implicit in our formulation, it is excluded from consideration. The term-by-term expression of \eqref{coordsing_1} is
\begin{equation}
    -\mathrm{i}m \Tilde{u}_r + \Tilde{u}_\phi = -\mathrm{i}\Tilde{u}_r + m \Tilde{u}_\phi = m \Tilde{u}_z = 0 ~~~ \text{as}~r\rightarrow 0.
    \label{coordsing_2}
\end{equation}
With the additional condition $d\tilde{u}_r /dr = d\tilde{u}_\phi /dr = 0$ for $|m|=1$ \citep{Mayer1992, Ash1995, Bolle2020}, which is independent of \eqref{coordsing_1} and from the regularity of the governing equations around $r=0$, the final formula is obtained.

In our numerical approach, the toroidal $\tilde{\psi} (r;m,\kappa)$ and poloidal $\tilde{\chi} (r;m,\kappa)$ streamfunctions are chosen as the state variables of the eigenvalue problem and are expanded by the mapped Legendre functions, both of which behave $O \left( r^{|m|+2s} \right)$ for a non-negative integer $s$ as $r \rightarrow 0$ \citep[see][]{Matsushima1995, Matsushima1997}. That is, in our numerical method, it is guaranteed that as $r\rightarrow 0$, these streamfunctions are expressed in power series as 
\begin{equation}
    \Tilde{\psi} (r;m,\kappa) = a_0  r^{|m|} + a_1 r^{|m|+2} + \cdots, ~~~ \Tilde{\chi} (r;m,\kappa) = b_0 r^{|m|} + b_1 r^{|m|+2} + \cdots,
    \label{tppower}
\end{equation}
where all coefficients are finite constants, as in \eqref{expand_near_0}. From the decomposition, it is known that
\begin{equation}
    \Tilde{u}_r = \frac{\mathrm{i}m}{r} \Tilde{\psi} + \mathrm{i}\kappa \frac{\partial \Tilde{\chi}}{\partial r},~~
    \Tilde{u}_\phi = -\frac{\partial \Tilde{\psi}}{\partial r} - \frac{\kappa m}{r} \Tilde{\chi},~~
    \Tilde{u}_z = -\frac{1}{r}\frac{\partial}{\partial r}\Bigl( r \frac{\partial \Tilde{\psi}}{\partial r}  \Bigl) + \frac{m^2}{r^2} \Tilde{\psi}.
    \label{tpprim}
\end{equation}
Therefore, our method ensures that as $r\rightarrow 0$,
\begin{equation}
\begin{aligned}
    \Tilde{u}_r = &~\bigl( \mathrm{i}a_0m + \mathrm{i}b_0 \kappa |m| \bigl) r^{|m|-1} + \bigl( \mathrm{i}a_1m + \mathrm{i}b_1 \kappa (|m|+2) \bigl) r^{|m|+1} + \cdots, \\
    \Tilde{u}_\phi = &~\bigl( -a_0|m| - b_0 \kappa m \bigl) r^{|m|-1} + \bigl( -a_1(|m|+2) - b_1 \kappa m \bigl) r^{|m|+1} + \cdots, \\
    \Tilde{u}_z = &~a_1\bigl( -(|m|+2)^2 +  m^2 \bigl) r^{|m|} + \cdots.
    \label{primpower}
\end{aligned}
\end{equation}
These power series satisfy \eqref{coordsing_2} for all $m$, which can be shown by simply putting \eqref{primpower} into \eqref{coordsing_2}. This verifies that the mapped Legendre expansion of the poloidal and toroidal streamfunctions, as in \eqref{tppower}, meets the centerline condition of the primitive velocity components, as in \eqref{bcs}.

The power series expansion in \eqref{primpower} ultimately stands for the analyticity at the origin, providing more accurate constraints for smoothness on the coordinate singularity. The typical centerline condition is not a sufficient condition for smoothness due to the lack of derivative constraints, as seen in \eqref{coordsing_1}, even requiring an additional condition for some cases. Correctly removing coordinate singularities in spectral methods has been known to be crucial for the accuracy of the spectral representation, which can be done by choosing appropriate basis spectral elements with regards to what coordinate singularity is in consideration \citep{Orszag1974,Bouaoudia1991,Matsushima1995,Matsushima1997}. General Chebyshev or Legendre spectral methods that do not implicitly take into account such analyticity issue, thus necessitating an explicit boundary condition to mimic the analyticity, might not be the suitable choice for systems with coordinate singularities to achieve fast spectral convergence \citep[see][]{Gottlieb1977, Boyd2000}. We note two papers \citep{Vasil2016,Vasil2019} that looked at a variety of spectral methods dealing with coordinate singularities and gave evidence to support the use of the mapped associated Legendre functions for the cylindrical coordinate singularity.

\bibliographystyle{jfm}
\bibliography{jfm_awv_p1}

\begin{thebibliography}{68}
\expandafter\ifx\csname natexlab\endcsname\relax\def\natexlab#1{#1}\fi
\def\au#1{#1} \def\ed#1{#1} \def\yr#1{#1}\def\at#1{#1}\def\jt#1{\textit{#1}}
  \def\bt#1{#1}\def\bvol#1{\textbf{#1}} \def\vol#1{#1} \def\pg#1{#1}
  \def\publ#1{#1}\def\arxiv#1{#1}\def\org#1{#1}\def\st#1{\textit{#1}}

\bibitem[Apkarian \& Noll(2020)]{APKARIAN2020}
{\sc \au{Apkarian, P.} \& \au{Noll, D.}} \yr{2020}  \at{Optimizing the {Kreiss}
  constant}.  \jt{SIAM Journal on Control and Optimization}  \bvol{58}~(6),
  \pg{3342--3362}.

\bibitem[Ash \& Khorrami(1995)]{Ash1995}
{\sc \au{Ash, R.~L.} \& \au{Khorrami, M.~R.}} \yr{1995}  \at{Vortex stability}.
   \bt{In {\em Fluid Vortices\/} (ed. \ed{S.~I. Green})},  \pg{pp. 317--372}.
  \publ{Dordrecht: Springer Netherlands}.

\bibitem[Batchelor(1964)]{Batchelor1964}
{\sc \au{Batchelor, G.~K.}} \yr{1964}  \at{Axial flow in trailing line
  vortices}.  \jt{Journal of Fluid Mechanics}  \bvol{20}~(4),  \pg{645--658}.

\bibitem[Batchelor \& Gill(1962)]{Batchelor1962}
{\sc \au{Batchelor, G.~K.} \& \au{Gill, A.~E.}} \yr{1962}  \at{Analysis of the
  stability of axisymmetric jets}.  \jt{Journal of Fluid Mechanics}
  \bvol{14}~(04),  \pg{529}.

\bibitem[B{\"{o}}lle {\em et~al.\/}(2021)B{\"{o}}lle, Brion, Robinet, Sipp \&
  Jacquin]{Bolle2020}
{\sc \au{B{\"{o}}lle, T.}, \au{Brion, V.}, \au{Robinet, J.-C.}, \au{Sipp, D.}
  \& \au{Jacquin, L.}} \yr{2021}  \at{On the linear receptivity of trailing
  vortices}.  \jt{Journal of Fluid Mechanics}  \bvol{908},  \pg{A8}.

\bibitem[Bouaoudia \& Marcus(1991)]{Bouaoudia1991}
{\sc \au{Bouaoudia, S.} \& \au{Marcus, P.~S.}} \yr{1991}  \at{Fast and accurate
  spectral treatment of coordinate singularities}.  \jt{Journal of
  Computational Physics}  \bvol{96}~(1),  \pg{217--223}.

\bibitem[Boyd(2001)]{Boyd2000}
{\sc \au{Boyd, J.~P.}} \yr{2001} {\em {Chebyshev and Fourier Spectral
  Methods}\/}, 2nd edn.  \publ{Mineola, NY: Dover Publications}.

\bibitem[Breitsamter(2011)]{Breitsamter2011}
{\sc \au{Breitsamter, C.}} \yr{2011}  \at{Wake vortex characteristics of
  transport aircraft}.  \jt{Progress in Aerospace Sciences}  \bvol{47}~(2),
  \pg{89--134}.

\bibitem[Bristol {\em et~al.\/}(2004)Bristol, Ortega, Marcus \&
  Sava\c{s}]{Bristol2004}
{\sc \au{Bristol, R.~L.}, \au{Ortega, J.~M.}, \au{Marcus, P.~S.} \&
  \au{Sava\c{s}, {\"{O}}.}} \yr{2004}  \at{On cooperative instabilities of
  parallel vortex pairs}.  \jt{Journal of Fluid Mechanics}  \bvol{517},
  \pg{331--358}.

\bibitem[Canuto {\em et~al.\/}(1988)Canuto, Hussaini, Quarteroni \&
  Zang]{Canuto1988}
{\sc \au{Canuto, C.}, \au{Hussaini, M.~Y.}, \au{Quarteroni, A.} \& \au{Zang,
  T.~A.}} \yr{1988} {\em {Spectral Methods in Fluid Dynamics}\/}, 1st edn.
  \publ{Berlin, Heidelberg: Springer Berlin Heidelberg}.

\bibitem[Case(1960)]{Case1960}
{\sc \au{Case, K.~M.}} \yr{1960}  \at{Stability of inviscid plane {Couette}
  flow}.  \jt{Physics of Fluids}  \bvol{3}~(2),  \pg{143--148}.

\bibitem[Chandrasekhar(1981)]{Chandrasekhar1981}
{\sc \au{Chandrasekhar, S.}} \yr{1981} {\em {Hydrodynamic and Hydromagnetic
  Stability}\/}, {D}over edn.  \publ{Mineola, NY: Dover Publications}.

\bibitem[Crow(1970)]{Crow1970}
{\sc \au{Crow, S.~C.}} \yr{1970}  \at{Stability theory for a pair of trailing
  vortices}.  \jt{AIAA Journal}  \bvol{8}~(12),  \pg{2172--2179}.

\bibitem[Crow \& Bate(1976)]{Crow1976}
{\sc \au{Crow, S.~C.} \& \au{Bate, E.~R.}} \yr{1976}  \at{Lifespan of trailing
  vortices in a turbulent atmosphere}.  \jt{Journal of Aircraft}
  \bvol{13}~(7),  \pg{476--482}.

\bibitem[Drazin \& Reid(2004)]{Drazin2004}
{\sc \au{Drazin, P.~G.} \& \au{Reid, W.~H.}} \yr{2004} {\em {Hydrodynamic
  Stability}\/}, 2nd edn.  \publ{Cambridge, UK: Cambridge University Press}.

\bibitem[Eisen {\em et~al.\/}(1991)Eisen, Heinrichs \& Witsch]{Eisen1991}
{\sc \au{Eisen, H.}, \au{Heinrichs, W.} \& \au{Witsch, K.}} \yr{1991}
  \at{Spectral collocation methods and polar coordinate singularities}.
  \jt{Journal of Computational Physics}  \bvol{96}~(2),  \pg{241--257}.

\bibitem[Fabre \& Jacquin(2004)]{fabre2004}
{\sc \au{Fabre, D.} \& \au{Jacquin, L.}} \yr{2004}  \at{Viscous instabilities
  in trailing vortices at large swirl numbers}.  \jt{Journal of Fluid
  Mechanics}  \bvol{500}~(500),  \pg{239--262}.

\bibitem[Fabre {\em et~al.\/}(2006)Fabre, Sipp \& Jacquin]{Fabre2006}
{\sc \au{Fabre, D.}, \au{Sipp, D.} \& \au{Jacquin, L.}} \yr{2006}  \at{Kelvin
  waves and the singular modes of the {Lamb-Oseen} vortex}.  \jt{Journal of
  Fluid Mechanics}  \bvol{551},  \pg{235--274}.

\bibitem[Feys \& Maslowe(2014)]{Feys2014}
{\sc \au{Feys, J.} \& \au{Maslowe, S.~A.}} \yr{2014}  \at{Linear stability of
  the {Moore-Saffman} model for a trailing wingtip vortex}.  \jt{Physics of
  Fluids}  \bvol{26}~(2),  \pg{024108}.

\bibitem[Feys \& Maslowe(2016)]{Feys2016}
{\sc \au{Feys, J.} \& \au{Maslowe, S.~A.}} \yr{2016}  \at{Elliptical
  instability of the {Moore-Saffman} model for a trailing wingtip vortex}.
  \jt{Journal of Fluid Mechanics}  \bvol{803},  \pg{556--590}.

\bibitem[Gallay \& Smets(2020)]{Gallay2020}
{\sc \au{Gallay, T.} \& \au{Smets, D.}} \yr{2020}  \at{Spectral stability of
  inviscid columnar vortices}.  \jt{Analysis \& PDE}  \bvol{13}~(6),
  \pg{1777--1832}.

\bibitem[Gottlieb \& Orszag(1977)]{Gottlieb1977}
{\sc \au{Gottlieb, D.} \& \au{Orszag, S.~A.}} \yr{1977} {\em {Numerical
  Analysis of Spectral Methods}\/}.  \publ{Philadelphia, PA: Society for
  Industrial and Applied Mathematics}.

\bibitem[Grosch \& Salwen(1978)]{Grosch1978}
{\sc \au{Grosch, C.~E.} \& \au{Salwen, H.}} \yr{1978}  \at{The continuous
  spectrum of the {Orr-Sommerfeld} equation. {Part 1.} {The} spectrum and the
  eigenfunctions}.  \jt{Journal of Fluid Mechanics}  \bvol{87}~(1),
  \pg{33--54}.

\bibitem[Hagan \& Priede(2013)]{Hagan2013}
{\sc \au{Hagan, J.} \& \au{Priede, J.}} \yr{2013}  \at{Capacitance matrix
  technique for avoiding spurious eigenmodes in the solution of hydrodynamic
  stability problems by {Chebyshev} collocation method}.  \jt{Journal of
  Computational Physics}  \bvol{238},  \pg{210--216}.

\bibitem[Hallock \& Holz{\"{a}}pfel(2018)]{Hallock2018}
{\sc \au{Hallock, J.~N.} \& \au{Holz{\"{a}}pfel, F.}} \yr{2018}  \at{A review
  of recent wake vortex research for increasing airport capacity}.
  \jt{Progress in Aerospace Sciences}  \bvol{98},  \pg{27--36}.

\bibitem[Heaton(2007)]{Heaton2007}
{\sc \au{Heaton, C.~J.}} \yr{2007}  \at{Centre modes in inviscid swirling flows
  and their application to the stability of the {Batchelor} vortex}.
  \jt{Journal of Fluid Mechanics}  \bvol{576},  \pg{325--348}.

\bibitem[Heaton \& Peake(2007)]{HEATON2007b}
{\sc \au{Heaton, C.~J.} \& \au{Peake, N.}} \yr{2007}  \at{Transient growth in
  vortices with axial flow}.  \jt{Journal of Fluid Mechanics}  \bvol{587},
  \pg{271--301}.

\bibitem[Howard \& Gupta(1962)]{Howard1962}
{\sc \au{Howard, L.~N.} \& \au{Gupta, A.~S.}} \yr{1962}  \at{On the
  hydrodynamic and hydromagnetic stability of swirling flows}.  \jt{Journal of
  Fluid Mechanics}  \bvol{14}~(3),  \pg{463--476}.

\bibitem[Ivers(1989)]{Ivers1989}
{\sc \au{Ivers, D.~J.}} \yr{1989}  \at{On generalised toroidal-poloidal
  solutions of vector field equations}.  \jt{The Journal of the Australian
  Mathematical Society. Series B. Applied Mathematics}  \bvol{30}~(4),
  \pg{436--449}.

\bibitem[Jacobs \& Durbin(1998)]{Jacobs1998}
{\sc \au{Jacobs, R.~G.} \& \au{Durbin, P.~A.}} \yr{1998}  \at{Shear sheltering
  and the continuous spectrum of the {Orr-Sommerfeld} equation}.  \jt{Physics
  of Fluids}  \bvol{10}~(8),  \pg{2006--2011}.

\bibitem[Jones(2008)]{JONES200845}
{\sc \au{Jones, C.~A.}} \yr{2008}  \at{Dynamo theory}.  \bt{In {\em
  Dynamos\/}},  \st{Les Houches},  \vol{vol.~88},  \pg{pp. 45--135}.
  \publ{Amsterdam, NL: Elsevier}.

\bibitem[Kelvin(1880)]{Kelvin1880}
{\sc \au{Kelvin, L.}} \yr{1880}  \at{Vibrations of a columnar vortex}.  \jt{The
  London, Edinburgh, and Dublin Philosophical Magazine and Journal of Science}
  \bvol{10}~(61),  \pg{155--168}.

\bibitem[Khorrami(1991)]{Khorrami1991}
{\sc \au{Khorrami, M.~R.}} \yr{1991}  \at{On the viscous modes of instability
  of a trailing line vortex}.  \jt{Journal of Fluid Mechanics}  \bvol{225},
  \pg{197--212}.

\bibitem[Khorrami {\em et~al.\/}(1989)Khorrami, Malik \& Ash]{Khorrami1989}
{\sc \au{Khorrami, M.~R.}, \au{Malik, M.~R.} \& \au{Ash, R.~L.}} \yr{1989}
  \at{Application of spectral collocation techniques to the stability of
  swirling flows}.  \jt{Journal of Computational Physics}  \bvol{81}~(1),
  \pg{206--229}.

\bibitem[{Le Diz{\`{e}}s}(2004)]{LeDizes2004}
{\sc \au{{Le Diz{\`{e}}s}, S.}} \yr{2004}  \at{Viscous critical-layer analysis
  of vortex normal modes}.  \jt{Studies in Applied Mathematics}
  \bvol{112}~(4),  \pg{315--332}.

\bibitem[{Le Diz{\'{e}}s} \& Lacaze(2005)]{LeDizes2005}
{\sc \au{{Le Diz{\'{e}}s}, S.} \& \au{Lacaze, L.}} \yr{2005}  \at{An asymptotic
  description of vortex kelvin modes}.  \jt{Journal of Fluid Mechanics}
  \bvol{542},  \pg{69--96}.

\bibitem[Leibovich(1978)]{Leibovich1978}
{\sc \au{Leibovich, S.}} \yr{1978}  \at{Structure of vortex breakdown}.
  \jt{Annual Review of Fluid Mechanics}  \bvol{10}~(1),  \pg{221--246}.

\bibitem[Leibovich \& Stewartson(1983)]{Stewartson1983}
{\sc \au{Leibovich, S.} \& \au{Stewartson, K.}} \yr{1983}  \at{A sufficient
  condition for the instability of columnar vortices}.  \jt{Journal of Fluid
  Mechanics}  \bvol{126},  \pg{335--356}.

\bibitem[Lessen {\em et~al.\/}(1974)Lessen, Singh \& Paillet]{Lessen1974}
{\sc \au{Lessen, M.}, \au{Singh, P.~J.} \& \au{Paillet, F.}} \yr{1974}  \at{The
  stability of a trailing line vortex. {Part} 1. {I}nviscid theory}.
  \jt{Journal of Fluid Mechanics}  \bvol{63}~(4),  \pg{753--763}.

\bibitem[Leweke {\em et~al.\/}(2016)Leweke, {Le Diz{\`{e}}s} \&
  Williamson]{Leweke2016}
{\sc \au{Leweke, T.}, \au{{Le Diz{\`{e}}s}, S.} \& \au{Williamson, C. H.~K.}}
  \yr{2016}  \at{Dynamics and instabilities of vortex pairs}.  \jt{Annual
  Review of Fluid Mechanics}  \bvol{48}~(1),  \pg{507--541}.

\bibitem[Lin(1955)]{lin1955theory}
{\sc \au{Lin, C.-C.}} \yr{1955} {\em {The Theory of Hydrodynamic Stability}\/},
  1st edn.  \publ{Cambridge, UK: Cambridge University Press}.

\bibitem[Lin(1961)]{Lin1961}
{\sc \au{Lin, C.-C.}} \yr{1961}  \at{Some mathematical problems in the theory
  of the stability of parallel flows}.  \jt{Journal of Fluid Mechanics}
  \bvol{10}~(3),  \pg{430--438}.

\bibitem[Lopez {\em et~al.\/}(2002)Lopez, Marques \& Shen]{Lopez2002}
{\sc \au{Lopez, J.~M.}, \au{Marques, F.} \& \au{Shen, J.}} \yr{2002}  \at{An
  efficient spectral-projection method for the {Navier–Stokes} equations in
  cylindrical geometries}.  \jt{Journal of Computational Physics}
  \bvol{176}~(2),  \pg{384--401}.

\bibitem[Mao \& Sherwin(2011)]{Mao2011}
{\sc \au{Mao, X.} \& \au{Sherwin, S.}} \yr{2011}  \at{Continuous spectra of the
  {Batchelor} vortex}.  \jt{Journal of Fluid Mechanics}  \bvol{681},
  \pg{1--23}.

\bibitem[Mao \& Sherwin(2012)]{Mao2012}
{\sc \au{Mao, X.} \& \au{Sherwin, S.~J.}} \yr{2012}  \at{Transient growth
  associated with continuous spectra of the {Batchelor} vortex}.  \jt{Journal
  of Fluid Mechanics}  \bvol{697},  \pg{35--59}.

\bibitem[Marcus {\em et~al.\/}(2015)Marcus, Pei, Jiang, Barranco, Hassanzadeh
  \& Lecoanet]{Marcus2015}
{\sc \au{Marcus, P.~S.}, \au{Pei, S.}, \au{Jiang, C.-H.}, \au{Barranco, J.~A.},
  \au{Hassanzadeh, P.} \& \au{Lecoanet, D.}} \yr{2015}  \at{Zombie vortex
  instability: {I.} {A} purely hydrodynamic instability to resurrect the dead
  zones of protoplanetary disks}.  \jt{The Astrophysical Journal}
  \bvol{808}~(1),  \pg{87}.

\bibitem[Maslowe(1986)]{Maslowe1986}
{\sc \au{Maslowe, S.~A.}} \yr{1986}  \at{Critical layers in shear flows.}
  \jt{Annual Review of Fluid Mechanics}  \bvol{1},  \pg{405--432}.

\bibitem[Matsushima \& Marcus(1995)]{Matsushima1995}
{\sc \au{Matsushima, T.} \& \au{Marcus, P.~S.}} \yr{1995}  \at{A spectral
  method for polar coordinates}.  \jt{Journal of Computational Physics}
  \bvol{120}~(2),  \pg{365--374}.

\bibitem[Matsushima \& Marcus(1997)]{Matsushima1997}
{\sc \au{Matsushima, T.} \& \au{Marcus, P.~S.}} \yr{1997}  \at{A spectral
  method for unbounded domains}.  \jt{Journal of Computational Physics}
  \bvol{137}~(2),  \pg{321--345}.

\bibitem[Maxworthy {\em et~al.\/}(1985)Maxworthy, Hopfinger \&
  Redekopp]{Maxworthy1985}
{\sc \au{Maxworthy, T.}, \au{Hopfinger, E.~J.} \& \au{Redekopp, L.~G.}}
  \yr{1985}  \at{Wave motions on vortex cores}.  \jt{Journal of Fluid
  Mechanics}  \bvol{151},  \pg{141}.

\bibitem[Mayer \& Powell(1992)]{Mayer1992}
{\sc \au{Mayer, E.~W.} \& \au{Powell, K.~G.}} \yr{1992}  \at{Viscous and
  inviscid instabilities of a trailing vortex}.  \jt{Journal of Fluid
  Mechanics}  \bvol{245},  \pg{91--114}.

\bibitem[McFadden {\em et~al.\/}(1990)McFadden, Murray \&
  Boisvert]{McFadden1990}
{\sc \au{McFadden, G.~B.}, \au{Murray, B.~T.} \& \au{Boisvert, R.~F.}}
  \yr{1990}  \at{Elimination of spurious eigenvalues in the {Chebyshev} tau
  spectral method}.  \jt{Journal of Computational Physics}  \bvol{91}~(1),
  \pg{228--239}.

\bibitem[Moore \& Saffman(1973)]{Moore1973}
{\sc \au{Moore, D.~W.} \& \au{Saffman, P.~G.}} \yr{1973}  \at{Axial flow in
  laminar trailing vortices}.  \jt{Proceedings of the Royal Society of London.
  A. Mathematical and Physical Sciences}  \bvol{333}~(1595),  \pg{491--508}.

\bibitem[Moore \& Saffman(1975)]{Moore1975}
{\sc \au{Moore, D.~W.} \& \au{Saffman, P.~G.}} \yr{1975}  \at{The instability
  of a straight vortex filament in a strain field}.  \jt{Proceedings of the
  Royal Society of London. A. Mathematical and Physical Sciences}
  \bvol{346}~(1646),  \pg{413--425}.

\bibitem[Orszag(1974)]{Orszag1974}
{\sc \au{Orszag, S.~A.}} \yr{1974}  \at{Fourier series on spheres}.
  \jt{Monthly Weather Review}  \bvol{102}~(1),  \pg{56--75}.

\bibitem[Press {\em et~al.\/}(2007)Press, Teukolsky, Vetterling \&
  Flannery]{press_numerical_2007}
{\sc \au{Press, W.~H.}, \au{Teukolsky, S.~A.}, \au{Vetterling, W.~T.} \&
  \au{Flannery, B.~P.}} \yr{2007} {\em {Numerical Recipes: The Art of
  Scientific Computing}\/}, 3rd edn.  \publ{Cambridge, UK: Cambridge University
  Press}.

\bibitem[Qiu {\em et~al.\/}(2021)Qiu, Cheng, Xu, Xiang \& Liu]{Qiu2021}
{\sc \au{Qiu, S.}, \au{Cheng, Z.}, \au{Xu, H.}, \au{Xiang, Y.} \& \au{Liu, H.}}
  \yr{2021}  \at{On the characteristics and mechanism of perturbation modes
  with asymptotic growth in trailing vortices}.  \jt{Journal of Fluid
  Mechanics}  \bvol{918},  \pg{A41}.

\bibitem[Roy \& Subramanian(2014)]{Roy2014}
{\sc \au{Roy, A.} \& \au{Subramanian, G.}} \yr{2014}  \at{Linearized
  oscillations of a vortex column: {The} singular eigenfunctions}.  \jt{Journal
  of Fluid Mechanics}  \bvol{741},  \pg{404--460}.

\bibitem[Saffman(1993)]{Saffman1993}
{\sc \au{Saffman, P.~G.}} \yr{1993} {\em {Vortex Dynamics}\/}, 1st edn.
  \publ{New York City, NY: Cambridge University Press}.

\bibitem[Sakurai \& Napolitano(2021)]{sakurai_modern_2021}
{\sc \au{Sakurai, J.~J.} \& \au{Napolitano, J.}} \yr{2021} {\em {Modern Quantum
  Mechanics}\/}, 3rd edn.  \publ{Cambridge, UK: Cambridge University Press}.

\bibitem[Smith(2003)]{smith_using_2003}
{\sc \au{Smith, D.~M.}} \yr{2003}  \at{Using multiple-precision arithmetic}.
  \jt{Computing in Science \& Engineering}  \bvol{5}~(4),  \pg{88--93}.

\bibitem[Spalart(1998)]{Spalart1998}
{\sc \au{Spalart, P.~R.}} \yr{1998}  \at{Airplane trailing vortices}.
  \jt{Annual Review of Fluid Mechanics}  \bvol{30}~(1),  \pg{107--138}.

\bibitem[Trefethen \& Embree(2005)]{trefethen_spectra_2005}
{\sc \au{Trefethen, L.~N.} \& \au{Embree, M.}} \yr{2005} {\em {Spectra and
  Pseudospectra: The Behavior of Nonnormal Matrices and Operators}\/}.
  \publ{Princeton, NJ: Princeton University Press}.

\bibitem[Tsai \& Widnall(1976)]{Tsai1976}
{\sc \au{Tsai, C.~Y.} \& \au{Widnall, S.~E.}} \yr{1976}  \at{The stability of
  short waves on a straight vortex filament in a weak externally imposed strain
  field}.  \jt{Journal of Fluid Mechanics}  \bvol{73}~(4),  \pg{721--733}.

\bibitem[Vasil {\em et~al.\/}(2016)Vasil, Burns, Lecoanet, Olver, Brown \&
  Oishi]{Vasil2016}
{\sc \au{Vasil, G.~M.}, \au{Burns, K.~J.}, \au{Lecoanet, D.}, \au{Olver, S.},
  \au{Brown, B.~P.} \& \au{Oishi, J.~S.}} \yr{2016}  \at{Tensor calculus in
  polar coordinates using {J}acobi polynomials}.  \jt{Journal of Computational
  Physics}  \bvol{325},  \pg{53--73}.

\bibitem[Vasil {\em et~al.\/}(2019)Vasil, Lecoanet, Burns, Oishi \&
  Brown]{Vasil2019}
{\sc \au{Vasil, G.~M.}, \au{Lecoanet, D.}, \au{Burns, K.~J.}, \au{Oishi, J.~S.}
  \& \au{Brown, B.~P.}} \yr{2019}  \at{Tensor calculus in spherical coordinates
  using {J}acobi polynomials. {Part-I:} {Mathematical} analysis and
  derivations}.  \jt{Journal of Computational Physics: X}  \bvol{3},
  \pg{100013}.

\bibitem[Widnall(1975)]{Widnall1975}
{\sc \au{Widnall, S.~E.}} \yr{1975}  \at{The structure and dynamics of vortex
  filaments}.  \jt{Annual Review of Fluid Mechanics}  \bvol{7}~(1),
  \pg{141--165}.

\bibitem[Zebib(1987)]{Zebib1987}
{\sc \au{Zebib, A.}} \yr{1987}  \at{Removal of spurious modes encountered in
  solving stability problems by spectral methods}.  \jt{Journal of
  Computational Physics}  \bvol{70}~(2),  \pg{521--525}.

\end{thebibliography}

\end{document}